\documentclass[rmp,nofootinbib,endfloats]{revtex4}
\usepackage{graphics}
\usepackage{epsfig}

\def\inbar{\,\vrule height1.5ex width.4pt depth0pt}
\def\IR{\relax{\rm I\kern-.18em R}}
\def\IC{\relax\hbox{$\inbar\kern-.3em{\rm C}$}}

\begin{document}
\title{The $s$ Process: Nuclear Physics, Stellar Models, Observations}

\author{F. K\"appeler}
\email{franz.kaeppeler@kit.edu}
\affiliation{Karlsruhe Institute of Technology, Campus Nord, Institut f\"{u}r Kernphysik, 76021 Karlsruhe, Germany}
\author{R. Gallino}
\email{gallino@ph.unito.it}
\affiliation{Dipartimento di Fisica Generale, Universit{\`a} di Torino I-10125 Torino, Italy, 
   and INAF-Osservatorio Astronomico di Teramo, I-64100 Teramo, Italy}
\author{S. Bisterzo}
\email{bisterzo@ph.unito.it}
\affiliation{Dipartimento di Fisica Generale, Universit{\`a} di
Torino, I-10125 Torino, Italy }
\author{Wako Aoki}
\email{aoki.wako@nao.ac.jp}
\affiliation{National Astronomical Observatory, Mitaka, Tokyo 181-8588, Japan}

\begin{abstract}  
Nucleosynthesis in the $s$ process takes place in the He 
burning layers of low mass AGB stars and during the He and C 
burning phases of massive stars. The $s$ process contributes about half of 
the element abundances between Cu and Bi in solar system material. Depending on stellar
mass and metallicity the resulting $s$-abundance patterns
exhibit characteristic features, which provide comprehensive 
information for our understanding of the stellar life cycle 
and for the chemical evolution of galaxies. The rapidly growing 
body of detailed abundance observations, in particular 
for AGB and post-AGB stars, for objects in binary systems, and 
for the very faint metal-poor population represents exciting 
challenges and constraints for stellar model calculations. 
Based on updated and improved nuclear physics data for the
$s$-process reaction network, current models are aiming at 
ab initio solution for the stellar physics related to convection
and mixing processes. Progress in the intimately related areas 
of observations, nuclear and atomic physics, and stellar modeling
is reviewed and the corresponding interplay is illustrated by the general 
abundance patterns of the elements beyond iron and by the effect of 
sensitive branching points along the $s$-process path. The strong variations of the $s$-process 
efficiency with metallicity bear also interesting consequences for 
Galactic chemical evolution.
\end{abstract}                                                                 
\maketitle
\tableofcontents

\section{Introduction \label{sec1}}

Neutron capture nucleosynthesis during stellar He burning contributes about half of 
the elemental abundances between Fe and Bi. Substantial progress in the quantitative 
description of this slow neutron capture ($s$) process has been achieved by an 
interdisciplinary approach involving improved nuclear physics input, advanced stellar 
model codes, and a wealth of data from astronomical observations and from the analysis 
of circumstellar dust grains. These three topics and their mutual connections are 
briefly described, before each topic is addressed in detail in Secs. II to IV. In the 
final section the synergies between the main topics are outlined with particular 
emphasis on the open quests and future prospects of this field. 

\begin{figure}[tb]
\includegraphics[width=12cm]{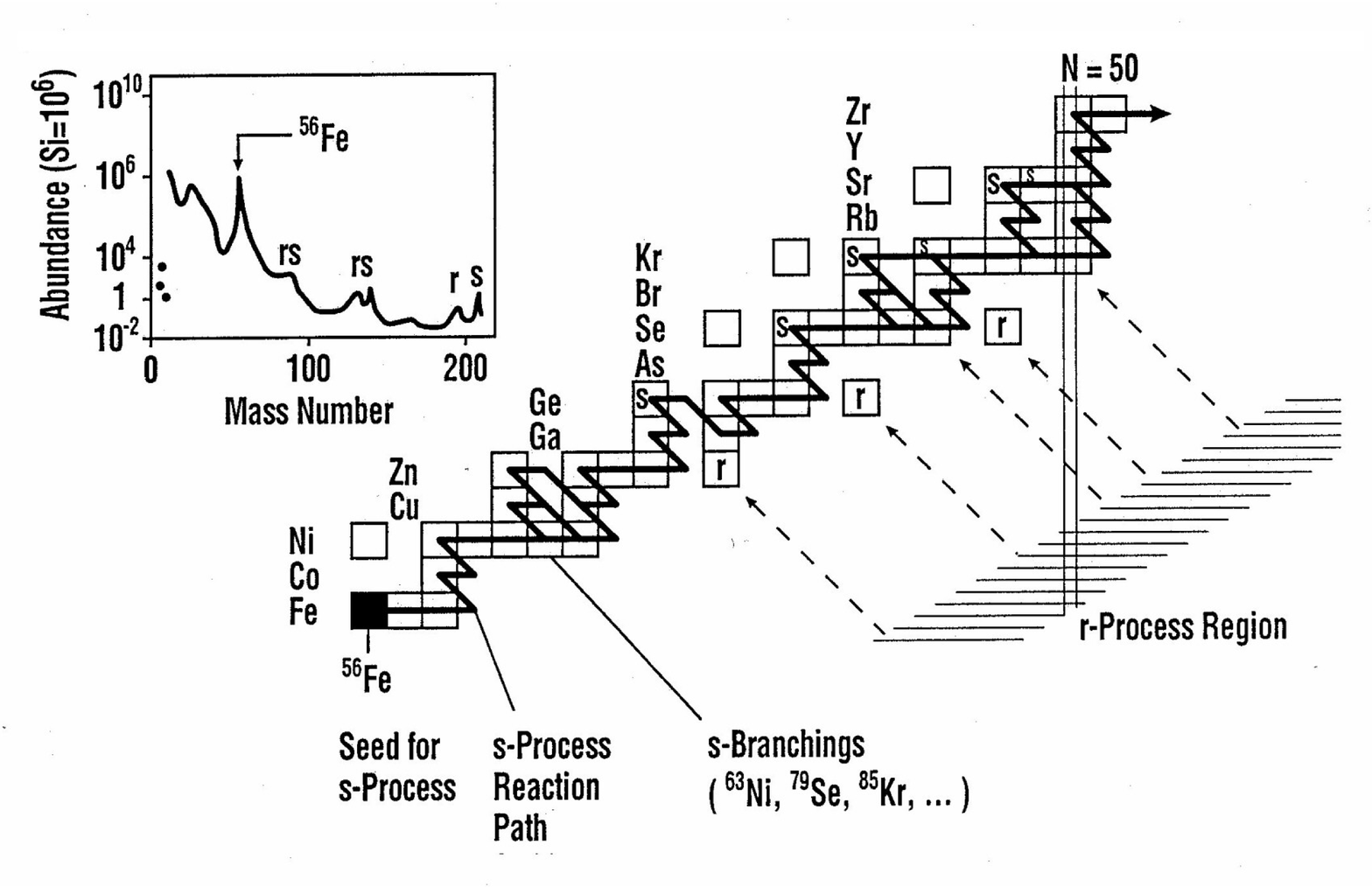}
      \caption{\label{fig:1} An illustration of the neutron capture processes 
            responsible for the formation of the nuclei between iron 
            and the actinides. 
            The observed abundance distribution in the inset shows 
            characteristic twin peaks. These result from the nuclear 
            properties where the $s$- and $r$-reaction paths encounter
	    magic neutron numbers. Note that a $p$ process has to be 
	    invoked for producing the proton rich nuclei that are not
	    reached by neutron capture reactions. (For details see
	    discussion in text.)}
\end{figure}

The phenomenological picture of the classical $s$ process was formulated about 50 years 
ago in the seminal papers of \textcite{BBF57} 
(hereafter referred to as B$^2$FH) and of \textcite{Cam57}, where the entire 
$s$-process panorama was already sketched in its essential parts. These ideas were 
worked out in the following decades by \textcite{CFH61}, \textcite{SFC65}, 
\textcite{ClR67}, and \textcite{ClW74}. The distinction of a slow and a rapid ($s$ and 
$r$) neutron capture process follows from the isotopic pattern in the chart of nuclides
(Fig. \ref{fig:1}), which shows that the $s$ process follows the stability valley 
because the neutron capture time scale is slower than that for $\beta$ decay. The 
neutron-rich isotopes outside the $s$ path are ascribed to the $r$ process, which 
occurs under explosive conditions, presumably in supernovae. The decay of the 
reaction products from the $r$-process path on the far neutron-rich side of the 
stability valley forms the $r$-only isotopes. It also contributes to most of the 
other isotopes, except for those, which are shielded by stable isobars. The
corresponding ensembles of $s$- and $r$-only isotopes are important for the
separation of the respective abundance distributions. The subset of stable 
isotopes on the proton-rich side are ascribed to the $p$ process, which is 
likely to occur also in supernova explosions \cite{ArG03}. With a few 
exceptions, the $p$ abundances are much smaller than the $s$ and $r$ 
components.

The decisive role of nuclear physics for a quantitative model
of the $s$ process was clearly expressed already by B$^2$FH.
In spite of the fact that neither the neutron source reactions 
nor the neutron capture cross sections in the astrophysically 
relevant energy range were known apart from some scatterted and 
uncertain information, all essential features had been inferred 
from this meager information: The product of the stellar ($n, 
\gamma$) cross sections and of the resulting $s$ abundances, 
$\langle \sigma \rangle N_s$, which represents the reaction flow, 
was found to be a smooth function of mass number $A$. From the 
composite slope of this function, the two different $s$ processes
had already been postulated. The steep decline between $A \approx 63$ 
and 100 was interpreted as the result of an $s$-process site 
with not enough neutrons available per $^{56}$Fe seed to build 
the nuclei to their saturation abundances. In the mass region 
beyond $A \approx 100$, the much smaller slope was suggesting that 
steady flow was achieved and that all of these nuclei reached their 
saturation abundances. It was concluded that "two different 
processes might have occurred in two different red-giant stars
(B$^2$FH)". 

In the 1990s, however, improvements in the accuracy of the nuclear
input data revealed that the classical $s$ process suffered from 
inconsistencies in the description of the abundance signatures in 
$s$-process branchings. Because such patterns are particularly 
sensitive to neutron density and temperature, this implied that
these parameters were not constant in time as assumed in the 
formulation of the classical model \cite{Kae99}. 

A few years before, stellar models of the He burning stages of 
stellar evolution started to provide an increasingly realistic 
picture of $s$-process nucleosynthesis. The prospects of this 
approach were clearly superior to that of the classical model 
because it could be directly linked to astronomical observations. 

A first generation of models \cite{HoI88,GBP88} was soon replaced 
by scenarios related to core He \cite{Heg06, Lim06} and shell C 
burning \cite{RBG91b,RGB93,LSC00} in massive stars for the weak 
$s$ process on the one hand, and to thermally pulsing low mass 
asymptotic giant branch (AGB) stars \cite{SGB95,GAB98,AKW99b}
for the main $s$ process on the other hand. The current status
of AGB evolution includes phenomena such as hot bottom burning, 
the ab initio treatment of third dredge up and related mixing
processes as well as the effect of metallicity and initial 
stellar mass \cite{Her05}. The latter aspects are particularly 
important for the discussion of the $s$-process 
component in galactic chemical evolution \cite{TGA04}. 

The success of the stellar models could be impressively verified by 
comparison with the solar $s$ component and with a large body of data 
obtained from analyses of presolar material in form of refractive
dust grains of circumstellar origin \cite{Zin98}.

With respect to the origin of the heavy elements, observations 
of $s$-process abundances in AGB stars began 1952 with the discovery 
of Tc lines in red giant stars of spectral type S by 
\textcite{Mer52b}. Ever since, spectral observations of peculiar red giants 
turned out to be a prolific source of $s$-process information for 
the He burning stage of stellar evolution \cite{Gus89}. Spectroscopy 
of astronomical objects has made spectacular progress
over the last decades by the deployment of new telescopes on the ground 
and in space and by the impressive increase in computing power, that 
led to enormous improvement in the modeling of stellar atmospheres 
and in synthetic spectrum calculations \cite{Asp05}. Our understanding 
of stellar and galactic evolution has been promoted accordingly, 
e.g. via refined studies along the AGB 
\cite{Lam91,LSB95,Her05}. Separation of the $s$ and $r$ components 
in solar material through careful evaluation of the $s$ 
abundances \cite{AKW99b} provided the key for the abundance 
distributions in the oldest, very metal-poor stars in the Universe
\cite{CoS06,CHB01,SCB98}, which were found to scale with the solar $r$-process 
distribution \cite{BeC05}. The composition of planetary nebulae \cite{PeB94} 
and circumstellar envelopes \cite{Hab96} could be investigated by 
IR observations, while the composition of interstellar matter
\cite{SaS96} is inferred from UV absorption line diagnostics. At 
higher energies, X-ray \cite{FRA97} and $\gamma$-ray astronomy 
\cite{Die98} have produced exciting new vistas of explosive 
nucleosynthesis \cite{CLT92,DBB97}. 
 
The three aspects of $s$-process research are addressed in the following 
sections with an attempt to illustrate their mutual connections.

\section{Nuclear physics \label{sec2}}

The discussion of the nuclear part concentrates on the neutron
capture reactions and $\beta$-decay rates needed to calculate the
$s$ abundances between Cu and Bi. For a summary of the charged 
particle reactions, which are of key importance for energy and 
neutron production at the various $s$-process sites and for further 
stellar evolution see the compiled data of the NACRE collaboration 
\cite{AAR99} and of \textcite{IDS01} as well as recent work on 
$^{12}$C \cite{AFL06,PHK05} and on the two major neutron source reactions 
$^{13}$C($\alpha, n$)$^{16}$O \cite{HKT08} and $^{22}$Ne($\alpha, 
n$)$^{25}$Mg \cite{JKM01}.

In their discussion of the $s$-abundance characteristics B$^2$FH  
noted that more detailed conclusions were impeded by the lack of reliable 
neutron capture cross sections and emphasized that "unambiguous 
results would be obtained by measuring the total absorption cross 
sections. It is our view that such measurements would serve as 
a crucial test of the validity of the $s$ process." 

\begin{figure}
\includegraphics[width=12cm]{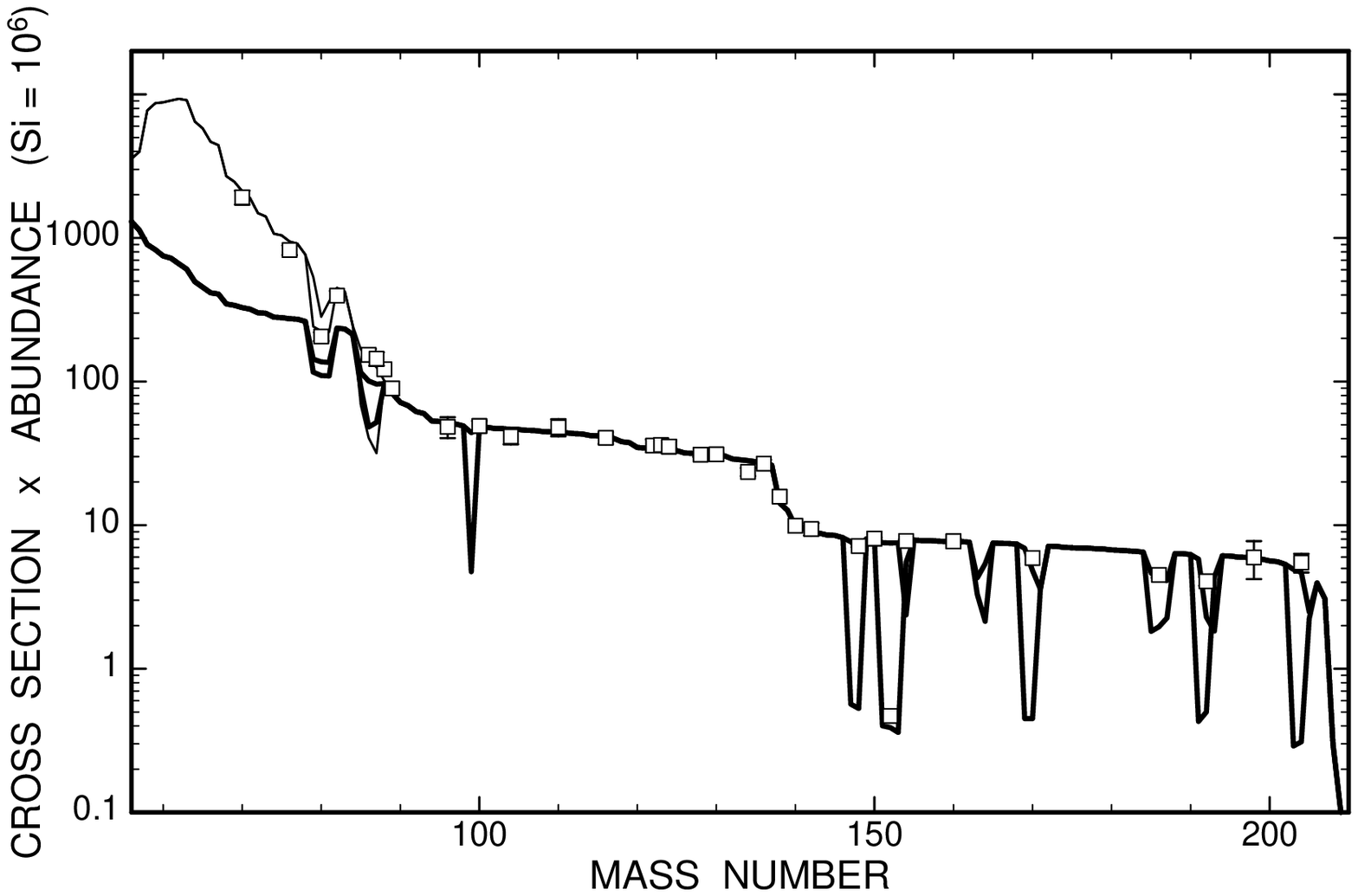}
\caption{\label{fig:2} The characteristic product of cross section times 
		    $s$-process abundance, $\langle \sigma \rangle N_s$, 
			plotted as a function of mass number. The thick solid line 
			represents the main component obtained by means of the 
			classical model, and the thin line corresponds to the weak 
			component in massive stars (see text). Symbols denote the 
			empirical products for the $s$-only nuclei. Some important 
			branchings of the neutron capture chain are indicated
            as well.}
\end{figure}

The importance of a complete set of experimental data for the 
reliable description of the $\langle \sigma \rangle N_s$-curve is 
illustrated in Fig. \ref{fig:2}, which corresponds to the situation 
obtained by the time of the cross section compilation of \textcite{BBK00}, 
when experimentally determined cross sections were available for the
majority of the involved isotopes.

Apart from the clear separation of the two $s$-process components,
Fig. \ref{fig:2} also shows the pronounced effect of $s$-process 
branchings, which could not be addressed by B$^2$FH simply because 
the data at the time were far too uncertain to reveal their signatures in the 
$\langle \sigma \rangle N_s$-curve. These branchings are the result 
of the competition between neutron capture and $\beta$-decay at 
unstable isotopes in the half-life range from a few weeks to a few 
years. For the same reason, also the effect of stellar temperature 
on the $\beta$-decay half-lives had not been anticipated.  
$^{79}$Se represents such an example, where the drastically reduced 
stellar half-life gives rise to a pronounced branching that can be
characterized by the strongly different $\langle \sigma \rangle 
N_s$-values of the $s$-only isotopes $^{80}$Kr and $^{82}$Kr
\citep{KlK88}.

This section on the nuclear physics of the $s$ process starts
with a summary of current techniques for the experimental
determination of stellar neutron capture rates (Sec. \ref{sec2A}), 
followed by the theoretical aspects, which have to be considered 
in the step from laboratory measurements to stellar applications 
(Sec. \ref{sec2B}). The problems related to the often dramatic 
enhancement of  $\beta$-decay rates under stellar conditions are 
discussed in Sec. \ref{sec2C}. The status of stellar ($n, 
\gamma$) rates and further improvements by new experimental facilities 
and advanced techniques are addressed in Sec. \ref{sec2D}.

\subsection{\label{sec2A}Measurement of neutron capture rates}

\subsubsection{Pulsed neutron sources \label{sec2A1}}

The laboratory neutron sources used in nuclear astrophysics measurements 
cover a variety of facilities, which differ in many aspects. The 
discussion presented here is focused on the main concepts and does, 
therefore, not include rarely used options such as filtered reactor 
beams \cite{BPS79} and radioactive sources \cite{Kno79}.

At small accelerators, neutrons are produced by nuclear reactions, 
such as the $^7$Li($p, n$)$^7$Be reaction, with the possibility of 
tailoring the neutron spectrum exactly to the stellar energy range 
between 0.3 and $\approx$500 keV. In many cases, the limited source 
strength can be compensated by low backgrounds and the use of 
comparably short neutron flight paths \cite{WiK81,NIM91,JaK96a}.  
Operated in DC mode this type of accelerator can also be used for 
the simulation of stellar neutron spectra, which are important in
applications of the activation technique (see below). 

Much higher intensities can be achieved via ($\gamma, n$) reactions 
at electron linear accelerators, such as GELINA at Geel, Belgium, 
and ORELA at Oak Ridge, USA, by bombarding heavy metal targets with 
electron beams of typically 50 to 100 MeV. These so-called white 
neutron sources provide continuous neutron spectra over an energy 
range from thermal to some tens of MeV. Measurements at these 
facilities need to be carried out at larger neutron flight paths 
because of the strong $\gamma$ flash from the impact of the lectron 
beam. In turn, the longer flight paths provide the 
possibility to study the resolved resonance region with high 
resolution (see for example \textcite{KSW96}).

Spallation reactions induced by energetic particle beams constitute the 
most prolific pulsed sources of fast neutrons suited for time-of-flight 
(TOF) measurements. Presently, two such spallation sources are in 
operation, LANSCE at Los Alamos \cite{LBR90} and the n$\_$TOF facility 
at CERN \cite{AAA03}. The main advantage of these facilities is the 
superb efficiency for neutron production due to the high primary 
proton beam energies of 800 MeV and 20 GeV at LANSCE and n$\_$TOF,
respectively. At n$\_$TOF, for example, 300 neutrons are produced per 
incident proton, which makes this facility the most luminous
white neutron source presently available. 

Due to their excellent efficiency, spallation sources can be operated 
at rather low repetition rates while still maintaining high 
average intensities. The situation at LANSCE is characterized by a 
comparably short flight path of 20 m, a time resolution of 250 ns,
and a repetition rate of 50 Hz, similar to the performance of the
SNS at Oak Ridge \cite{SNS04} and at J-PARC in Japan \cite{JPC04}. 
The n$\_$TOF facility at CERN represents a complementary approach
aiming at higher resolution (185 m flight path, 7 ns pulse width)
and even lower repetition rates of typically 0.4 Hz \cite{AAA03}.

The astrophysics options at various white neutron sources have been 
compared by \textcite{Koe01b} with respect to measurements on 
radioactive samples. As expected, spallation sources are unique 
for their superior peak neutron fluxes in the astrophysically 
relevant keV region. However, only the n$\_$TOF facility exhibits 
a neutron energy resolution comparable to that of electron linear 
accelerators. 

\subsubsection{Time-of-flight methods \label{sec2A2}}

The aim of the energy-differential TOF methods is to measure the 
neutron capture cross sections over a sufficiently large neutron 
energy range that Maxwellian averaged cross sections (MACS) can 
be determined from these data for any stellar temperature of 
interest. Recent developments and improvements in pulsed neutron 
sources and detection techniques have led to ($n, \gamma$) cross 
section measurements with improved accuracy, in many cases with 
uncertainties of a few percent. This progress is essential for
obtaining the $s$ abundances accurately enough to infer the physical 
conditions at the stellar site by analysis of the abundance patterns 
of  $s$-process branchings either in solar material or in presolar 
grains.

\centerline{(i)Total absorption calorimeters}
The energy sum of the $\gamma$-ray cascade emitted in the decay of the 
compound nucleus corresponds to the binding energy of the captured 
neutron. Therefore, this neutron separation energy represents the 
best signature of a capture event. Hence, $4\pi$ detectors with an 
efficiency close to 100\% are the most direct way to unambiguously 
identify ($n, \gamma$) reactions and to determine capture cross 
sections. This calorimetric approach started with the use of large 
liquid scintillator tanks, which are meanwhile replaced by arrays 
of BaF$_2$ crystals because of their superior resolution in 
$\gamma$-ray energy and their correspondingly lower backgrounds. 
A detector of this type consisting of 42 modules was developed at 
Karlsruhe~\cite{WGK90b} and is also in use at the n\_TOF facility 
at CERN \cite{HRF01}. In this design the BaF$_2$ crystals are shaped 
as truncated pyramids, forming a fullerene-type geometry where each 
module covers the same solid angle with respect to the sample. A 
somewhat simpler approach was chosen at ORNL \cite{GSK97b} and at 
FZ Rossendorf \cite{KAB07}, where cylindrical BaF$_2$ arrays  
have been constructed with hexagonal crystals. 

Recent examples of accurate cross section measurements made with 
the Karlsruhe 4$\pi$ detector comprise the unstable branch point 
isotope $^{151}$Sm \cite{WVK06c} and the Lu- and Hf-isotopes
\cite{WVK06a,WVK06b}. These results are essential for constraining
the temperature at the $s$-process site via the branchings at 
$A=151$, 175 and 179. 

A higher segmentation of such a 4$\pi$ detector is of advantage 
for separating true capture events from backgrounds and for 
handling the data rates in measurements on radioactive samples.
The state-of-the-art in this respect is the DANCE array with 
162 BaF$_2$ modules that is operated at the LANSCE facility in 
Los Alamos~\cite{RBA04}.
 
The high efficiency of $4\pi$ arrays in combination with intense
pulsed neutron sources provides the possibility for measurements on
very small samples. In general, this is important for any samples, 
where only small quantities are available, and in particular for radioactive 
isotopes, where the background from the activity of the sample needs
to be kept at minimum. An illustrative example for the first aspect 
is the keV ($n, \gamma$) cross section of $^{180}$Ta where the sample
consisted of 6.7 mg of $^{180}$Ta immersed in 145 mg of $^{181}$Ta 
\cite{WVA04}. Even smaller samples of about 200--400 $\mu$g have
recently been used in the DANCE array to determine the keV ($n, \gamma$) 
cross sections of actinide samples \cite{JBB08,ERB08}. A detailed 
survey for future measurements of neutron capture cross sections on 
radioactive isotopes with special emphasis on branching points along 
the $s$-process path is given by \textcite{CoR07}.

The main problem in using 4$\pi$ arrays arises from their response
to neutrons scattered in the sample. Although the scintillator is
selected to consist of nuclei with small ($n, \gamma$) cross sections,
about 10\% of the scattered neutrons are captured in the scintillator.
The resulting background can be attenuated by an absorber shell around 
the sample, preferentially consisting of $^6$LiH \cite{RBA04} or a $^6$Li 
containing compound \cite{HRF01}. Such an absorber is not required
in the setup at Karlsruhe if the neutron spectrum is limited
to energies below 225 keV, which allows one to separate the 
background from sample scattered neutrons via TOF.   

This type of background becomes crucial in measurements on
neutron magic nuclei and on light isotopes, where neutron scattering
dominates the rather weak capture by orders of magnitude.
Therefore, large detector arrays are less suited for these 
isotopes, which are of fundamental importance because they act 
as bottlenecks in the $s$-process path or as potential neutron poisons.

The potential of a 4$\pi$ BaF$_2$ array was extensively used at Karlsruhe 
for determination of accurate MACS values for an almost complete set 
of lanthanide isotopes between $^{141}$Pr \cite{VWA99} and $^{176}$Lu 
\cite{WVK06a} including all $s$-only nuclei (for a complete list of 
references see http://www.kadonis.org). The group of the lanthanide 
isotopes are particularly suited for a precise test of $s$-process 
nucleosynthesis concepts, because the relative abundances of the 
lanthanides are very well known \cite{AGS09} so that the $s$process 
reaction chain and the associated branchings can be consistently followed. 
In fact, the failure of the classical $s$ process and the success
of the stellar $s$ process in thermally pulsing low mass AGB stars 
\cite{AKW99b} was possible after accurate cross sections for the Nd 
isotopes and in particular for the $s$-only nucleus $^{142}$Nd became 
available \cite{WVK98a} (Sec. \ref{sec3A}).

Comprehensive measurements were also performed at Karlsruhe for the long isotope 
chains of Cd \cite{WVK02}, Sn \cite{WVT96a}, Te \cite{WVK92}, and Ba 
\cite{VWG94b} to provide detailed information for studying the full 
mass range of the main $s$-process component with well-defined MACS 
data. Reliable cross sections are also instrumental for defining the 
strength of the branchings in the reaction path, where the specific
abundance patterns yield constraints for important parameters of 
the stellar plasma, i.e. neutron density, temperature, pressure, and 
mixing phenomena. Such examples are
the branchings at A=122/123, 128, and 147/148, which represent sensitive 
tests for the quasi-equilibrium of the $s$-process reaction flow
\cite{WVK92} and for details of the stellar $s$-process conditions,
i.e. for the convective velocities \cite{RKV04} and the neutron density
\cite{WGV93} during He shell flashes in thermally pulsing low-mass 
AGB stars. Complementary information on the $s$-process temperature 
can be obtained from the branchings at $^{151}$Sm \cite{AAA04c,WVK06c,MAA06} 
and at $^{175}$Lu \cite{WVK06a,WVK06b}. These important aspects of the 
$s$-abundance distribution are the subject of Secs. \ref{sec3C}
and \ref{sec4C}.

\centerline{(ii) Detectors with low neutron sensitivity}

Originally, the neutron sensitivity problem led to the development of  
Moxon-Rae type detectors~\cite{MoR63}. The idea was to design a 
$\gamma$-ray detector with an efficiency proportional 
to the energy deposited. With this feature, the 
probability for detecting a capture event becomes 
\begin{equation}\label{eq:yield}
\varepsilon_{casc} =  \sum_{i=1}^{m} \varepsilon_{i} (E_{\gamma}^i) 
                   =  \sum_{i=1}^{m} k \times E_{\gamma}^i 
				   = k\times E_{\gamma}^{tot}
\end{equation}
independent of the cascade multiplicity $m$ and of the $\gamma$ 
energies. To avoid systematic uncertainties, the efficiency of
Moxon-Rae detectors had to be small enough that no more than one 
$\gamma$ ray was detected per cascade. 

In order to improve the overall efficiency, the principle of Moxon-Rae 
detectors was generalized by introducing the pulse height weighting 
technique (PHWT)~\cite{MaG67b, Rau63}, where the proportionality
between deposited energy and $\gamma$-ray efficiency is achieved a 
posteriori by an off-line weighting function applied to the detector 
signals. 

The PHWT technique was first used in experiments with C$_6$F$_6$ 
liquid scintillators. Although smaller than for scintillators 
containing hydrogen, the neutron sensitivity of C$_6$F$_6$ detectors 
gave still rise to large systematic uncertainties as
illustrated by \textcite{KWG00} and \textcite{GLS05,GLS05b}. 
This problem was reduced in a second generation of detectors, 
which are based on deuterated benzene (C$_6$D$_6$) because of the 
smaller capture cross section of deuterium. Further improvement was 
achieved by minimizing the construction materials and by replacing 
aluminum and steel by graphite or carbon fiber, resulting in a
solution, where the background due to scattered neutrons is practically 
negligible \cite{PHK03}.

The accuracy of the PHWT has been an issue for a long time. When the 
technique was proposed, the uncertainties introduced by the weighting
function (WF) were about 20\% in some particular cases
\cite{Mac87}. Dedicated measurements and Monte Carlo (MC) calculations
\cite{CPL88,PJG88} have led to gradually improved WFs.
With present advanced MC codes, realistic detector response functions 
and WFs could be determined by means of precise and detailed computer 
models of the experimental setup \cite{KSW96,TGC02,AAA04a,BAG05b}. A 
dedicated set of measurements at the n\_TOF facility confirmed that 
WFs obtained by such refined simulations allows one to determine 
neutron capture cross sections with a systematic accuracy of better 
than 2\%~\cite{AAA04a}. 

\begin{figure}
\includegraphics[width=8.5cm]{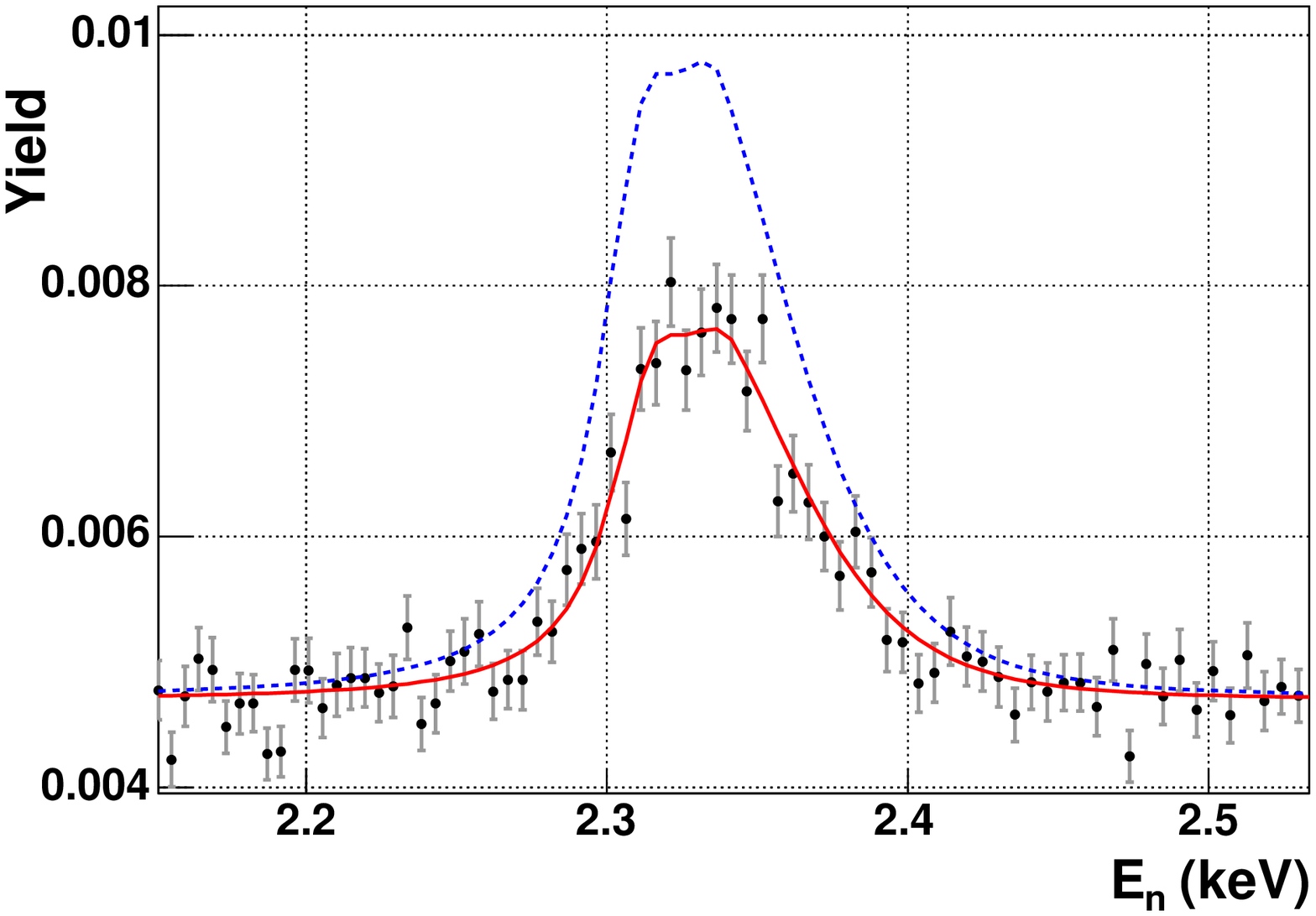}
\caption{\label{fig:3} (Color online) R-matrix analysis of
the second resonance in bismuth. The dashed
line corresponds to the yield calculated with the
resonance parameters from the ENDF/B-VI.8 evaluation
which exhibits the effect of neutron sensitivity
in previous data.}
\end{figure}

Recent applications of the improved PHWT technique with optimized
C$_6$D$_6$ detectors are the measurements of ($n, \gamma$) cross sections
on isotopes at or near magic neutron numbers, which are characterized
by small capture/scattering ratios.  Examples for such measurements are
the studies of the n\_TOF collaboration on $^{209}$Bi \cite{DAA06a} and
on a sequence of stable Pb isotopes \cite{DAA06b,DAA07a,DAA07b}. 
Several of the involved resonances show the effect of neutron 
sensitivity as illustrated in Fig. \ref{fig:3} for two cases 
in the $^{209}$Bi cross section, where the resonance yields obtained 
from previous data are clearly overestimated.

Other neutron magic isotopes, which were recently studied with improved 
accuracy, are $^{139}$La \cite{TAA07} and $^{90}$Zr \cite{TFM08}. The
importance of $^{139}$La results from the fact that it is abundantly 
produced by the $s$ process (e.g. 70\% of solar La). Because La is easily 
detectable by stellar spectroscopy, it can be used as an indicator for
the onset of the $s$ process in low-metallicity stars, which were 
formed early in galactic history. It is interesting to note 
that the $^{139}$La cross section was accurately confirmed by  
independent activation measurements \cite{BDH03,WDH06}. The position
of $^{90}$Zr in the $s$-process reaction chain is exactly located
at the matching point between the mass regions dominated by the weak 
and main $s$ process. Therefore, this isotope assumes a key role 
among the isotopes with N = 50. As in the Bi and in the Pb isotopes, 
the new data are significantly smaller than measured previously
due to neutron sensitivity problems in the past.

The potential of TOF measurements is illustrated in Table \ref{tab1}
at the example of the important pairs of $s$-only isotopes, which 
can be used to define branching ratios 
$$B=\frac{\langle \sigma \rangle_p\,N_p}{\langle \sigma \rangle_f\,N_f}$$
where $\langle \sigma \rangle_p$ denotes the MACS of the partially 
bypassed isotope and $N_p$ its isotopic abundance. The corresponding 
values for the heavier isotope, which experiences the full reaction 
flow, are indicated by index $f$. One such pair are the isotopes 
$^{80}$Kr and $^{82}$Kr, which characterize the branching at $^{79}$Se 
as indicated in Fig. \ref{fig:1}. Pairs of $s$-only isotopes are
particularly valuable because these branchings can be evaluated
without interference by the uncertainties related to elemental 
abundance values. For this reason, the MACSs of these isotopes 
have been determined with the highest possible accuracy. The 
uncertainties of the respective branching ratios in the last 
column of Table \ref{tab1} are refer only the MACS uncertainties 
although isotopic abundance ratios exhibit also non-negligible 
uncertainties \cite{RoT98}. For example, a 6\% uncertainty 
has been attributed to the abundance ratio of $^{122}$Te and 
$^{124}$Te. The quoted values, which include small corrections 
due to minor deviations from reaction flow equilibrium 
\citep{AKW99b}, were estimated via the classical approach. Note 
that weak branchings can only be analyzed with confidence if the 
MACS are accurately known as in case of the $^{148,150}$Sm pair.  

\subsubsection{Data acquisition and analysis techniques \label{sec2A3}}

Modern electronic techniques have led to substantial improvements in 
data acquisition and analysis. At the front end, flash analog-to-digital 
converters (FADCs) provide the fastest and safest way to record the 
complete information contained in the analogue detector output by 
digitizing the entire wave-form of the signals. 

This has the obvious
advantage that the raw data are preserved for repeated and refined 
off-line analysis, which allows for the efficient identification 
and correction of baseline shifts, pile-up, and noise, resulting
in a rigorous assessment of systematic uncertainties. Specifically 
designed pulse shape analysis (PSA) algorithms can be used for
evaluating the relevant signal parameters, i.e. time information, 
amplitude, area, and shape,  for each type of detector. In case that 
these algorithms are improved at some point, one can always return 
to the original information and repeat the analysis of a particular
experiment.

Another advantage is that accidental errors due to the failure of an 
electronic module are minimized, simply because there is much less
electronics needed to run an experiment. It is sufficient, for example, 
to connect the anode signal of a photomultiplier tube with an 
FADC-channel, which is then read by a computer.

These features were recently used in ($n, \gamma$) studies with 
4$\pi$ BaF$_2$ arrays, where flexible algorithms have been employed 
to reduce background events in the scintillator via 
$n/\gamma$-identification~\cite{MBB06} and to suppress the intrinsic 
$\alpha$ background in BaF$_2$ crystals~\cite{RBA04}. 

A good example is the data acquisition system at n\_TOF~\cite{AAA05b}, 
which is based on 8-bit FADC modules, with sampling rates up to 2~GHz 
and 8 or 16~Mbyte memory. The low repetition rate of $\sim$0.4~Hz 
leaves enough time to digitize and store all the raw FADC information 
accumulated during each neutron bunch and for all detectors employed. 
Although peak rates of 8~Mbyte are reached per burst and per detector 
due to the very high instantaneous neutron flux, the data acquisition 
system works practically dead-time free, except for a narrow interval 
of 15-20~ns that is needed to separate two consecutive signals 
unambiguously. 

The use of FADCs in TOF measurements at facilities with repetition rates 
above about 1 kHz is hampered by rapidly increasing dead times caused
by the transfer of the large amount of data to the storage medium, a 
problem that is presently studied by the EFNUDAT collaboration \cite{Pla09}. 

The main drawback of a FADC-based acquisition system is the huge 
amount of accumulated data, which demands large storage capabilities 
and high data transfer rates. While this difficulty can be mitigated by 
applying a zero suppression algorithm on the fly~\cite{AAA05b}, the 
enormous improvement in computing power and in the capacity of storage 
media was essential for the handling and analysis of Terabytes of 
data taken with FADC systems.   

Another general aspect of data analysis techniques is the increasing
importance of Monte Carlo simulations, which are becoming standard tools 
for planning of measurements and for analyzing experimental data. 
The efficient application of the GEANT \cite{Gea03}and MCNP \cite{MCN07} 
software packages has also been favored by the recent advances in 
computing power. 

\subsubsection{Activations \label{sec2A4}} 
 
Activation in a quasi-stellar neutron spectrum provides a completely
different approach for the determination of stellar ($n, \gamma$) rates.
Apart from the fact that the method is restricted to cases, where neutron 
capture produces an unstable nucleus, it has a number of appealing 
features.
\begin{itemize}
\item It was found that stellar neutron spectra can be very well 
approximated in the laboratory so that MACS measurements can be directly 
performed by irradiation and subsequent determination of the induced 
activity.
\item Technically, the method is comparably simple and can be performed 
at small electrostatic accelerators with standard equipment for $\gamma$ 
spectroscopy.
\item The sensitivity is orders of magnitude better than for TOF experiments
because the accelerator can be operated in DC mode and because the sample 
can be placed directly at the neutron production target in the highest 
possible neutron flux. This feature opens the possibility for measurements 
on sub-$\mu$g samples and on rare isotopes, an important advantage if one 
deals with radioactive materials. 

\item In most cases the induced activity can be measured via the $\gamma$
decay of the product nucleus. This implies favorable signal/background
ratios and unambiguous identification of the reaction products. The
excellent selectivity achieved in this way can often be used to study 
more than one reaction in a single irradiation, either by using 
elemental samples of natural composition or suited chemical compounds.  
\end{itemize} 

In an astrophysical environment with temperature $T$, interacting 
particles are quickly thermalized by collisions in the stellar plasma. 
The neutron energy distribution corresponds to a Maxwell-Boltzmann 
spectrum, $$\Phi = dN/dE_n \sim \sqrt{E_n}{\rm exp}(-E_n/kT).$$
So far experimental neutron spectra, which simulate the energy dependence 
of the flux $v \Phi \sim E_n {\rm exp}(-E_n/kT)$, have been produced by three 
reactions. The $^7$Li($p, n$)$^7$Be reaction allows one to simulate the 
spectrum for a thermal energy of $kT$ = 25 keV \cite{BeK80,RaK88} very 
close to the 23 keV effective thermal energy in He shell flashes of 
low mass AGB stars, where neutrons are produced via the $^{22}$Ne($\alpha, 
n$)$^{25}$Mg reaction. More recently, the $^{18}$O($p, n$)$^{18}$F reaction
has been shown to provide a spectrum for $kT=5$ keV \cite{HDJ05},
which is well suited for $s$-process studies of the main 
neutron source in these stars, the $^{13}$C($\alpha, n$)$^{16}$O 
reaction that operates at 8 keV thermal energy.

While these two quasi-stellar spectra allow one to determine the MACSs 
necessary for studies of the main $s$ component, the weak component 
associated with massive stars is characterized by higher temperatures, 
i.e. 26 keV thermal energy during core He burning and about 90 keV in the 
shell C burning phase. The situation during core He burning is again well 
described by the  $^7$Li($p, n$)$^7$Be reaction, but the high temperatures 
during shell C burning are only roughly represented by means of the 
$^3$H($p, n$)$^3$He reaction, which provides a spectrum for $kT$ = 52 keV 
\cite{KNA87}. In this case, the measured MACSs have to be extrapolated by 
statistical model calculations.
 
Because the proton energies for producing these quasi-stellar spectra
are only slightly higher than the reaction thresholds, all neutrons 
are emitted in forward direction. In this way, the samples are exposed 
to the full spectrum and backgrounds from scattered neutrons are negligible.
With a proton beam current of 100 $\mu$A on target intensities of the 
order of 10$^9$, 10$^8$, and 10$^5$ s$^{-1}$ can be achieved for the 
($p, n$) reactions on $^7$Li, $^3$H, and $^{18}$O with present electrostatic 
accelerators. Future developments, however, will provide much higher beam 
currents and correspondingly higher neutron fluxes (Sec. \ref{sec2A4}).
 
Already at present, the neutron intensities for activation measurements 
exceed the fluxes obtainable in TOF measurements by orders of magnitude. 
For example, the highest neutron flux reached at an experimental TOF setup 
is 5$\times$10$^5$ s$^{-1}$ at the DANCE array in Los Alamos. 
Accordingly, activation represents the most sensitive method for ($n, \gamma$)
measurements in the astrophysically relevant energy range. This feature is 
unique for the possibility to measure the MACSs of neutron poisons, abundant 
light isotopes with very small cross sections, as well as for the use of 
extremely small sample masses.
 
The latter aspect is most important for the determination of MACSs of 
unstable isotopes, which are needed for investigating unstable nuclei 
of relevance for $s$-process branchings. In most cases TOF measurements 
on unstable branch point isotopes are challenged by the background 
due to the sample activity (Sec. \ref{sec2A4}). Illustrative 
examples in this respect are the successful measurements of the MACS of 
$^{60}$Fe \cite{URS09} and $^{147}$Pm \cite{RAH03}. In the first case, the 
sample \cite{SND10} consisted of $1.4\times 10^{16}$ atoms or 1.4 
$\mu$g and the activation was complicated by the 6 min half life of the 
$^{61}$Fe, which required 47 repeated irradiations, and by the small 
capture cross section of 5.7 mb. Note that the number of atoms and the 
cross section result reduced by a factor 1.75 compared to the original
paper \cite{URS09} because of a new precise half-life determination for 
$^{60}$Fe \cite{RFK09}. The second experiment was performed with an 
even smaller sample of only 28 ng or $1.1\times10^{14}$ atoms in order 
to keep the $^{147}$Pm activity ($t_{1/2}=2.6$ yr) at a reasonable value.
In this case, the small sample mass could be used because of the half 
live of $^{148}$Pm and the cross section were conveniently large.    
 
Another advantage of the activation method is that it is insensitive 
to the reaction mechanism. In particular, it includes the contributions 
from Direct Capture (DC), where the neutron is captured directly into a bound state. 
The DC component, which contributes substantially to the ($n, \gamma$) cross sections 
of light nuclei, is extremely difficult to determine in TOF measurements 
(for an exception see the specialized setup used by \textcite{INM95}).
 
Apart from measurements on unstable isotopes as discussed below, the 
excellent sensitivity of the activation technique has been extensively
used for the determination of cross section with a non-negligible DC 
component, for the determination of partial cross sections for the 
population of isomeric states, for the measurement of small cross 
sections in general. 

\begin{figure}
\includegraphics[width=8.5cm,angle=270]{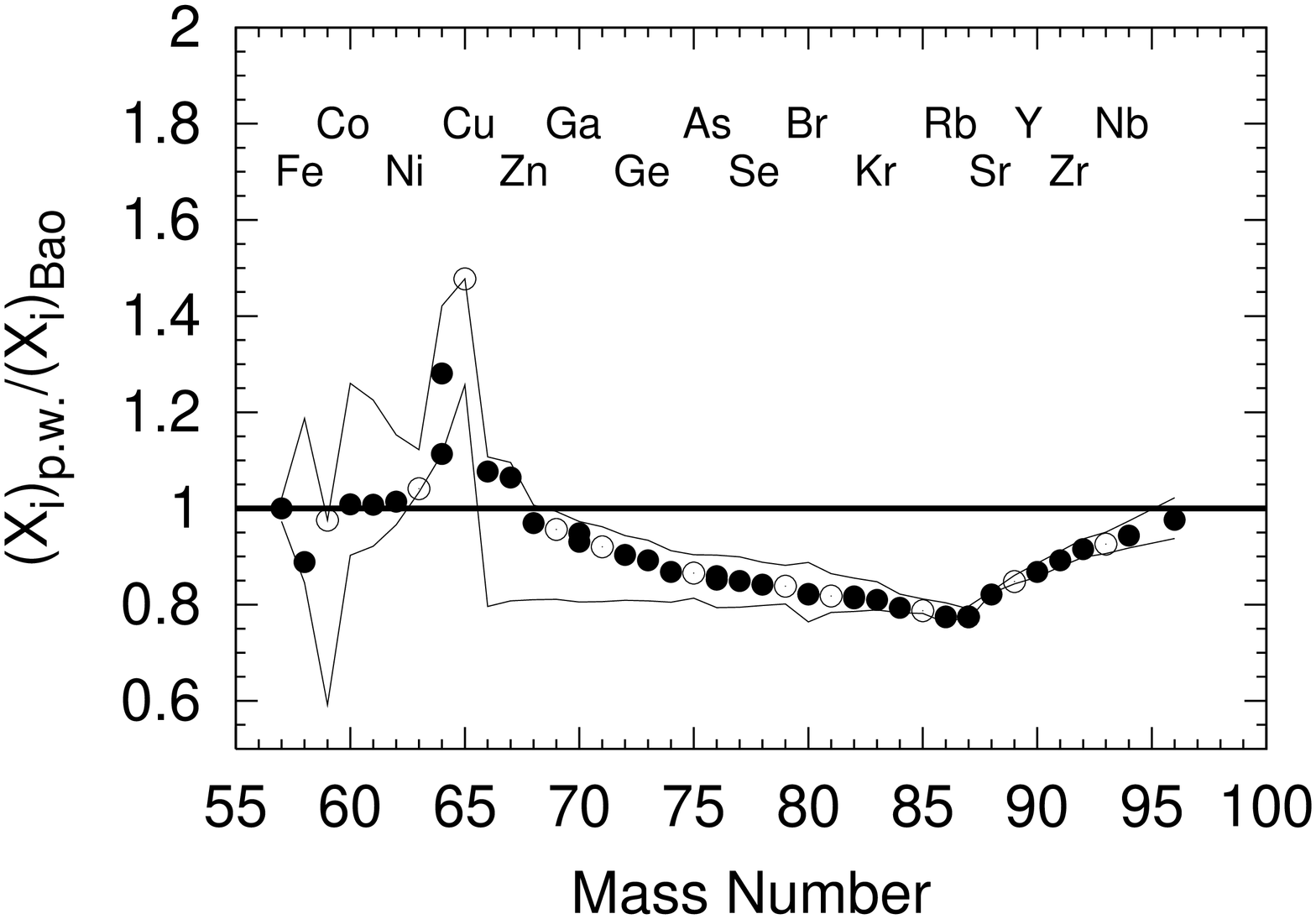}
\caption{\label{fig:4} Nucleosynthesis yields between Fe and Nb 
illustrating the final $s$-process yields after shell C burning for a 
25 $M_\odot$ star with solar metallicity. To illustrate the combined 
effect of new cross sections for $^{58}$Fe, $^{59}$Co, $^{64}$Ni, 
$^{63,65}$Cu the yield is plotted relative to the case obtained with 
the previous cross sections of \textcite{BBK00}. Even and odd Z elements 
are distinguished by black and open symbols, respectively. The thin 
lines demonstrate the uncertainties stemming from the extrapolation
of the measured cross sections to higher and lower energies.
}
\end{figure}

Prominent examples of the latter type are the 
series of measurements between Fe and Sr, which are related to the 
reaction flow of the weak component \cite{RDF07,HKU08a,HKU08b,MDD09a}.
These data were consistently smaller than previous TOF results, which 
evidently suffered from an underestimated neutron sensitivity. Fig.
\ref{fig:4} shows that these changes gave rise to strong
propagation effects in the abundance distribution. These effects are 
to a large part originating from the high-temperature phase during 
shell C burning that operates at $kT=90$ keV. In order to reduce the 
uncertainties in the extrapolation from the measured MACS at 25 keV 
complementary TOF measurements on the stable Fe and Ni isotopes are 
under way at CERN.

For some branchings the population of long-lived isomers plays an
important role \cite{War77}. In most cases the respective partial
cross sections feeding the isomers have to be determined in
activation measurements because this information is very difficult 
to obtain via the TOF technique. Important isomers are those in 
$^{176}$Lu and in $^{180}$Ta for example. The isomer in $^{176}$Lu 
at 123 keV is the key for the interpretation of the mother/daughter 
ratio of the $s$-only isotopes $^{176}$Lu and $^{176}$Hf as an 
$s$-process thermometer for the He shell flashes in low mass AGB 
stars \cite{KKB91b,DBJ99}. At temperatures above about 150 MK 
the initial population probabilities of isomer ($t_{1/2}=3.68$ h)
and ground state ($t_{1/2}=37.5$ Gyr) \cite{WVK06a,HWD08} are 
altered by a delicate interplay between neutron density and 
temperature, depending on subtle details in the nuclear structure
of $^{176}$Lu \cite{MBG09}. The case of $^{180}$Ta is of interest
because this is the only isotope in nature, which is (almost) stable 
in its isomeric state. Again, the survival of $^{180}$Ta$^{\rm m}$ 
depends on the ($n, \gamma$) cross section \cite{WVA01,WVA04,KAH04} 
and on the effect of the high temperatures at the $s$-process site 
\cite{BAB99,BAB02} in depopulating the isomer to the ground state.
 
A few years ago, the potential of the activation technique was 
considerably extended where activation products are counted directly 
by Accelerator Mass Spectrometry (AMS) instead of the induced 
activity (Sec. \ref{sec2D2}).

The complementarity between TOF and activation measurements is illustrated 
in Table \ref{tab2} for examples between Fe and Au. Overall, there is good
agreement between the results obtained with both methods. As far as the related
uncertainties are concerned one could be tempted to assume that the small, resonance-dominated 
cross sections of the Fe group and at magic neutron numbers are more accurately 
determined by activation, whereas the smooth cross sections of the heavier nuclei are 
generally better determined in TOF experiments. However, this is no general rule 
as indicated by the cases of $^{88}$Sr and $^{197}$Au, but the quality of the data 
depends on individual details of the respective experiments. 

\subsubsection{Studies on radioactive isotopes \label{sec2A5}}

Although many cross sections of the stable isotopes 
are still rather uncertain and need to be improved, a major challenge 
of future experiments is to extend such measurements to the largely 
unexplored subset of unstable branch point nuclei. 

The concept of $s$-process branchings was formulated by \textcite{WNC76} 
to obtain information on {\it average} $s$-process neutron densities 
and temperatures. With the advent of quantitative stellar models 
the abundance patterns of the branchings were understood to represent 
sensitive tests for the time-dependence of these parameters during the 
various $s$-process episodes in thermally pulsing low mass AGB stars 
\cite{GAB98,AKW99b} and in massive stars \cite{RBG91b}. 

Branchings in the $s$ path occur whenever $\lambda_n \approx \lambda_\beta$.
The $\beta$ decay rate is $\lambda_\beta=ln2/t_{1/2}$, whereas the 
neutron capture rate $\lambda_n = n_n\,\langle\sigma\,v\rangle\,v_T$
is the product of neutron density, MACS, and mean thermal velocity. 
Depending on neutron density and MACS of the unstable branch point isotope, 
the stellar half lives of branch point nuclei may range between several 
days and several years. In addition, stellar half lives can be strongly 
enhanced compared to the corresponding terrestrial values as discussed in 
Sec. \ref{sec2C}. 

The branch point isotope $^{79}$Se in Fig. \ref{fig:1} represents 
such a case, where the decay is accelerated by thermal population of a 
short-lived excited state. In contrast, the other two branch points 
indicated in Fig. \ref{fig:1}, $^{63}$Ni and $^{85}$Kr, are not or 
only weakly altered by temperature effects. All three branch points 
fall in the mass range of the weak component associated with massive 
stars.

Measurements of the MACS for the branch points of the weak component 
are hampered by the high specific activity of $^{85}$Kr and by the lack
of sample material in case of $^{79}$Se. Only for $^{63}$Ni a TOF 
measurement has been recently performed with the DANCE detector at Los 
Alamos using a mildly enriched nickel sample containing 11\% $^{63}$Ni 
\cite{Cou09}. 

For the main component experimental information could be obtained 
for some important branch point isotopes. With one exception these 
measurements were performed via the activation technique, where
the activity problem was relaxed because the high sensitivity of the 
method allows one to use $\mu$g or even sub-$\mu$g samples. Results
were reported for $^{135}$Cs \cite{PAK04}, $^{147}$Pm \cite{RAH03}, 
$^{163}$Ho \cite{JaK96a}, $^{155}$Eu \cite{JaK95b}, and $^{182}$Hf 
\cite{VDH07}. As noted before the data obtained via activation usually 
represent the respective MACS values at $kT=25$ keV and have to be 
extrapolated to higher and lower temperatures by means of theoretical 
data (Sec. \ref{sec2B}).

TOF measurements on branch point isotopes of the main component have 
been reported for the quasi-stable nuclei $^{99}$Tc \cite{Mac82b,WiM87}, 
$^{107}$Pd \cite{Mac85a}, and $^{129}$I \cite{Mac83}. So far $^{151}$Sm 
is the only branch point with a half-life shorter than 100 yr where 
the ($n, \gamma$) cross section has been studied by means of the TOF
technique over a wide energy range. Combination of the accurate and 
comprehensive data measured at Karlsruhe \cite{WVK06c} and CERN \cite{AAA04c} 
provided a full set of MACS values for $^{151}$Sm. 

The measurement of the $^{14}$C($n, \gamma$)$^{15}$C cross section 
\cite{RHF08} was of interest because this reaction determines whether 
the neutron balance of the $s$ process could be affected by neutron 
induced CNO cycles \cite{WGS99}, it contributes to the reaction flow 
in neutrino driven wind scenarios of the $r$ process \cite{TSK01}, and
is important for validating the ($n, \gamma$) cross sections calculated 
via detailed balance from the inverse Coulomb dissociation reaction 
\cite{WGT90,TBD06}. 

The possibility to complement ($n, \gamma$) experiments by studies of 
the inverse ($\gamma, n$) reactions has been invoked by \citet{SMV03} and \citet{MSU04}
for the branch point isotope $^{185}$W. Other examples of ($\gamma, n$) 
measurements refer to applications in the $p$ rather than in the $s$ process 
\cite{VMB01,SVG04,SVG05}.

Another indirect approach for obtaining information on ($n, \gamma$) cross 
sections of unstable nuclei is the surrogate method \cite{DiE07,EAB05},
which uses the assumption that the reaction of interest proceeds via the 
formation of a compound nucleus and that formation and decay of the 
compound state can be separated, provided that both steps are independent 
of each other. In many cases, the formation cross section can be 
calculated reasonably well by using optical potentials, but theoretical
decay probabilities are often quite uncertain. In the surrogate approach
the compound nucleus is produced via an alternative direct reaction and its 
decay probability is then measured. There are several challenges of the 
surrogate method, in particular the "J population mismatch", which means 
that in the desired reaction different compound states might be populated, the 
difficulty to convert the experimental observables into decay probabilities, 
the role of pre-equilibrium reactions, where the intermediate configuration 
decays before a compound nucleus is formed, and the role of projectile 
break-up, which may disturb the proper identification of the surrogate 
reaction \cite{FAB05}. A recent example for the application of this method 
is the work of \textcite{BDW06}.

Although the indirect approaches rely in essential parts on theory and 
are therefore limited in accuracy, they often provide valuable information, 
which is impossible to obtain otherwise.

\subsection{\label{sec2B}Cross section calculations}

As already pointed out in the previous sections, measurements
cannot be performed at all energies and for all relevant isotopes.
In addition, the reaction rates in a stellar environment require
estimation of reaction processes for nuclei in their excited states,
which are impossible to measure under laboratory conditions.
Therefore, a close collaboration between experiment and theory
remains crucial for establishing the complete nuclear physics input 
for $s$-process studies. On the other hand, experimental information
is also mandatory for guiding and testing developments in theory 
in the region of unstable nuclei, a necessary step towards 
quantitative models of explosive nucleosynthesis.

\subsubsection{Statistical model \label{sec2B1}}

The key approach for the calculation of stellar $s$-process reaction 
rates is based on the Hauser-Feshbach statistical model (HFSM), which 
has been formulated over 30 years ago \cite{Mol75}. The model relies 
essentially on two basic assumptions, the validity of the compound 
nucleus reaction mechanism and a statistical distribution of nuclear 
excited states. With these assumptions, the reaction cross section 
(e.g. for neutron capture) can be written in terms of model parameters
such as the energy-dependent neutron transmission functions $T_{n,ls}$
and the $\gamma$-ray transmission functions $T_{\gamma,J}$. The 
general expression reads
\begin{equation}
\sigma_{n,\gamma}(E_{n}) = \frac{\pi}{k_{n}^2} \sum_{J,\pi} g_{J}
\frac{\sum_{ls}T_{n,ls} T_{\gamma,J}}{\sum_{ls} T_{n,ls} + \sum_{ls} T_{n',ls} + T_{\gamma,J}} 
W_{\gamma,J}
\label{HFSM}
\end{equation}
where $E_{n}$ is the incident neutron energy, $k_{n}$ the wave 
number, $s=1/2$ the intrinsic spin of the incident particle, and 
$l$ the orbital angular momentum of neutron and nucleus. The $g_{J} 
= (2J+1)(2s+1)^{-1}(2I+1)^{-1}$ is a statistical weight factor 
for target nuclei of spin $I$ and compound states of total angular 
momentum $J$ compatible with spin and parity conservation laws. 
The width fluctuation factor $W_{\gamma}$ takes the different 
statistical properties of the $\gamma$-decay channel and of the 
competing neutron elastic ($n, n$) and inelastic ($n, n'$) channels 
into account. The various HFSM approaches differ by the particular 
nuclear structure and de-excitation models adopted for calculating 
the nuclear quantities in Eq. \ref{HFSM}. 

Examples of widely used HFSM approaches for applications in nuclear 
astrophysics are those of \textcite{HWF76},  
\textcite{Har81}, and the latest "NON-SMOKER" \cite{Rau01}, "MOST" 
\cite{Gor05}, and "TALYS" \cite{KHD05} versions. Most of the quoted 
references include also HFSM computer codes for calculation of reaction 
cross sections. A repository of parameters and systematics
of nuclear structure quantities can be found in the "RIPL" initiative 
\cite{Bel05}. Additional model codes have been used for individual 
reaction rate calculations.

\subsubsection{Maxwellian averaged cross sections \label{sec2B2}}

The neutron spectrum typical of the various $s$-process sites is 
described by a Maxwell-Boltzmann distribution, because neutrons are 
quickly thermalized in the dense stellar plasma. The effective 
stellar reaction cross sections are therefore obtained by 
averaging the experimental data over that spectrum. The resulting
Maxwellian averaged cross sections (MACS)
\begin{equation}\label{eq:macs}
\langle \sigma \rangle_{kT}=\frac{2}{\sqrt{\pi}}\frac{\int_0^\infty 
  \sigma(E_n)~E_n~{\rm e}^{-E_n/kT}~dE_n}{\int_0^\infty 
  E_n~{\rm e}^{-E_n/kT}~dE_n}.
\end{equation}
are commonly compared for a thermal energy of $kT=30$ keV, but
for realistic $s$-process scenarios a range of thermal energies 
has to be considered, from about 8 keV in the $^{13}$C pocket of 
thermally pulsing low mass AGB stars to about 90 keV during
carbon shell burning in massive stars. To cover this full range, 
energy-differential cross sections $\sigma(E_n)$ are needed
in the energy region $0.1 \leq E_n \leq 500$ keV. Whenever
experimental data are available only for part of this range,
cross section calculations are required for filling these 
gaps. Correspondingly, calculated cross sections are needed 
to extrapolate the results from activation measurements to the 
entire temperature range of the various $s$-process sites
(see Ref. \cite{HKU08a}). 

As a service to the community a complete set of MACS data is 
available from the continuously updated compilation described 
in Sec. \ref{sec2D1}.

\subsubsection{Stellar enhancement factors \label{sec2B3}}

Apart from the need of cross section calculations for filling the 
gaps in experimental data, theory is indispensable for adapting 
the experimental data to the stellar environment. With respect
to MACSs, this refers to the fact that excited nuclear states are 
populated under stellar conditions due to interactions with the 
hot thermal photon bath. Because of the high photon intensity, 
all states with excitation energies $E_i$ are in thermal equilibrium 
with population probabilities 
\begin{equation}
p_i = \frac{(2J_i + 1)\,{\rm exp}(-E_i/kT)}{\sum_m (2J_m + 1)\,{\rm exp}(-E_m/kT)}
\end{equation}
\noindent
where $J$ denotes the level spin and the denominator represents the 
nuclear partition function. 

The capture cross section of excited states can be modeled as for 
ground states. However, in reactions on excited states an additional
possibility for inelastic scattering must be considered, the so-called 
"super-elastic" channel, in which the incident neutron gains in energy,
leaving the target nucleus in a lower state. This particularly relevant 
process has to be taken into account in the HFSM equation (Eq. \ref{HFSM})
by adding the transmission coefficients $T_{n',ls}$ for the open 
super-elastic channels. 

Possibilities for testing the calculations of the stellar MACS are 
the comparison of measured and calculated inelastic scattering cross 
sections, which provides a good benchmark for the neutron-nucleus 
interaction needed to obtain the transmission functions in such 
calculations and the comparison with the experimental capture cross 
section for the ground state. Unfortunately, experimental data for 
the inelastic channel in the astrophysically relevant energy range 
are rather scarce.
 
In practice, these effects are taken into account by the so-called 
stellar enhancement factor  
$$SEF = \frac{\langle\sigma\rangle^{*}}{\langle\sigma\rangle^{lab}}$$ 
\noindent
where the MACS labeled by $*$ and $lab$ indicate the stellar average 
over the thermally populated states and for the laboratory (ground state) 
cross section, respectively. On average, these factors are below 10\%
at the $s$-process temperatures in low mass AGB stars, but can reach 
values of more than 40\% during shell C burning in massive stars, 
especially for the heavy odd isotopes with low lying excited states.  

An important example for the role of SEFs is the Re/Os nuclear 
cosmochronology \cite{Cla64}, where the MACSs of the 
$s$-only isotopes $^{186}$Os and $^{187}$Os are of key 
importance. The aspects related to the neutron physics of this 
clock has been studied by the n\_TOF collaboration in a series of 
three papers dealing with cross section measurements 
in  the astrophysically relevant energy range \citep{MFM10,MHK10}
and the SEF calculations based on the HFSM approach with model 
parameters tuned to reproduce the experimental ($n, \gamma$) and
($n, n'$) cross sections \citep{FMM10}.

\begin{figure}
\includegraphics[width=8.5cm]{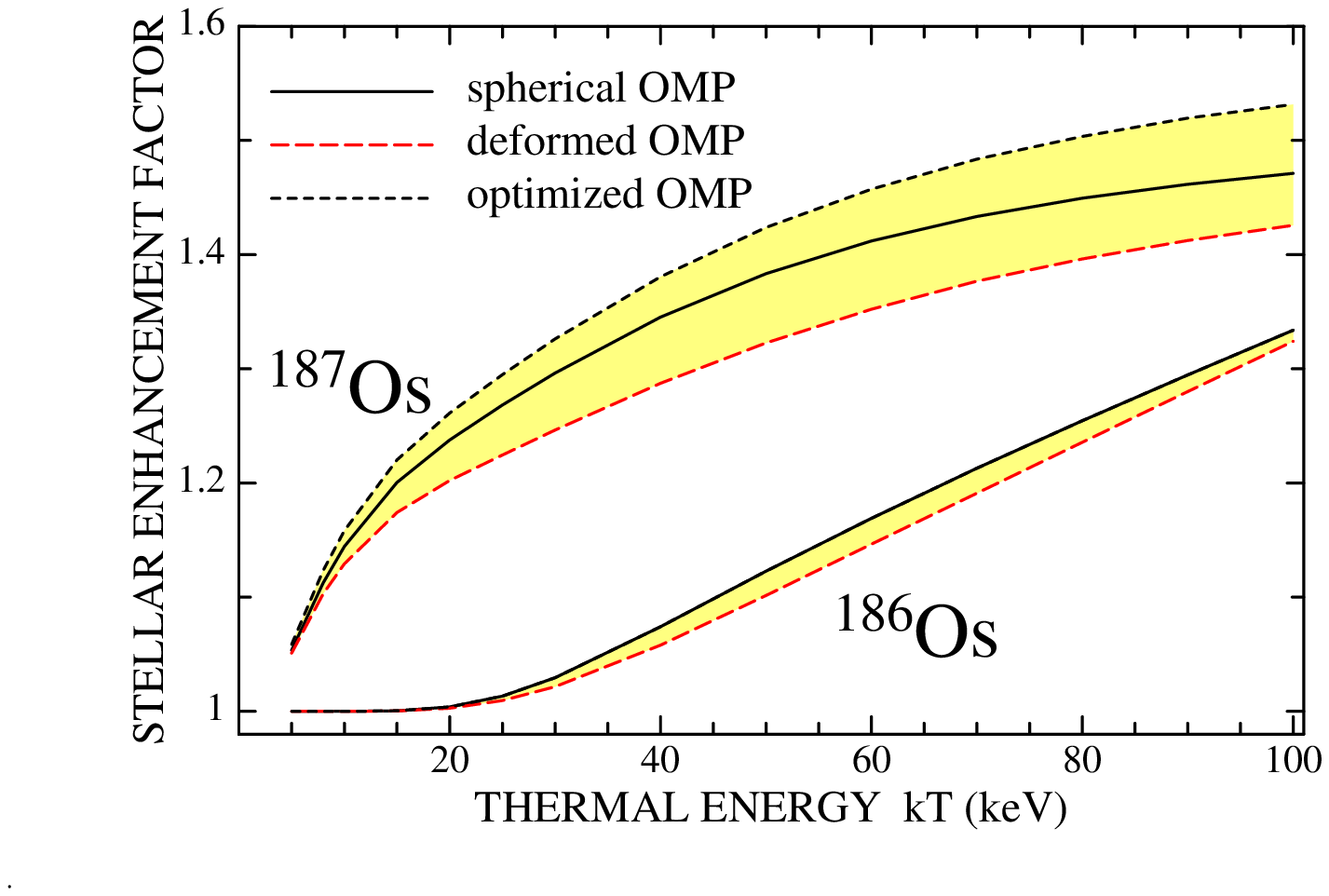}
\caption{\label{fig:5} (Color online) Stellar enhancement factor (SEF) for the $^{187}$Os($n,
\gamma$)$^{188}$Os reaction. The first ten excited states in $^{187}$Os, 
up to an excitation energy of 260 keV, have been included in the calculation. 
The difference between the results obtained with a spherical and a deformed 
optical model potential for the neutron-nucleus interaction could be used for 
estimating the uncertainties of the SEF results.}
\end{figure}

The SEF corrections are particularly relevant for $^{187}$Os, where
the ground state is populated by only about 30\% at $kT=30$ keV, 
while 70\% of the nuclei exist in excited states, 47\% alone in 
the first excited state at 9.75 keV, the state which strongly dominates the 
competition by inelastic and superelastic scattering. The comparison of 
the SEF values for $^{186}$Os and $^{187}$Os in Fig.~\ref{fig:5} 
underlines the importance of this correction for $^{187}$Os in 
the relevant range of thermal energies around $kT=25$ keV. 
A relatively small uncertainty of $\pm$4\% could be estimated for 
the SEF of $^{187}$Os from the difference between 
the results obtained with a spherical and a deformed optical model 
potential for the neutron-nucleus interaction.

The impact of the estimated SEF uncertainties and of other relevant 
nuclear physics input on the galactic age (or on the duration of 
nucleosynthesis) have been estimated by means of the simple model 
proposed by \textcite{FoH60} to be less than 1 Gyr \citep{FMM10}.
This means that uncertainties of the experimental data are no 
longer limiting a revision of the Re/Os chronometer. Further
work can now concentrate of the related astrophysical issues, 
which are mostly due to the time-dependence of the production 
rate of $^{187}$Re and the related problem of astration, i.e. that 
$^{187}$Re is partially destroyed in later stellar generations. 

The example of $^{187}$Re shows that uncertainties of $\approx$4 - 5\% 
can be obtained in SEF calculations, provided that the measured capture 
cross section for the ground-state is known with sufficient accuracy and 
that the other parameters in HFSM calculations can be derived from measured 
quantities.

\subsection{\label{sec2C}Beta decay under stellar conditions}

A detailed evaluation of beta decay rates for $s$-process 
analyses has been presented by \textcite{TaY87}
on the basis of a thorough classification of possible
contributions from thermally excited states and by 
considering the relevant effects related to the high degree of 
ionization in the stellar plasma. The most spectacular 
consequence of ionization is the enormous enhancement of decays
with small Q$_{\beta}$ values, where the decay electrons can be 
emitted into unoccupied atomic orbits. This bound beta decay 
was eventually confirmed in storage ring experiments with 
fully stripped $^{163}$Ho and $^{187}$Re atoms at GSI Darmstadt 
\cite{JBB92, BFF96}.

The quantitative assessment of the temperature dependent decay 
rates of the key branch point isotopes requires more
experimental information on log$_{10}$ ft values for the decay of 
excited states as well as more storage ring experiments to 
expand our knowledge of bound beta decay rates. Experimental 
possibilities have been discussed \cite{Kae99}, but
must be extended by the successful recent application of 
($d, ^2$He) reactions \cite{Fre04}.

\subsection{\label{sec2D} Status and Prospects}

\subsubsection{Compilations of stellar ($n, \gamma$) cross 
sections and further requirements \label{sec2D1}}

Stellar neutron capture cross sections have first been compiled
in 1971 by  \textcite{AGM71}, who presented
a set of recommended ($n, \gamma$) cross sections averaged over 
a Maxwell-Boltzmann distribution for a thermal energy of $kT$ = 
30 keV. This first collection of Maxwellian averaged cross 
sections (MACS) comprised already 130 experimental cross sections 
with typical uncertainties between 10 and 25\%. These data were 
complemented by 109 semi-empirical values estimated from the 
cross section trends with neutron number of neighboring nuclei 
to provide a full set of nuclear data for quantitative studies 
of the $s$-process. 

The next compilation of experimental and theoretical stellar neutron 
cross sections for $s$-process studies, which was published 16 years 
later by \textcite{BaK87}, included cross sections 
for ($n, \gamma$) reactions between $^{12}$C and $^{209}$Bi, some 
($n, p$) and ($n, \alpha$) reactions (from $^{33}$Se to $^{59}$Ni), 
and also ($n, \gamma$) and ($n, f$) reactions for long-lived 
actinides. Also in this version MACSs were given at a single 
thermal energy of $kT = 30$ keV, sufficient for studies with the 
canonical $s$-process formulated by 
\textcite{SFC65} for a constant temperature and neutron density 
scenario. A major achievement, however, was the significant 
improvement of the accuracy, which was reaching the 1 - 2\%
level for a number of important $s$-process isotopes.

Meanwhile, the canonical or "classical" approach had been 
challenged by refined stellar  models, which indicated different 
sites for the $s$-process, from He shell burning in thermally 
pulsing low mass AGB stars \cite{GBP88,HoI88} to shell C burning 
in massive stars \cite{RBG91a,RBG91b}, where 
($\alpha, n$) reactions on $^{13}$C and $^{22}$Ne were identified 
as the dominant neutron sources, respectively. The fact that the 
temperatures at the various sites require MACS data for thermal 
energies between 8 and 90 keV was taken into account in the  
compilation of \textcite{BVW92}, which listed 
values in the range $5 \leq kT \leq 100$ keV.

The following compilation of \textcite{BBK00} was  
extended to cover a network of 364 ($n, \gamma$) reactions, including
also relevant partial cross sections. This work presents detailed 
information on previous MACS results, which were eventually 
condensed into recommended values. Again, data are given for 
thermal energies from 5 to 100 keV. For isotopes without experimental 
cross section information, recommended values were derived from
calculations with the Hauser-Feshbach statistical model code 
{\sc NON-SMOKER} \cite{RaT00}, which were empirically corrected 
for known systematic deficiencies in the nuclear input of the 
calculation. For the first time, stellar enhancement factors 
(SEF), which take the effect of thermally excited nuclear states 
into account, were included as well. 

For easy access, the compilation of \textcite{BBK00} 
was published in electronic form via the \textsc{KADoNiS} project 
(http://www.kadonis.org) \cite{DHK05}.
The current version \textsc{KADoNiS} v0.3 \cite{DPK09} 
is already the third update and includes -- compared to the Bao {\it et al.} 
compilation \cite{BBK00} -- recommended values for 38 improved and 14 new cross 
sections. In total, data sets are available for 356 isotopes, including 77 
radioactive nuclei on or close to the $s$-process path. For 13 of 
these radioactive nuclei, experimental data is available, i.e. for $^{14}$C, 
$^{60}$Fe, $^{93}$Zr, $^{99}$Tc, $^{107}$Pd, $^{129}$I, $^{135}$Cs, $^{147}$Pm, 
$^{151}$Sm, $^{155}$Eu, $^{163}$Ho, $^{182}$Hf, and $^{185}$W. The 
remaining 64 radioactive nuclei are not (yet) measured in the stellar 
energy range and are represented only by empirically corrected Hauser-Feshbach 
rates with typical uncertainties of 25 to 30\%. Almost all of the 
($n, \gamma$) cross sections of the 277 stable isotopes have been measured. 
The few exceptions are $^{17}$O, $^{36,38}$Ar, $^{40}$K, $^{50}$V, $^{70}$Zn, $^{72,73}$Ge, 
$^{77,82}$Se, $^{98,99}$Ru, $^{131}$Xe, $^{138}$La, $^{158}$Dy, and $^{195}$Pt,
which lie mostly outside the $s$-process path in the proton-rich $p$-process 
domain. These cross sections are difficult to determine because 
they are often not accessible by activation measurements or not available in 
sufficient amounts and/or enrichment for time-of-flight measurements.

\begin{figure}
\includegraphics[width=12cm]{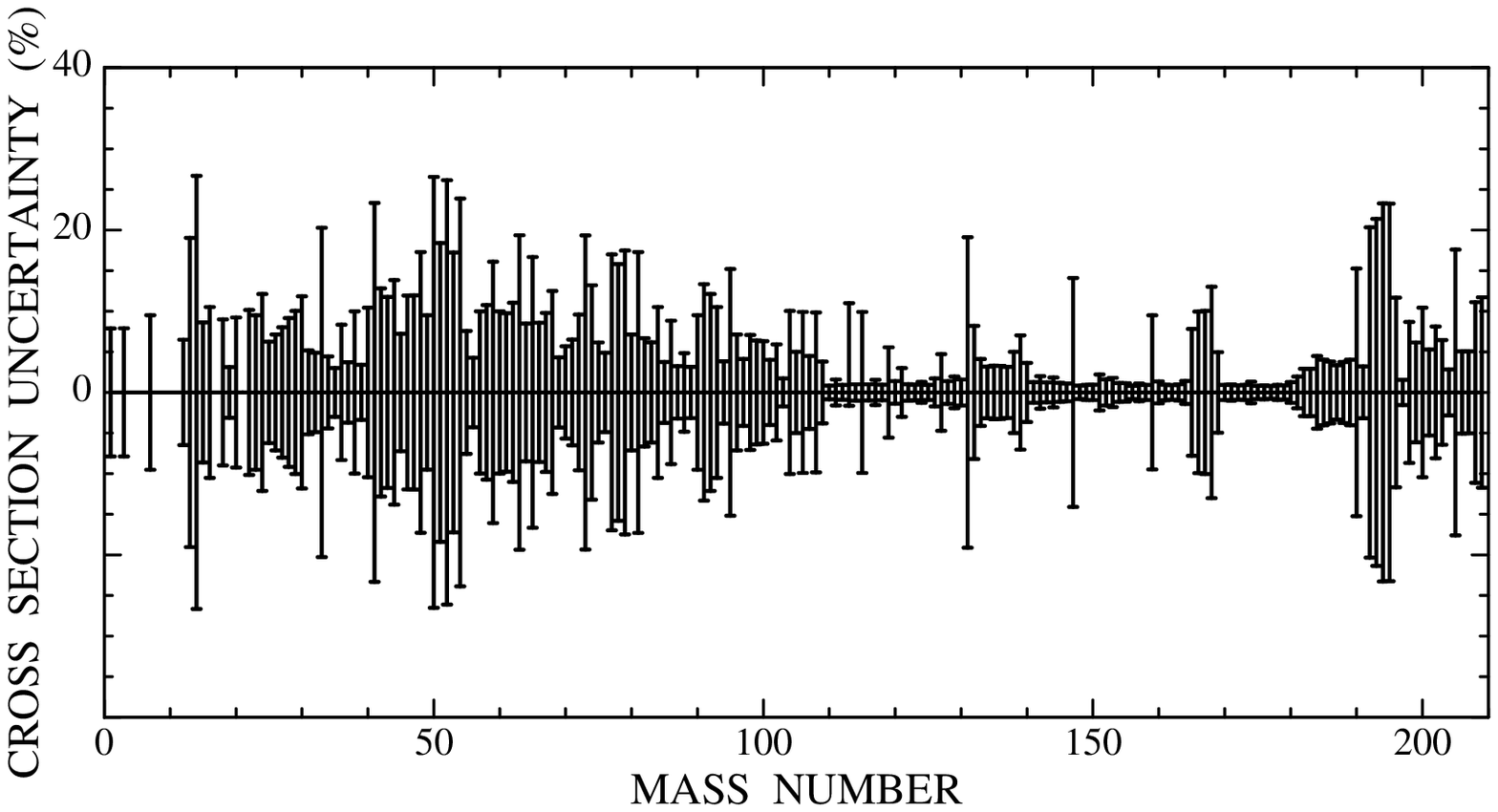}
\caption{\label{fig:6} Uncertainties of the stellar ($n, \gamma$)
            cross sections for $s$-process nucleosynthesis. 
            These values refer to a thermal energy of $kT$ = 30 keV, 
			but may be considerably larger at lower and higher temperatures.}
\end{figure}

The actual status of the ($n, \gamma$) cross sections for $s$-process 
nucleosynthesis calculations is summarized in Fig. \ref{fig:6}, which shows the 
respective uncertainties as a function of mass number. Though the necessary 
accuracy of 1 to 5\% has been locally achieved, further improvements are clearly 
required, predominantly in the mass region below $A$ = 120 and above $A$ = 180.

Further efforts in this field are the more important as Fig. \ref{fig:6} reflects 
only the situation for a thermal energy of 30 keV. In most cases, however, 
extrapolation to lower and higher temperatures implies still larger 
uncertainties.  

The lack of accurate data is particularly crucial for the weak $s$-process 
in massive stars, which is responsible for most of the $s$ abundances between 
Cu and Sr. Since the neutron exposure of the the weak $s$ process is not
sufficient for achieving flow equilibrium, cross section uncertainties may
affect the abundances of a sequence of heavier isotopes (see Sec. \ref{sec3B}).

The present version of \textsc{KADoNiS} consists of two parts: the $s$-process 
library and a collection of available experimental $p$-process reactions. The 
$s$-process library will be complemented in the near future by some ($n, p$) and 
($n, \alpha$) cross sections measured at $kT$ = 30 keV, as it was already included 
in \cite{BaK87}. The $p$-process database will be a collection of all available 
charged-particle reactions measured within or close to the Gamow window of the 
$p$ process ($T_9$= 2-3 GK).

In a further extension of \textsc{KADoNiS} it is planned to include more radioactive 
isotopes, which are relevant for $s$-process nucleosynthesis at higher neutron 
densities (up to 10$^{11}$ cm$^{-3})$ \cite{CGS06}. Since these isotopes are more 
than one atomic mass unit away from the "regular" $s$-process path on the neutron-rich side 
of stability, their stellar ($n, \gamma$) values have to be extrapolated 
from known cross sections by means of the statistical Hauser-Feshbach model. 
The present list covers 73 new isotopes and is available on the \textsc{KADoNiS} 
homepage.

\subsubsection{Measurements on rare and unstable samples \label{sec2D2}}

The continuous development and optimization of techniques and facilities 
remains a most vital aspect of the field, especially with respect to 
unstable isotopes. This concerns the production of higher fluxes, i.e. by 
means of shorter flight paths at existing spallation sources, e.g. a 20
m station at n\_TOF, or beam lines for keV neutrons at the new generation of
high intensity accelerators such as the japanese proton accelerator complex 
J-PARC \cite{JPC04}. 

A completely new approach is presently developed at the University 
of Frankfurt \cite{RCM07}. The Frankfurt Neutron source at 
the Stern-Gerlach-Zentrum (FRANZ) will provide short neutron pulses 
by bombardment of a $^7$Li target with an intense proton beam.
The proton energy range is limited to 2.0$\pm$0.2 MeV and the 
pulse rate will be typically 250 kHz. 
 
\begin{figure}
\includegraphics[width=12cm]{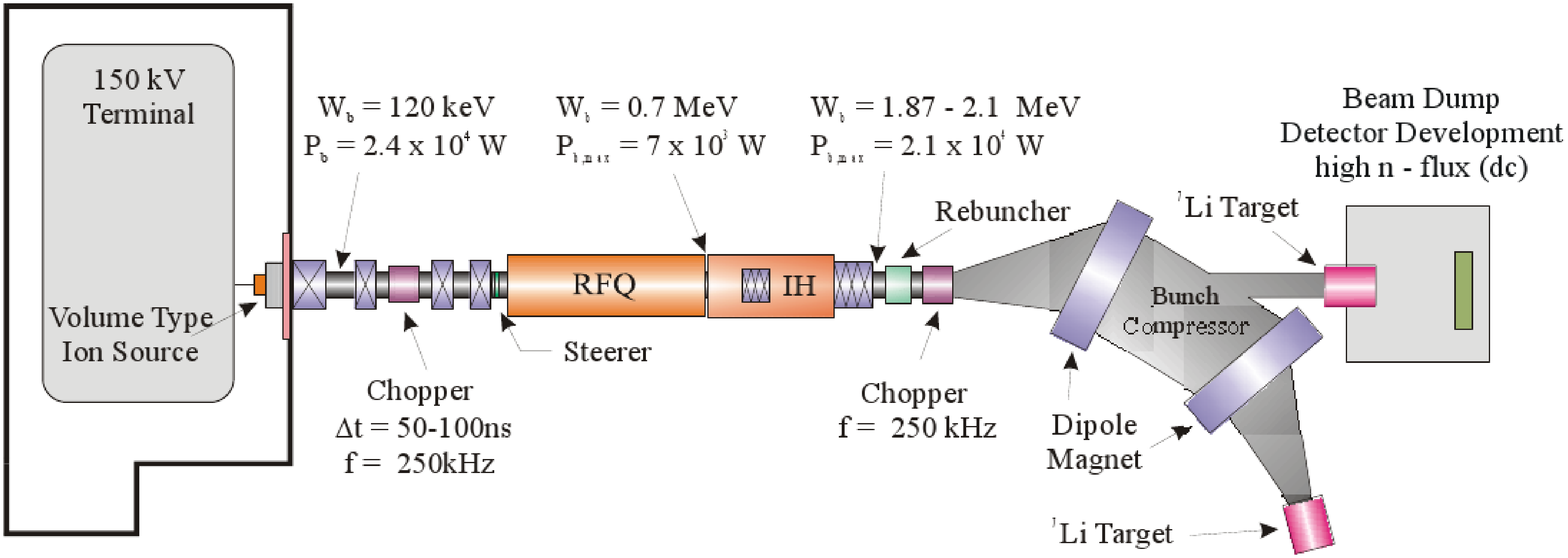}
\caption{\label{fig:7} (Color online)
            Schematic layout and main parameters of the Frankfurt 
			intense neutron source (see text).}
\end{figure}

The scheme of the accelerator is sketched in Fig. \ref{fig:7},
starting with a volume-type proton source on a 150 kV high 
voltage platform, followed by a 100 ns chopper in the low energy 
beam transport line to the acceleration stage consisting 
of an RFQ with an exit energy of 700 keV and IH-DTL structure
with an effective energy gain of 1.4 MeV. The macro pulses at 
the exit contain up to 10 rf bunches, which are compressed into 
1 ns pulses for neutron production in the Li target. This 
is achieved by a second chopper, which deflects the bunches 
to traces of different path length in a Mobley-type buncher 
system \cite{CMR06,MCM06,RCM07}. 

In this way, intense pulses of up to $5 \times 10^{10}$ protons 
can be focused onto the Li target within 1 ns. The average beam 
current is expected to reach 2 mA, corresponding to a neutron flux 
of about 10$^5$ n/cm$^{2}$/s/keV at a distance 
of 80 cm from the target, more than an order of magnitude higher 
than is achieved at present spallation sources. Since the amount of 
sample material can be reduced by the gain in flux, TOF
measurements at FRANZ appear to be feasible with samples 
of 10$^{14}$ to 10$^{16}$ atoms. The use of such minute amounts represents 
a break-through with respect to the production of unstable 
samples, because beam intensities of the order of 10$^{10}$ to 
10$^{12}$ s$^{-1}$ are expected at future Rare Isotope Facilities 
such as RIKEN~\cite{Tan98}, \citet{Fai07}, or \citet{Fri10}.

The gain in beam intensity and the related reduction in
sample mass implies that TOF measurements can be carried out 
on unstable nuclei with correspondingly higher specific activities,
including a number of branch point isotopes, which are not
accessible by activation techniques. The concise list of important 
branch point isotopes in Table \ref{tab3} shows that most of these
samples are accessible to TOF measurements at FRANZ thanks 
to the excellent sensitivity of that facility. It should be
noted, however, that Table \ref{tab3} includes a number of cases, 
which could, in principle, be studied at existing facilities. Such 
measurements are impeded, however, because isotopically enriched 
samples are not available in sufficient amounts. Among the 20 listed 
unstable isotopes in Table \ref{tab3} there are only three cases,
where the specific $\gamma$ activities seem to be too high for
a promising experiment in the foreseeable future.

In addition to the gain in TOF sensitivity, FRANZ is also perfectly 
suited for the simulation of stellar neutron spectra via $^7$Li($p, n$)$^7$Be 
reaction (Sec. \ref{sec2A4}). In fact the larger dispersion 
in proton energy will result in a closer approximation of the 
stellar spectrum as was pointed out by \textcite{MMP09}. Similar to
the situation sketched for TOF measurements, the higher flux
will enable one to apply the activation method to a largely
extended set of shorter-lived unstable nuclei, an important aspect 
for investigating the $s$-process during shell C-burning in
massive stars, where the reaction path is shifted by a few mass 
units from the valley of stability due to neutron densities in 
excess of $10^{12}$ cm$^{-3}$ \cite[e.g. ][]{PGH10}. In combination
with AMS it will even be possible to use the double neutron capture
method to obtain the MACS for such crucial cases as $^{59}$Fe, 
$^{125}$Sn, and $^{181}$Hf.

Apart from their importance for specific problems and their direct use 
in $s$-process networks, MACSs of unstable isotopes represent also most 
valuable information in a wider sense, particularly for testing and 
improving statistical model calculations in areas, which are not 
accessible to experiments in the near future.

\section{$s$-Process Models \label{sec3}}

Stellar models for the He burning stage of stellar evolution have been 
worked out in great detail over the past decade, both for low mass 
stars in the AGB phase while suffering recurrent thermal pulses in 
the He shell, and for massive stars. There, besides central He burning 
in the convective core, neutrons are released in the subsequent 
convective shell C burning phase, which involves a large fraction 
of the final ejected mass in the supernova event. Accounting for a 
continuous updated network of neutron captures and charged particle 
reaction rates, the $s$ process taking place in both low-mass stars
and massive stars will be discussed.

\subsection{Classical approach \label{sec3A}}

Shortly after stellar spectroscopy of the unstable element Tc provided 
evidence for active neutron capture nucleosynthesis in red giant stars 
\cite{Mer52b}, the canonical or classical model of the $s$ process was 
suggested  by B$^2$FH \cite{BBF57}. Although it was argued that the
He-burning zones of red giants were the most promising site of the $s$ process, 
the lack of detailed stellar models led to a phenomenological solution. 
Within this approach it is empirically assumed that a certain fraction 
$G$ of the observed $^{56}$Fe abundance was irradiated by an exponential 
distribution of neutron exposures \cite{SFC65}. In this case, an analytical 
solution can be obtained if a possible time dependence of the neutron 
capture rates, $\lambda_n  = n_n \langle\sigma\rangle v_T$ is neglected. 
In other words, it is assumed that temperature and neutron density, $n_n$, 
are constant. Then, the product of stellar cross section and resulting 
$s$ abundance, which characterizes the reaction flow, can be given by 

\begin{equation}
\langle \sigma \rangle_{(A)} N_{s(A)} = \frac{G \cdot N^{\odot}_{56}}{\tau_0} 
       \prod_{i=56}^{A} (1 + \frac{1}{\tau_0 \langle \sigma \rangle_i})^{-1}.
      \end{equation}

Apart from the two parameters $G$ and $\tau_0$ (which are adjusted by 
fitting the abundances of the $s$-only nuclei), the only remaining input 
for this expression are the stellar (n,$\gamma$) cross sections 
$\langle\sigma\rangle$ \cite{KBW89,WIP97}.

Given the very schematic nature of this classical approach, it surprisingly 
provided an excellent description of the $s$-process abundances in the solar 
system (Fig. \ref{fig:2}). One finds that equilibrium in the neutron 
capture flow was obtained between magic neutron numbers, where the 
$\langle \sigma \rangle N_{s}$-curve is almost constant. The small cross 
sections of the neutron magic nuclei around A$\sim$88, 140, and 208 act as 
bottlenecks for the capture flow, resulting in distinct steps in the 
$\langle \sigma \rangle N_{s}$-curve.

The global parameters, $G$ and $\tau_0$, which determine the overall shape 
of this curve, represent a first constraint for the stellar $s$-process 
site with respect to the required seed abundance and total neutron exposure. 
It is found that 0.04\% of the $^{56}$Fe abundance observed in solar system 
material is a sufficient seed, and that on average about 15 neutrons are 
captured by each seed nucleus \cite{KGB90}. These numbers refer to the {\it 
main} $s$-process component, which dominates the $s$ abundances for A$>$90. 
The rather steep increase of the $\langle \sigma \rangle N_{s}$-curve below 
A$=$90 requires an additional component, the {\it weak} component. 

The weak component is not firmly described by the classical analysis,
because there are only six $s$-only isotopes below $A=90$ ($^{70}$Ge, 
$^{76}$Se, $^{80,82}$Kr, $^{86,87}$Sr), which are also partly produced by 
the main component and possibly even by the $p$ process. Moreover, 
$^{80}$Kr and $^{86,87}$Sr are affected by branchings in the $s$-process 
path. The temperature and neutron density determining these branchings 
are to be treated as free parameters.  Accordingly, it is difficult 
within the classical approach to distinguish between a single 
component or an exponential distribution of neutron exposures \citep{Bee86, 
BeM89,KBW89}.

For about 40 years the classical model was quite successful in describing 
the solar $s$-process abundances \cite{KBW82,KGB90}. In fact, the empirical 
$\langle \sigma \rangle N_{s}$-values of the $s$-only isotopes  
that are not affected by branchings, are reproduced with a mean square 
deviation of only 3\% \cite{KGB90} as illustrated in Fig. 
\ref{fig:2}. About 20 years ago the development of new experimental techniques 
led to a set of accurate neutron capture cross sections, which ultimately
revealed that the classical $s$ process suffered from inherent inconsistencies. 
This was convincingly demonstrated at the example of the neutron-magic $s$-only 
isotope $^{142}$Nd \cite{AKW99b}. At this point, the classical approach 
was replaced by a first generation of stellar $s$-process models, where such 
problems could be successfully avoided \cite{GAB98, AKW99b}. 

\begin{figure}[tb]
\includegraphics[width=14cm]{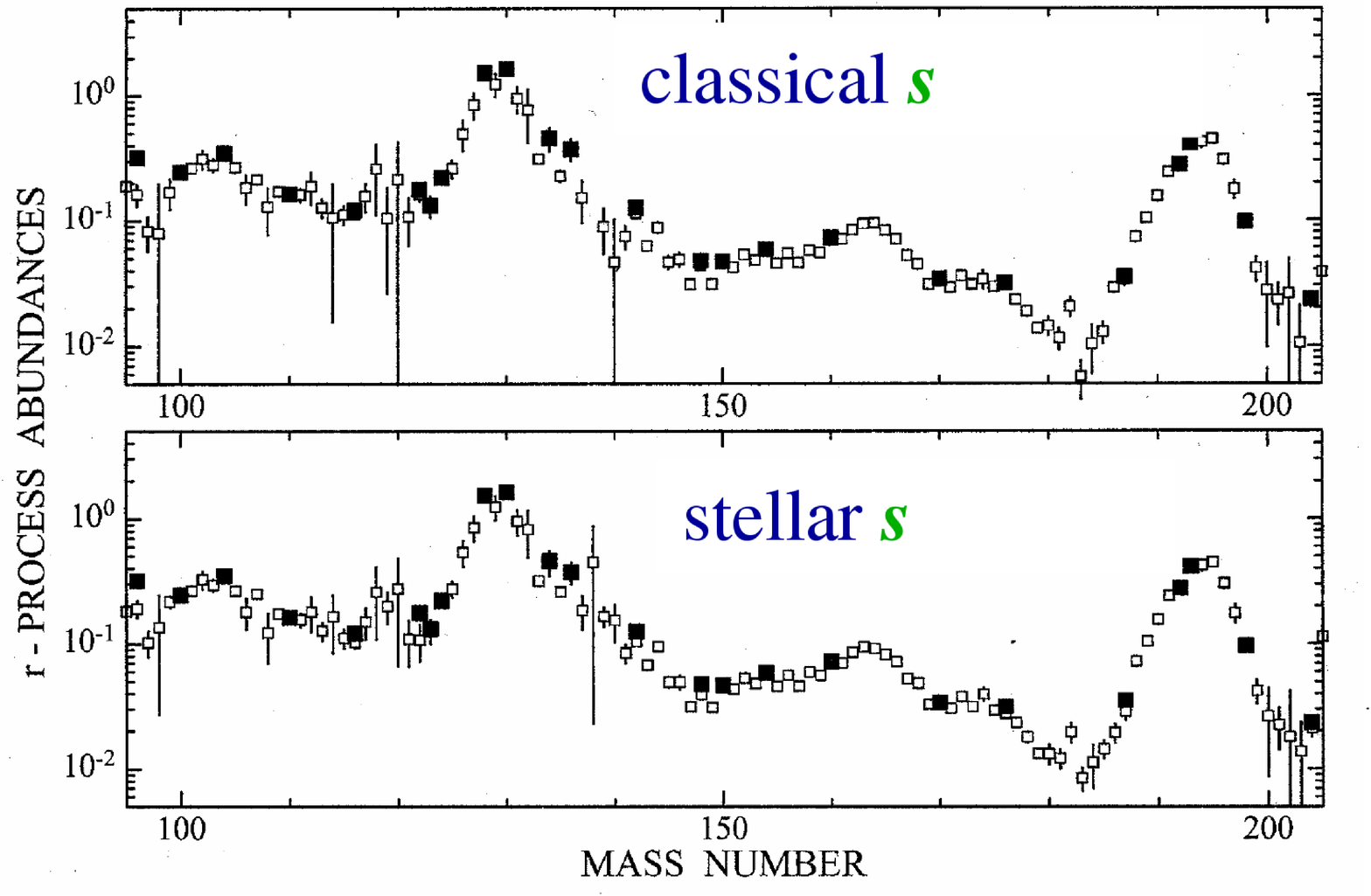}
 \caption{ (Color online) The $r$-process abundances (open squares) obtained via the 
		   $r$-process residual method, N$_r$ = N$_{\odot}$ - N$_s$, 
		   using the classical and stellar $s$-process model \citep{AKW99b}.
		   The $r$-only nuclei are represented by full squares.
		   (N$_i$ relative to Si $\equiv$ 10$^6$). \label{fig:8}}
\end{figure}

Nevertheless, the classical model provides still a useful tool for estimating 
the $s$ abundances, in particular in mass regions between magic numbers, where 
the cross sections are large enough that reaction flow equilibrium was actually 
reached. This is illustrated by the example of the solar $r$-process 
distribution obtained with the $r$-residual method,
$$N_r = N_{\odot} - N_s,$$
which is obtained by the difference between the solar abundances and the 
corresponding $s$-process yields. The $r$ distributions obtained via the 
classical approach or via stellar models are quite similar and match very 
well with the abundances of the $r$-only isotopes as shown in Fig. \ref{fig:8}. 

The separation of the solar abundance distribution into the $s$ and $r$ 
components became also important for the interpretation of abundance 
patterns of ultra-metal-poor stars \cite{WSG00,HCB02,CSB02,SCL03}, which 
turned out to agree very well with the scaled solar $r$ component for 
the elements heavier than Ba.

\subsection{Massive stars} \label{sec3B}

As discussed at the end of the previous Sec. \ref{sec3A}, the classical 
analysis of the $s$ process fails to reproduce the cosmic abundances 
of the $s$ isopotes below A = 90. The same problem is found for stellar 
models of low-mass AGB stars, as described in detail in Sec. \ref{sec3C}. 
Actually, a complementary weak $s$ process occurs
in massive stars ($M > 8 M_\odot$), which explode as supernovae of type 
II. Slow neutron captures are driven by the reaction $^{22}$Ne($\alpha, 
n$)$^{25}$Mg during convective core He burning at temperatures around 
3 $\times$ 10$^8$ K as well as in the subsequent convective shell C 
burning at 1 $\times$ 10$^9$ K \citep{CSA74,LHT77,KBW89,PHN90,BWK92b,RGB93}. 
The available $^{22}$Ne is produced via the reaction sequence 
$^{14}$N($\alpha, \gamma$)$^{18}$F($\beta^+ \nu$)$^{18}$O($\alpha, 
\gamma$)$^{22}$Ne, where $^{14}$N derives from the CNO cycle in the 
previous H-burning phase. Consequently, this weak $s$-process component 
produced in massive stars is secondary like, decreasing with metallicity.
At He exhaustion, not all the $^{22}$Ne is consumed
\cite[e.g. ][]{PHN90}, and neutron production via $^{22}$Ne($\alpha, 
n$)$^{25}$Mg continues during shell C-burning by means of the
$\alpha$-particles provided by the reaction channel $^{12}$C($^{12}$C, 
$\alpha$)$^{20}$Ne \citep{ArT69}. 

In the ejecta of SN II, the chemical composition of the core up to a mass 
of ~ 3.5 $M_\odot$ (for a star of 25 $M_\odot$) is modified by explosive 
nucleosynthesis, which destroys any previous $s$-process signature. However,
the ejecta still contain also an important mass fraction of ~ 2.5 $M_\odot$, 
which preserves the original $s$-process abundances produced by the 
hydrostatic nucleosynthesis phases of the presupernova evolution.

This scenario was confirmed by post-process models and full stellar 
models describing the evolution of massive stars up to the final burning 
phases and the SN explosion \citep{RGB93,WoW95,LSC00,WHW02,TEM07,ETM09,PGH10}. 

In contrast to the main $s$-process component, the neutron fluence in 
the weak $s$ process is too low for achieving reaction flow equilibrium. 
This has the important consequence that a particular MACS does not only 
determine the abundance of the respective isotope, but affects also
the abundances of all heavier isotopes as well \citep{PGH10}. This 
propagation effect is particularly critical for the abundant isotopes near 
the iron seed. A prominent example is the case of the $^{62}$Ni($n, 
\gamma$)$^{63}$Ni cross section, where the effect was discussed first 
\citep{RHH02,RaG02,RaG05}. This problem has triggered a series of experimental 
studies on that isotope \citep{NPA05,TTS05,ABE08,LBC10} and on other 
key reactions of the weak $s$ process between Fe and Sr, which could be
considerably improved \cite[e.g. ][]{HKU08a,HKU08b}. For a full account see 
the recent update of the KADoNiS library \citep{DPK09}. 

The cumulated uncertainties of the propagation effect are significant even
for the heavier isotopes of the weak $s$ process, up to Kr and Sr, with
a possible, minor contribution to the Y and Zr abundances \citep{PGH10}.
The corresponding uncertainties will be partly solved once the neutron 
capture cross sections of the isotopes between Fe up to Sr will be measured 
with an accuracy of 5\% (Fig. \ref{fig:6}).

Apart from these problems with the neutron capture
cross sections, it has been pointed out that the weak $s$ process is also still affected by large 
uncertainties of several charged-particle reactions during He and C 
burning \cite[e.g. ][]{TEM07,BHP10}.

\subsection{\label{sec3C}AGB stars} 

\begin{figure}
  \centering
\includegraphics[angle=0,width=10cm]{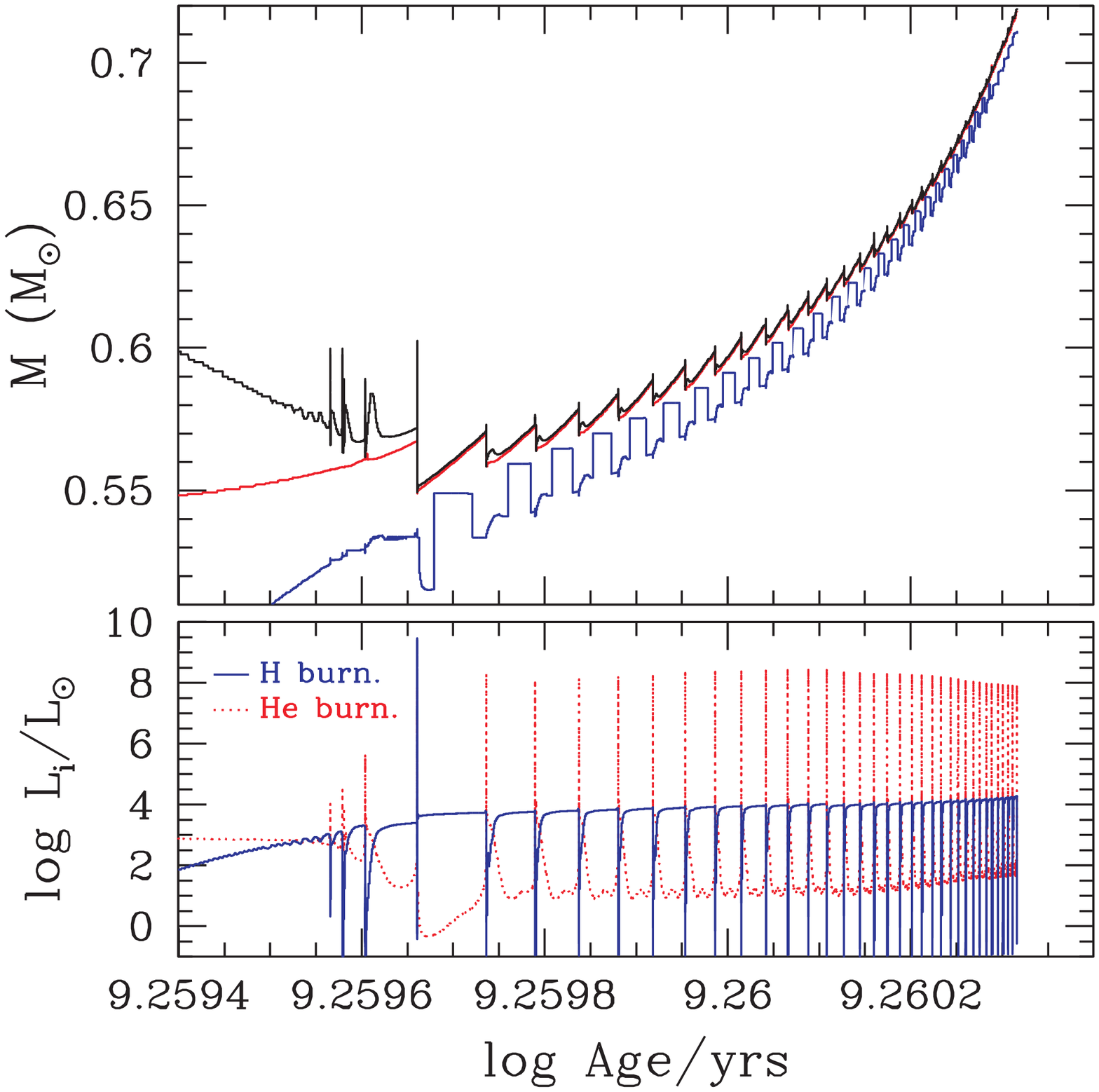}
\caption{(Color online) Structural characteristics versus age of an AGB model with 
initial mass $M^{\rm AGB}_{ini}$ = 2 $M_\odot$\ and solar metallicity.
Upper panel, top to bottom: Temporal evolution of the mass coordinates 
of the inner border of the convective envelope, the mass location of
maximum energy production in the H-burning shell, and the maximum 
energy production within the H-depleted core. During each interpulse 
period, the flat segment of the lowest (blue) line corresponds to 
the location of the radiative burning of the $^{13}$C($\alpha, 
n$)$^{16}$O reaction in the pocket. Lower panel: Temporal evolution 
of the H-burning and He-burning contributions to luminosity (from 
\citealt{CPS09a}).}
\label{fig:9}
\end{figure} 

\begin{figure}
  \centering
\includegraphics[angle=0,width=13cm]{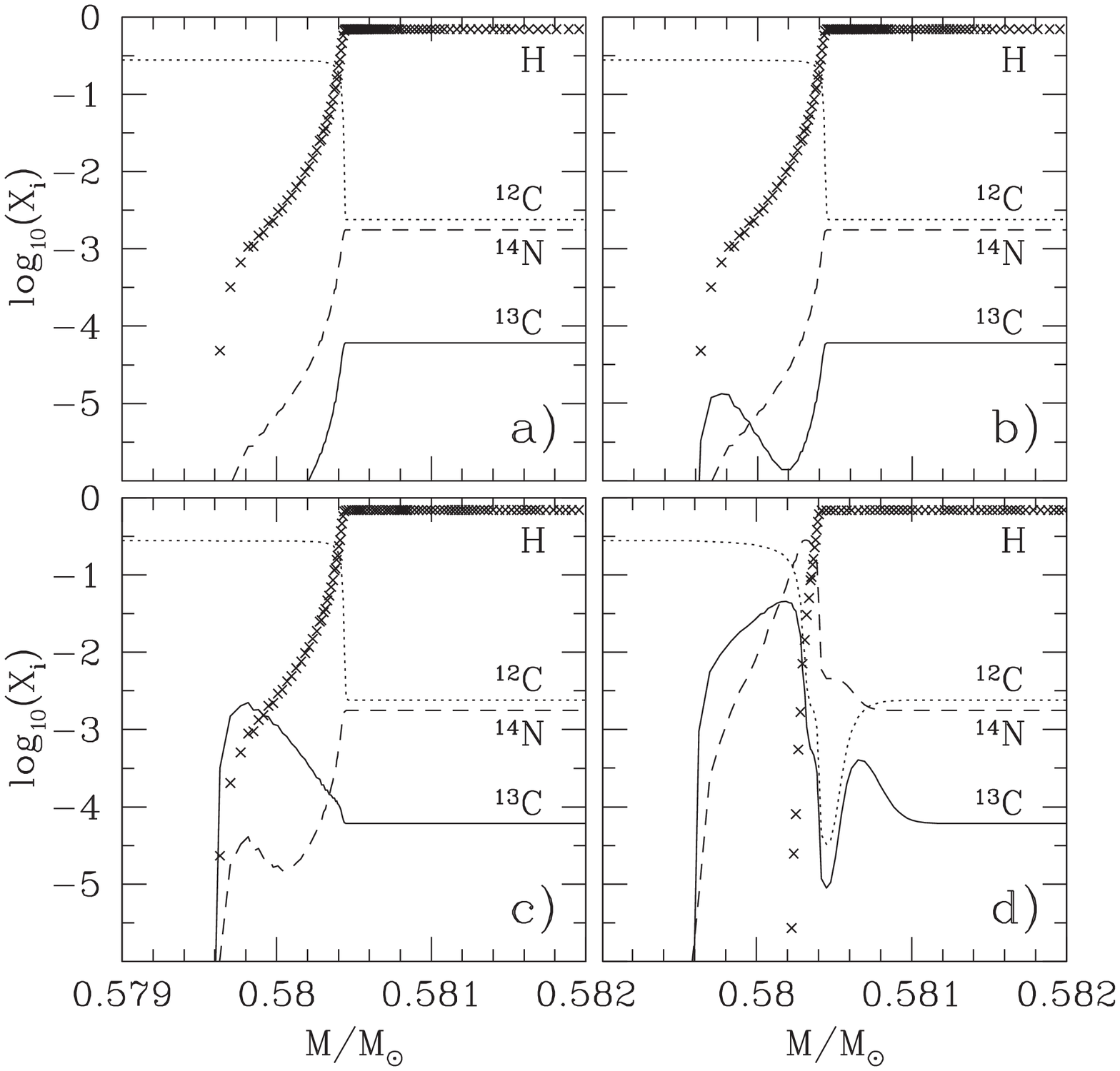}
\caption{The formation of  the $^{13}$C pocket according to 
\citet{SCG09}. The sequence of the panels shows the 
evolution of the chemical composition in the transition zone 
between the H-rich envelope and the H-exhausted core.
The various lines represent the abundances of H (crosses), 
$^{12}$C (dotted), $^{13}$C (solid), and $^{14}$N (dashed). 
Panel (a): The TDU just occurred and the convective envelope 
is receding. Panel (b): Production of $^{13}$C by proton capture 
on the abundant $^{12}$C starts in the hotter region. Panel (c): 
Some $^{14}$N is also produced, and panel (d): the
$^{13}$C (and $^{14}$N) pocket is fully developed.}
\label{fig:10}
\end{figure} 

During the AGB phase, the H- and He-burning shells are activated 
alternately on top of the degenerate C-O core. These two shells 
are separated by a thin zone in radiative equilibrium, the so-called 
He-intershell, enriched in He and C. The H-burning shell erodes the 
bottom layers of the envelope and produces He. The He-intershell 
grows in mass and is progressively compressed and heated until He 
burning is triggered in a quasi explosive way (thermal pulse, TP; 
\citealt{ScH65,Wei66}). For a general discussion see 
\citet*{IbR83,BGW99,Her04,SGC06,SCG08}. 

The sudden release of energy due to a TP drives convection in the 
whole intershell for a short period of time. During a TP, partial 
He burning occurs producing a large amount of $^{12}$C. The envelope 
expands and the H shell is temporarily extinguished. He shell burning 
continues radiatively for another few thousand years, and then H shell 
burning starts again. After a limited number of TPs, when the mass of 
the H-exhausted core reaches $\sim$ 0.6 $M_\odot$ and the H-shell is 
inactive, the convective envelope penetrates into the top region of 
the He-intershell and mixes newly synthesised material to the surface 
(third dredge-up, TDU). 

The star undergoes recurrent TDU episodes, whose occurrence and efficiency 
depend on the physical and numerical treatment of the convective borders.
The TDU is influenced by the parameters affecting the H-burning rate, such 
as the metallicity, the mass of the H-exhausted core, and the mass of the 
envelope, which in turn depends on the choice of the mass loss rate by 
stellar winds \cite[see discussion in ][]{SGC06}. The upper panel 
of Fig.~\ref{fig:9} illustrates the changes of the structural 
characteristics of a given AGB model with time 
\citep{CSG09,CPS09a}, the position in mass coordinates of the
inner border of the convective envelope, the maximum energy production of 
the H-burning shell and the maximum energy production within the H-depleted 
core. During each interpulse period, the flat segment of the lowest (blue) line
corresponds to the radiative burning of the $^{13}$C($\alpha, n$)$^{16}$O
reaction. The lower panel shows the H-burning and He-burning
contributions to the luminosity.

During the TP-AGB phase, the envelope becomes progressively enriched in 
primary $^{12}$C and in $s$-process elements. TDU drives 
a chemical discontinuity between the H-rich envelope and the He-intershell, 
where a few protons likely penetrate into the top layers of the He intershell. 
At hydrogen reignition, these protons are captured by the abundant $^{12}$C 
forming $^{13}$C via $^{12}$C($p, \gamma$)$^{13}$N($\beta^+ \nu$)$^{13}$C 
in a thin region of the He-intershell ($^{13}$C pocket). Neutrons are 
released in the pocket under radiative conditions by the $^{13}$C($\alpha, 
n$)$^{16}$O reaction at $T \sim 0.9 \times 10^8$ K. This neutron exposure 
lasts for about 10,000 years with a relatively low neutron density of 
10$^{6}$ to 10$^{8}$ cm$^{-3}$. The pocket, strongly enriched in 
$s$-process elements, is then engulfed by the subsequent convective TP. 
Models including rotation \citep{LHW99} or gravity waves
\citep{DeT03} have obtained a partial mixing zone at the base of 
the convective envelope during the TDU episodes, which leads to the 
formation of a $^{13}$C-rich layer of limited mass extension. 
\citet{HBS97} and \citet{Her00,Her04}, guided by dynamical 
simulations of \citet{FLS96}, introduced an exponential diffusive 
overshoot at the borders of all convective zones. A formally similar 
algorithm based on a non-diffusive mixing scheme 
has been proposed by \citet{SGC06}. Applying the Schwarzschild 
criterion to determine the border of the H-rich convective envelope 
and the inner He-rich and C-rich intershell, a thermodynamical 
instability would ensue. In order to handle this instability, \citet{SGC06} 
assumed an exponentially decaying profile of the convective velocity
$$ v = v_{bce} {\rm exp}(-d/\beta H_P)$$
where $d$ is distance from the convective boundary, $v_{\rm bce}$ the 
average element velocity at the convective boundary (as derived by means
of the mixing length theory), $H_P$ the pressure scale height at the 
convective boundary, and $\beta$ a free parameter (for a proper choice 
see \citealt{CSG09}). Fig.~\ref{fig:10} illustrates the formation 
of the $^{13}$C pocket according to the full evolutionary model described 
by \citet{SCG09}. The hydrogen profile adopted in the pocket 
and the consequent amount of $^{13}$C (and of $^{14}$N) determines the final 
$s$-abundance distribution.

At the maximum extension of the convective TP, when the temperature 
at the base of the convective zone exceeds $2.5 \times 10^{8}$ K, a 
second neutron burst is powered for a few years by the marginal activation
of the $^{22}$Ne($\alpha, n$)$^{25}$Mg reaction. This neutron burst is 
characterized by a low neutron exposure and a high neutron density 
up to 10$^{10}$ cm$^{-3}$, depending on the maximum temperature reached 
at the bottom of the TP. The dynamical conditions in which the two neutron sources 
operate are defining the final abundances of nuclei involved in 
branchings along the $s$-process path. A comparison of 
low mass AGB models computed with different evolutionary codes has 
been discussed by \citet{LHL03}.

In AGB stars of intermediate mass ($4 < M/M_{\odot} < 8$) the maximum 
temperature during a TP reaches about $3.5 \times 10^8$ K, leading 
to a substantial neutron production via the $^{22}$Ne($\alpha, 
n$)$^{25}$Mg reaction. However, both the mass of the He-intershell 
and the TDU efficiency are much smaller in these stars than in 
low-mass AGBs. Consequently, the predicted $s$-process abundances 
in the envelope are fairly low.


The production of the $s$ elements at very low metallicity ([Fe/H] 
$<$ $-$2.5) may be affected by a new phenomenon, which is limited to 
AGB stars of the lowest mass leading to TDU and takes place only 
during the first fully developed TP. There, 
the reduction of CNO catalysts is compensated by an increase of 
the temperature in the H-shell and, consequently, the entropy 
in the H-shell decreases. Under these conditions, the first convective 
He instability may expand over the H-shell, thus engulfing protons from 
the envelope, which are instantly captured by the abundant $^{12}$C 
in the convective region. The $^{13}$C($\alpha, n$)$^{16}$O reaction 
is now occurring during the TP, in competition with the $^{22}$Ne($\alpha, 
n$)$^{25}$Mg reaction at the bottom of the TP.
  
This complex feature has been found by many authors despite the 
different physics adopted in the various works 
\citep{HIF90,FII00,IKM04,SCG04,CSL07,SKY08,CaL08,WHP08,LST09}.
First calculations of the consequences of this mechanism for the 
$s$ process have been made by \citet{CPS09a} for an AGB model of 
an initial mass $M = 1.5 M_\odot$, a metallicity of [Fe/H] = 
$-$2.6, and no enhancement of the $\alpha$ elements. The convective 
shell was found to split into two sub-shells: 
the lower one boosted by the $^{13}$C($\alpha, n$)$^{16}$O reactions 
and the upper one by the CNO cycle. Once the splitting has occurred, 
the nucleosynthesis in the two shells exhibits a completely different 
behavior. 
In the upper shell, the very large $^{13}$C abundance is marginally 
consumed via $^{13}$C($\alpha$, n)$^{16}$O, leading to the production 
of a corresponding amount of ls elements (For the definition of ls and 
hs elements see Sec. \ref{sec3D} below). This peculiar phase is followed 
by a deep third dredge-up episode, which mixes freshly synthesized $^{13}$C, 
$^{14}$N, and ls elements into the envelope. The second TDU carries 
a large amount of hs elements and Pb to the surface, which was 
previously synthesized by $^{13}$C($\alpha$, n)$^{16}$O reactions 
in the lower splitted shell. After this initial event, the 
subsequent series of TPs and TDUs follow the standard pattern. The 
whole problem and its consequences on the surface abundances are 
currently a matter of intense study. This phenomenon may be of interest 
for the analysis of some Carbon-Enhanced Metal-Poor 
(CEMP) stars showing $s$-process enhancement (CEMP-$s$ stars) 
(Secs. \ref{sec3D} and \ref{sec4B2}).

\subsection{Theoretical AGB results \label{sec3D}} 

The $s$ process in AGB stars is not a unique process, but depends on the 
initial mass, metallicity, the strength of the $^{13}$C pocket, and the 
choice of the mass loss rate. Important observational constraints derive 
from the abundances of elements belonging to the three $s$-process peaks 
located at the magic neutron numbers N = 50, 82, and 126. The peaks occur
because the low neutron capture cross sections of Sr, Y, Zr (light $s$-process 
elements, ls), Ba, La, Ce, Nd, Sm (heavy $s$-process elements, hs), and 
Pb act as bottlenecks for the $s$-process reaction path. For a given 
metallicity, a spread in the three $s$-process peaks is observed in stars 
of spectral type MS, S, C(N) and Ba stars of the Galactic disk (see 
\citealt{BLB95,BGL01,ABG01,ADG02,GWB05}). At low metallicities, 
high-resolution spectroscopic measurements of CEMP-$s$ stars showed an 
even larger spread \citep{ISG05,ABG06,TIB08,RFS08,SCG08}. The observed 
spread can be interpreted by assuming different $s$-process efficiencies 
of these stars, corresponding to a change in the amount of 
$^{13}$C in the pocket.

We discuss here theoretical results from models based on the 
FRANEC code (Frascati Raphson-Newton Evolutionary Code, \citealt*{ChS89}),
coupled with a post-process code that includes a full $s$-process network 
up to Bi \citep{BGS10}. In the post-process code, the prescriptions 
for the amount of the dredged-up mass, the number of TDUs, the choice of 
the mass loss rate as well as for the temporal history of the temperature 
and density during the TPs were adopted from \citet{SDC03,SGC06}.
The $^{13}$C pocket is artificially introduced starting from the ST case 
adopted by \citet{GAB98} and \citet{AKW99b}, which was shown to reproduce 
the solar main $s$-process component as the average of the 1.5 and 3.0 
$M_{\odot}$ models at half solar metallicity (Sect.\ref{sec3E}). 
The $^{13}$C (and $^{14}$N) abundances in the pocket were then multiplied 
by different factors. A minimum $^{13}$C pocket may be defined as the one 
that affects the final $s$-process distribution. Higher $^{13}$C-pocket 
efficiencies than case ST $\times$ 2 would not produce a correspondingly 
higher abundance of $^{13}$C, because of the increasing competition by 
$^{13}$C($p, \gamma$)$^{14}$N reactions. For any given model, the efficiency
of the $^{13}$C pocket is assumed to be constant for all TPs.

\begin{figure}
  \centering
\includegraphics[angle=-90,width=13cm]{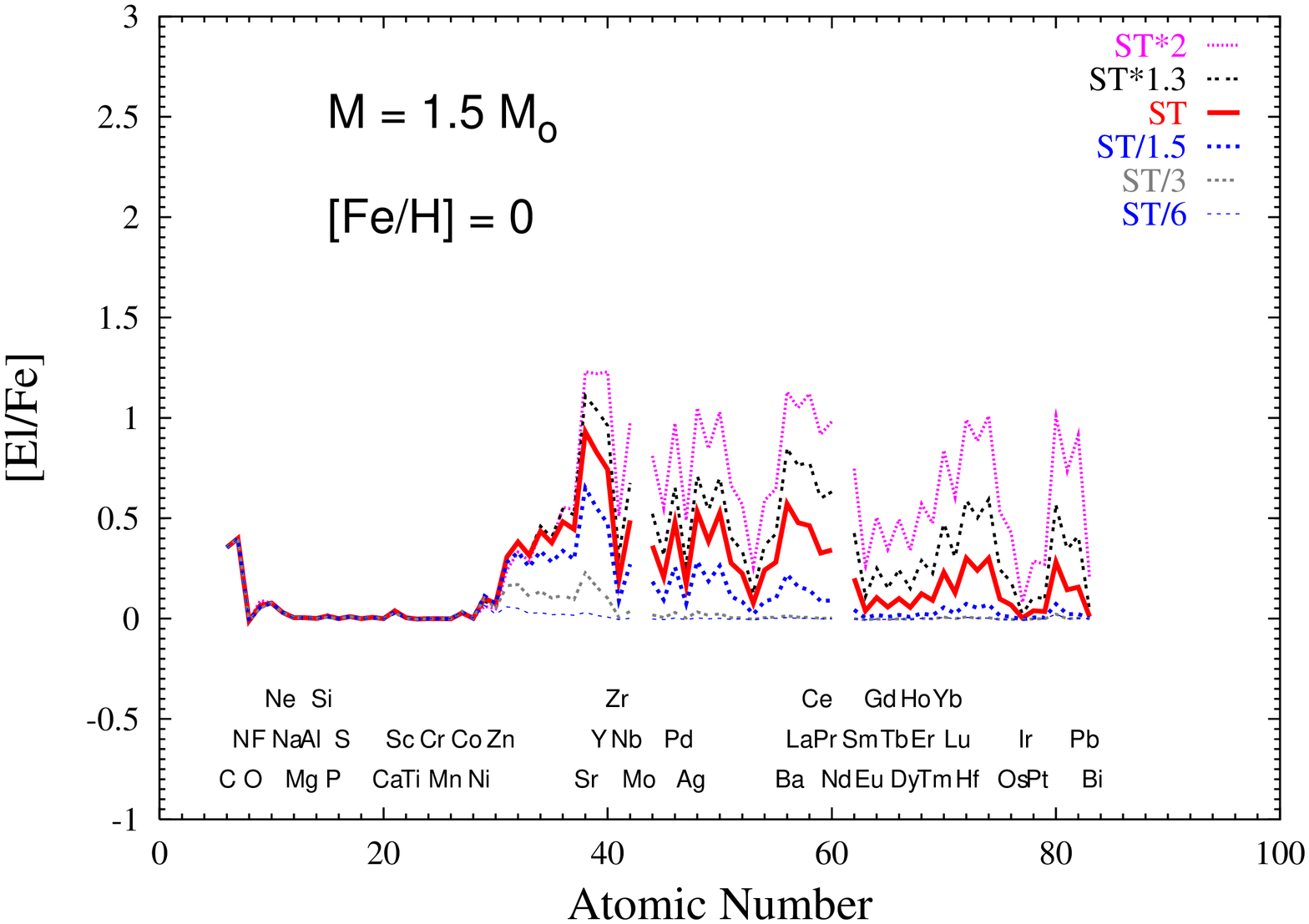}
\includegraphics[angle=-90,width=13cm]{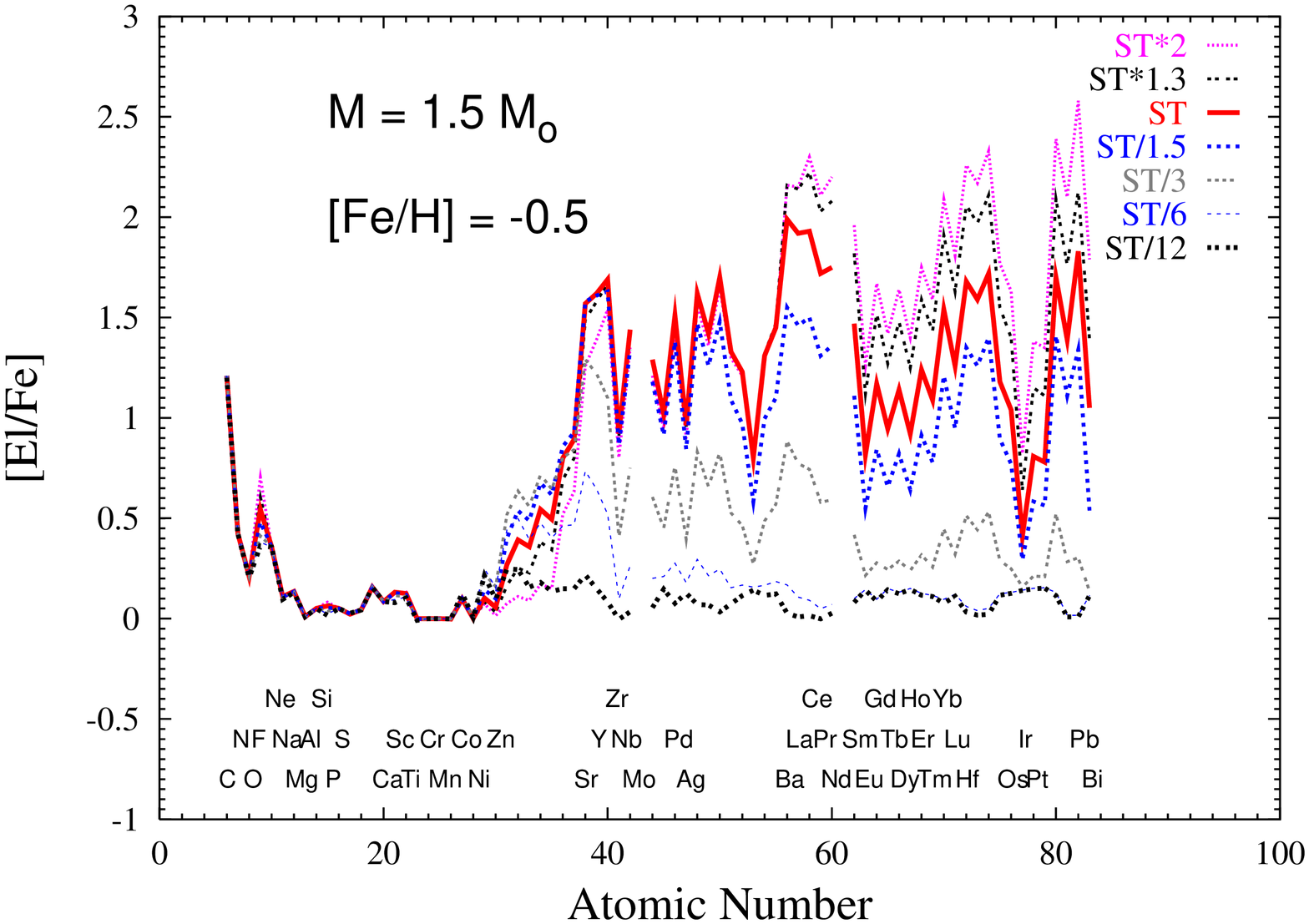}
\caption{(Color online) Top panel: Theoretical results of [El/Fe] versus 
atomic number at the last TDU episode in the envelope of an AGB star with 
initial mass $M$ = 1.5 $M_{\odot}$ and [Fe/H] = 0. Bottom panel: The 
same as top panel but [Fe/H] = $-$0.5. Adapted from \citet{BGS10} 
and \citet{HGB09}.}
\label{fig:11}
\end{figure}

The theoretical predictions in the envelope for elements from C to 
Bi ([El/Fe]) versus atomic number $Z$ are shown in Fig.~\ref{fig:11} 
for an AGB star with an initial mass $M$ = 1.5 $M_{\odot}$ and for a range 
of $^{13}$C pockets (from case ST down to ST/12) at solar metallicity (top 
panel) and at [Fe/H] = $-$0.5 (bottom panel) (see \citealt{BGS10}; \citealt{HGB09}).
With decreasing metallicity, the $s$-process distribution is shifted toward 
heavier elements. This is the consequence of the primary nature of the $^{13}$C 
neutron source: while $^{56}$Fe (the seed of the $s$-process) decreases with 
metallicity, the number of neutrons available per iron seed increases 
\citep{Cla88}. Hence, at halo metallicities, the neutron fluence 
overcomes the first two peaks, directly feeding $^{208}$Pb (\citealt{GAB98,GoM00,TGB01b}; 
see Sect.\ref{sec3E}).

\begin{figure}
  \centering
\includegraphics[angle=-90,width=10cm]{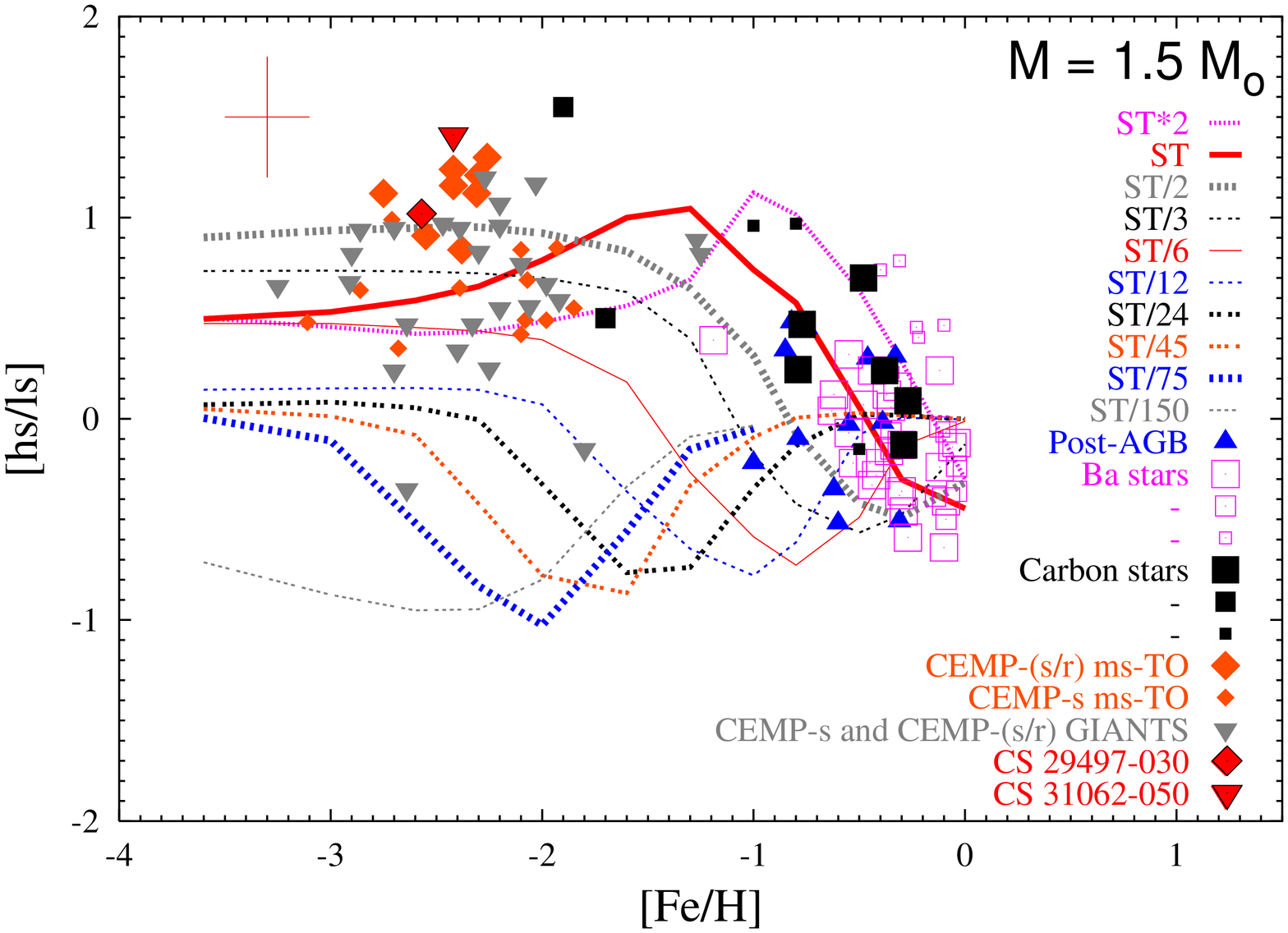}
\includegraphics[angle=-90,width=10cm]{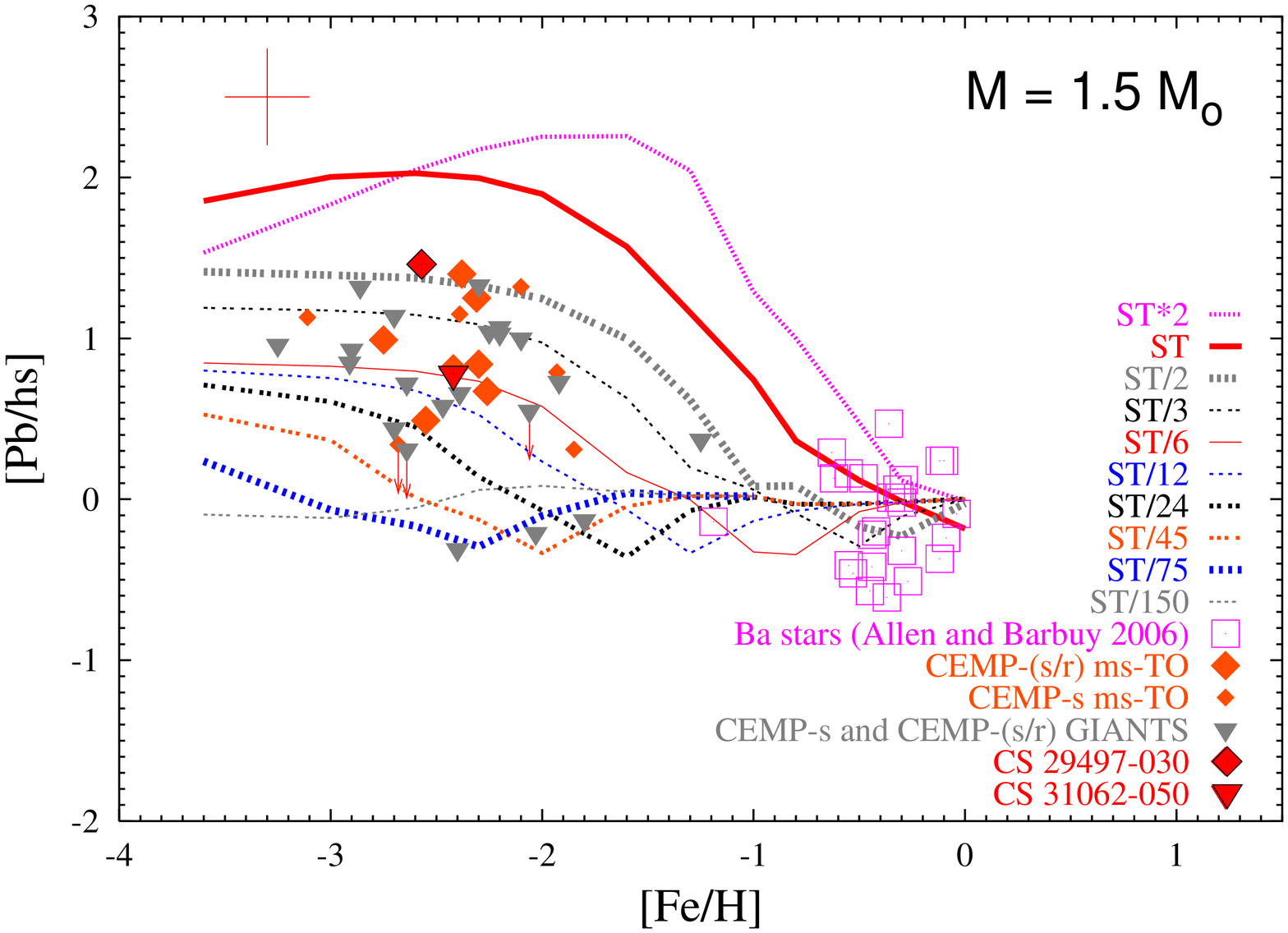}
\caption{(Color online) Top panel: Theoretical results of the $s$-process
index [hs/ls] as a function of [Fe/H] for AGB models of initial mass 
$M$ = 1.5~$M_\odot$\ and a wide range of $^{13}$C-pocket efficiencies.
Spectroscopic observations are plotted for \textbf{Post-AGB stars} (filled blue triangles, 
\citealt{RLG02}; \citealt{VaR00}; \citealt{RVG04,RAV07}), 
\textbf{Ba stars} (large open squares, \citealt{AlB06}; medium 
open squares, \citealt{SPS07}; small open squares, \citealt{LLD09}), 
\textbf{Carbon stars} (large full squares, \citealt{ZAP09}; medium 
full squares \citealt{ALW08}; small full squares \citealt{DAD06}),
\textbf{CEMP-$s$ and CEMP-$s/r$ stars} (full diamonds are main-sequence
turn-off stars; full gray triangles are giants stars, see text 
for references). Bottom panel: Theoretical results of the $s$-process 
index [Pb/hs] versus [Fe/H]. Spectroscopic observations: CEMP-$s$, CEMP-$s/r$
and Galactic Ba stars for which Pb abundances have been reported.
Symbols are the same as in top panel. Typical error bars are indicated 
in the top left corner of the panels.}
\label{fig:12}
\end{figure} 

In Fig.~\ref{fig:12}, the relative behaviour of the three $s$-process 
peaks ls, hs, and Pb are analyzed using the definitions [{\rm ls/Fe}] = 
1/2([{\rm Y/Fe}] + [{\rm Zr/Fe}]) and [{\rm hs/Fe}] = 1/3([{\rm La/Fe}] 
+ [{\rm Nd/Fe}] + [{\rm Sm/Fe}]). In general, the choice of the specific 
elements considered in the average ls and hs abundances varies from 
author to author and depends on the quality of the spectra available.
Our choice is made because, at disk metallicity, Sr and Ba have few and 
saturated lines (see \citealt{BLB95,BGL01}) and may be affected by non-local 
thermodynamical equilibrium (NLTE) corrections, in particular at low 
metallicities (\citealt{ASK09} and references therein). 

The [hs/ls] and [Pb/hs] ratios are indexes of the whole $s$-process 
distribution. They remain unchanged both in the envelope of the 
AGB companion (now a white dwarf) and in the envelope of the observed 
star after accretion of AGB winds. The top panel of Fig.~\ref{fig:12} 
shows [hs/ls] predictions for AGB stars of an initial 
mass of 1.5~$M_\odot$ as a function of [Fe/H] for different 
$^{13}$C-pocket efficiencies. The figure also contains representative abundance 
data discussed in Sec. \ref{sec4}.

Considering the ST case as a function of metallicity in the top panel 
of Fig.~\ref{fig:12}, one finds that the [hs/ls] ratio first increases 
with decreasing [Fe/H], reaching a maximum at [Fe/H]~=~$-$1.3 and then 
decreases again. This behaviour is due to the progressive build-up of hs 
elements and, subsequently, of the third $s$-peak at $^{208}$Pb. Note 
that $^{208}$Pb becomes dominant over the ls and hs components already at 
a metallicity of [Fe/H]~=~$-$1. In the bottom panel of Fig.~\ref{fig:12} 
theoretical predictions for [Pb/hs] versus [Fe/H] are compared with 
spectroscopic observations of Ba stars, CEMP-$s$ and CEMP-$s/r$ stars.
For halo metallicities, the spread in [Pb/hs] is found to be about 2 dex.

A sample of a hundred CEMP-$s$ stars have been observed (see 
\citealt*{SCG08} and references therein) including main-sequence stars, 
subgiants, or giants, far from the AGB phase where the $s$-process is 
taking place. Moreover, to be observed today, these stars 
have long lifetimes and correspondingly low initial masses 
($M$ $\leq$ 0.9~$M_\odot$). Therefore, the hypothesis of mass 
accretion of $s$-rich material from a more massive AGB companion
becomes essential to explain the overabundances detected in their
spectra. The spectroscopic $s$-process abundances in CEMP-$s$
stars depend on the fraction of the AGB mass transferred
by stellar winds, whereas the $s$-process indexes [hs/ls] and
[Pb/hs] remain unchanged. The mass transfer can be simulated by
introducing a dilution factor between the accreted AGB mass 
and the original envelope of the observed star. The dilution
factor $dil$ can be defined as the logarithmic ratio between the
mass of the convective envelope of the observed star before 
the mixing, and the total transferred AGB mass.

About half of the known CEMP-$s$ stars are also $r$-process-rich 
with [Eu/Fe] comparable to [La/Fe]. Among the sample of stars 
reported in the literature with Eu measurements, 
six stars show an $r$-process enrichment of $\sim$2~dex, 
among which are HE 2148-1247, the first CEMP-$s/r$ star discovered 
\citep{CCQ03}, CS 29497-030 \citep{ISG05}, HE~0338-3945 
\citep{JBG06}, and CS~22898-027 \citep{ABC07}. The $s$ and 
$r$ processes occur in completely different astrophysical scenarios, 
the $s$ process in low mass AGB stars and the $r$ process
during explosive nucleosynthesis in massive stars.
While Eu is a typical $r$-process element (about 94\% of solar 
Eu is of $r$ origin), the elements of the hs peak, as Ba and La, 
are mainly synthesized by the $s$ process. In particular, 70\% 
of solar La is produced by the $s$ process. AGB $s$-process
predictions give [La/Eu]$_{\rm s}$ 
$\sim$ 1. Consequently, spectroscopic observations of enhanced 
[La/Fe] and comparable [Eu/Fe] cannot be explained by $s$-process AGB 
models alone. \citet{CVH97} and \citet{VaC98} showed through 
numerical simulations how the supernova ejecta at high velocities 
interact with a nearby molecular cloud inducing Rayleigh-Taylor 
instabilities in the cloud, polluting it with freshly 
synthesized material, and at the same time triggering the 
condensation of a low mass binary system. This scenario may 
explain the high fraction of CEMP-$s/r$ stars.
Other explanations for these peculiar CEMP-$s/r$ stars have also been 
advanced, e.g. by \citet{CCQ03} and \citet{WNI06}. On the other hand, 13  
CEMP-$s$ stars do not show any $r$-process-enhancement, i.e.
CS~22880-074 \citep{ABC07}, HE~0024-2523 \citep{LGC03},
HE~2158-0348 \citep{CMS06}, and HE~0202-2203 \citep{BCB05}.

\begin{figure}
  \centering
\includegraphics[angle=-90,width=10cm]{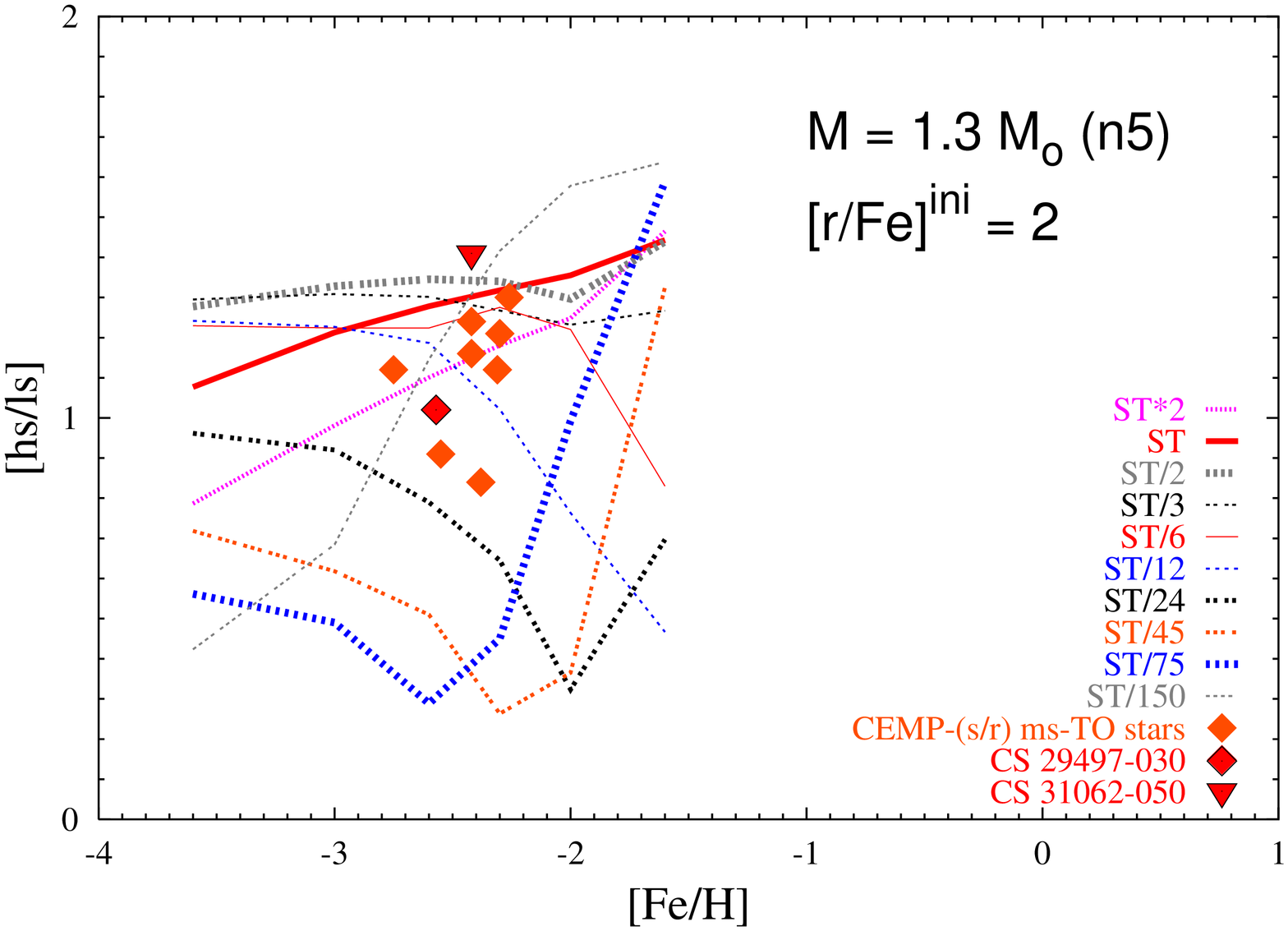}
\caption{(Color online)
Theoretical predictions of the $s$-process index [hs/ls] vs [Fe/H] for 
AGB models with initial mass $M$ = 1.3 $M_\odot$, a range of $^{13}$C-pocket 
efficiencies, and an initial $r$-process enrichment [r/Fe]$^{ini}$ = 2.0. 
The symbols of the spectroscopic observations are as in Fig. \ref{fig:12}.}
\label{fig:13}
\end{figure}

In Fig. \ref{fig:13} we show theoretical predictions of [hs/ls] vs 
[Fe/H] for AGB models with initial mass $M = 1.3 M_{\odot}$, 
different $^{13}$C-pocket efficiencies, and an initial $r$-process 
enrichment [r/Fe]$^{ini}$ = 2.0. The label "n5" indicates the 
number of TDU episodes suffered by the 1.3 M$_\odot$ model. For 
a given element, the initial $r$ enrichment is determined via 
the solar $r$-process contribution per isotope using the $r$-residual 
method \citep{AKW99b}. Fig. \ref{fig:13} shows also abundance data of 
all main-sequence/turnoff CEMP-$s/r$ stars (big full red diamonds). 
The two CEMP-$s/r$ stars CS 29497-030 and CS 31062-050, which will 
be discussed in detail in Sec. \ref{sec4B}, are indicated by a 
big filled diamond and a full rotated triangle, respectively. Due 
to the small number of TDUs in the 1.3 $M_\odot$ AGB model, 
the final surface distribution is affected by the initial $r$ 
enrichment. In particular, at low metallicities, the predicted 
[hs/ls] ratios reach values as high as 1.3 dex. In the top panel 
of Fig.~\ref{fig:12}, however, the predictions reach a maximum 
value [hs/ls]=1, independent of whether an $r$ enhancement of 2 dex
is included or not. Therefore, the most important message is that 
theoretical predictions given in Fig. \ref{fig:13} match the observations  
of main sequence/turn-off CEMP-$s/r$ stars quite well.

\begin{figure}
  \centering
\includegraphics[angle=-90,width=10cm]{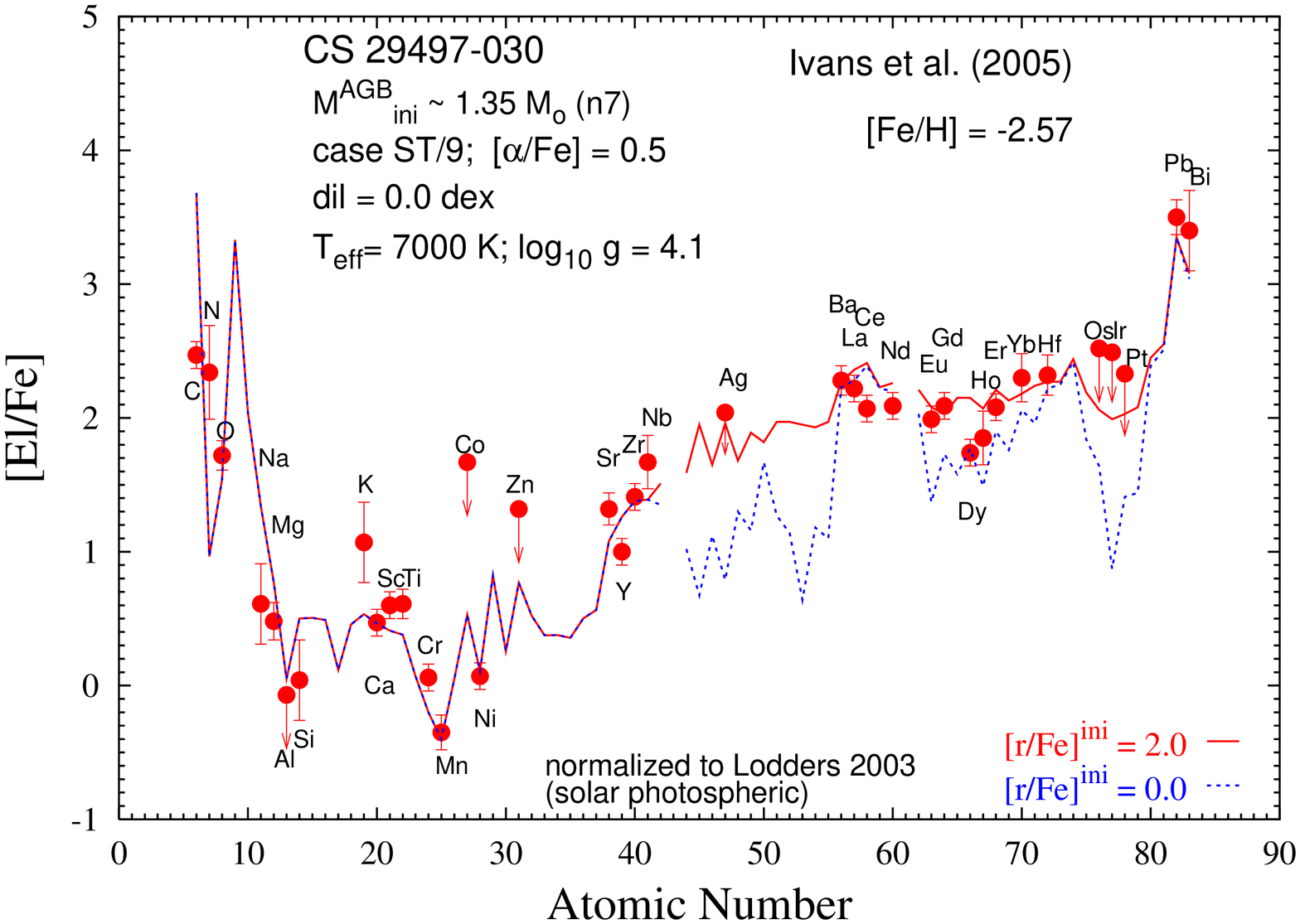}
\includegraphics[angle=-90,width=10cm]{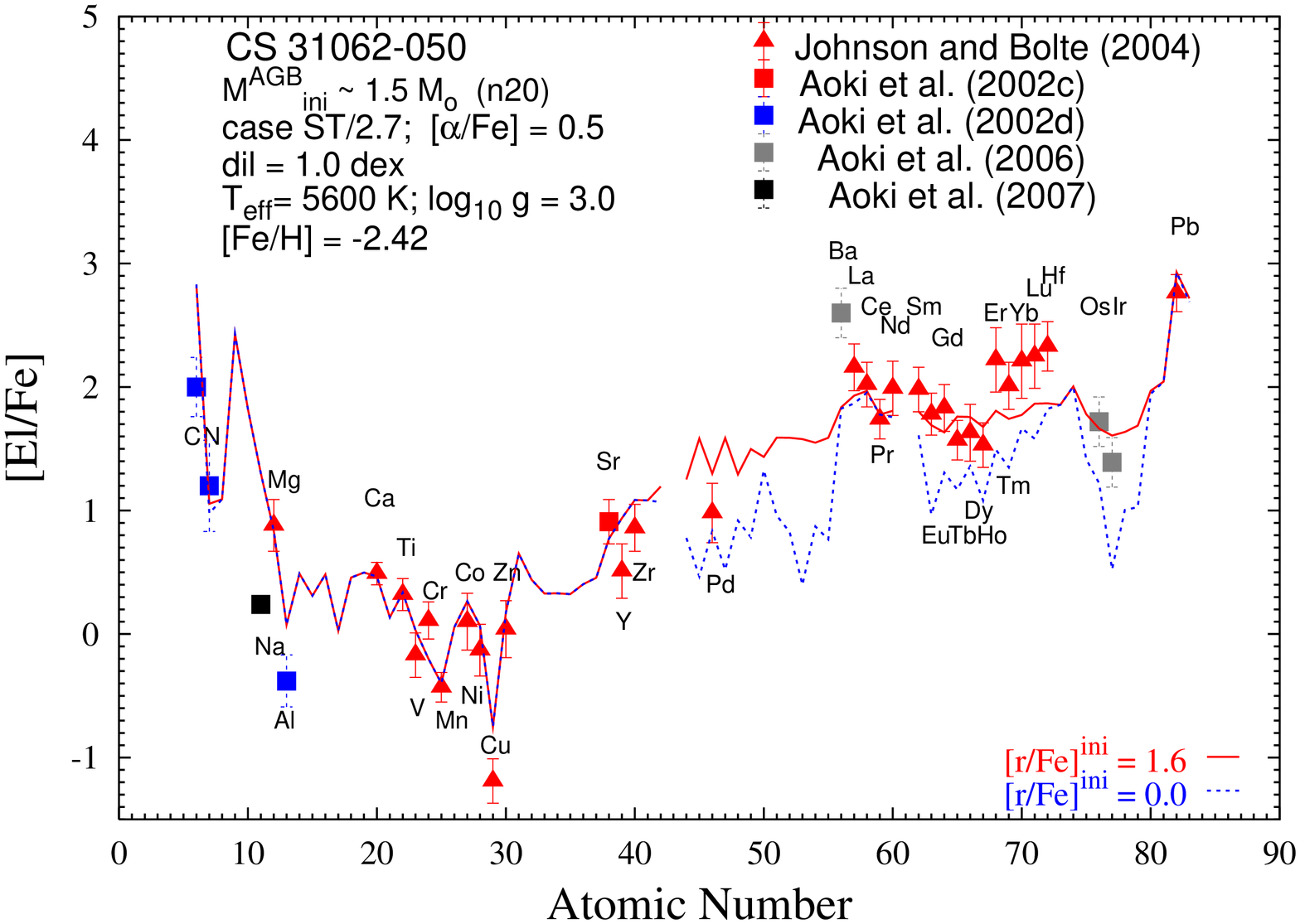}
\caption{(Color online) Abundance patterns of the CEMP-$s/r$ stars 
CS~31062–-050 and CS~29497–-030 compared with AGB models. Both stars 
show large excesses of heavy neutron-capture elements. The data and 
the theoretical expectations are normalized to the  solar photospheric 
abundances of \citet{Lod03}. Top panel: CS~29497–-030 compared with AGB 
models of 1.35 ~$M_\odot$\, case ST/9 and no dilution. In this case an 
initial $r$-process enrichment of [r/Fe]$^{ini}$ = 2.0 was adopted.
Bottom panel: CS~31062–-050 compared with AGB models of 1.5 ~$M_\odot$\, 
case ST/2.7 and $dil$ = 1.0 dex, with and without initial $r$-process 
enrichment ([r/Fe]$^{ini}$ = 1.6 and 0.0, respectively).}
\label{fig:14}
\end{figure} 

As an example, we show in Fig. \ref{fig:14} the 
two CEMP-$s/r$ stars CS~29497–-030 \citep{ISG05} and CS~31062–-050 
(\citealt{JoB04}, \citealt{ANR02b,ARN02,ABG06,ABC07}), which are 
spectroscopically well studied. A more detailed discussion is 
presented in Sec. \ref{sec4B}.

While the AGB model calculations are admittedly quite complex and
subject to a number of free and model-dependent parameters, they
have been successful in reproducing the abundance patterns of
the CEMP-$s$ stars. 

\subsection{The main $s$ component and the role of GCE\label{sec3E}} 

The $s$-process abundance distribution of the heavy isotopes beyond A 
$\sim$ 90 must be considered as the result of all previous generations 
of AGB stars that were polluting the interstellar medium before the 
formation of the Solar System. Therefore, the cosmic $s$-process 
abundances have to be explained by means of a general Galactic Chemical
Evolution (GCE) model. 

\begin{figure}
  \centering
\includegraphics[angle=-90,width=10cm]{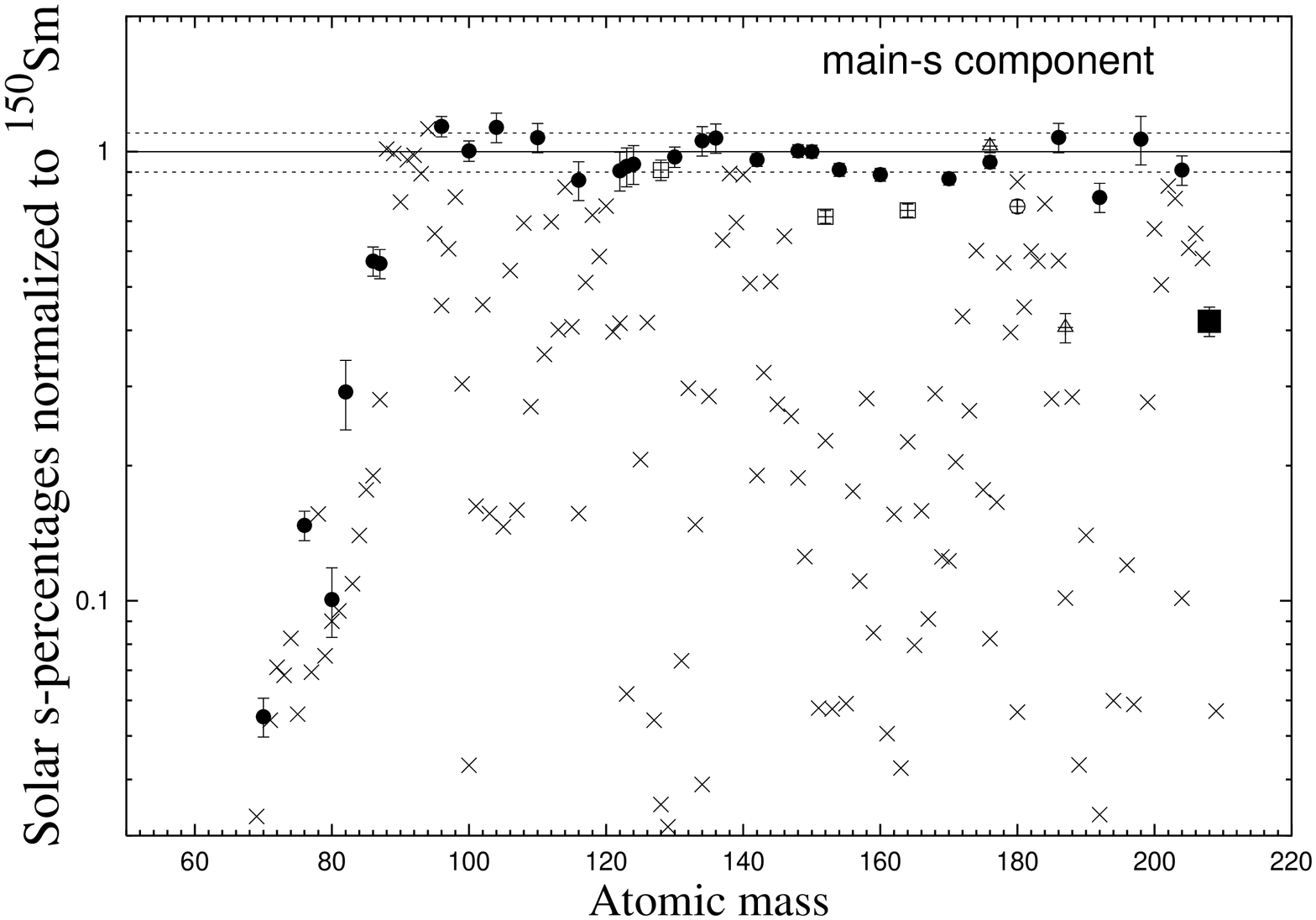}
\includegraphics[angle=-90,width=10cm]{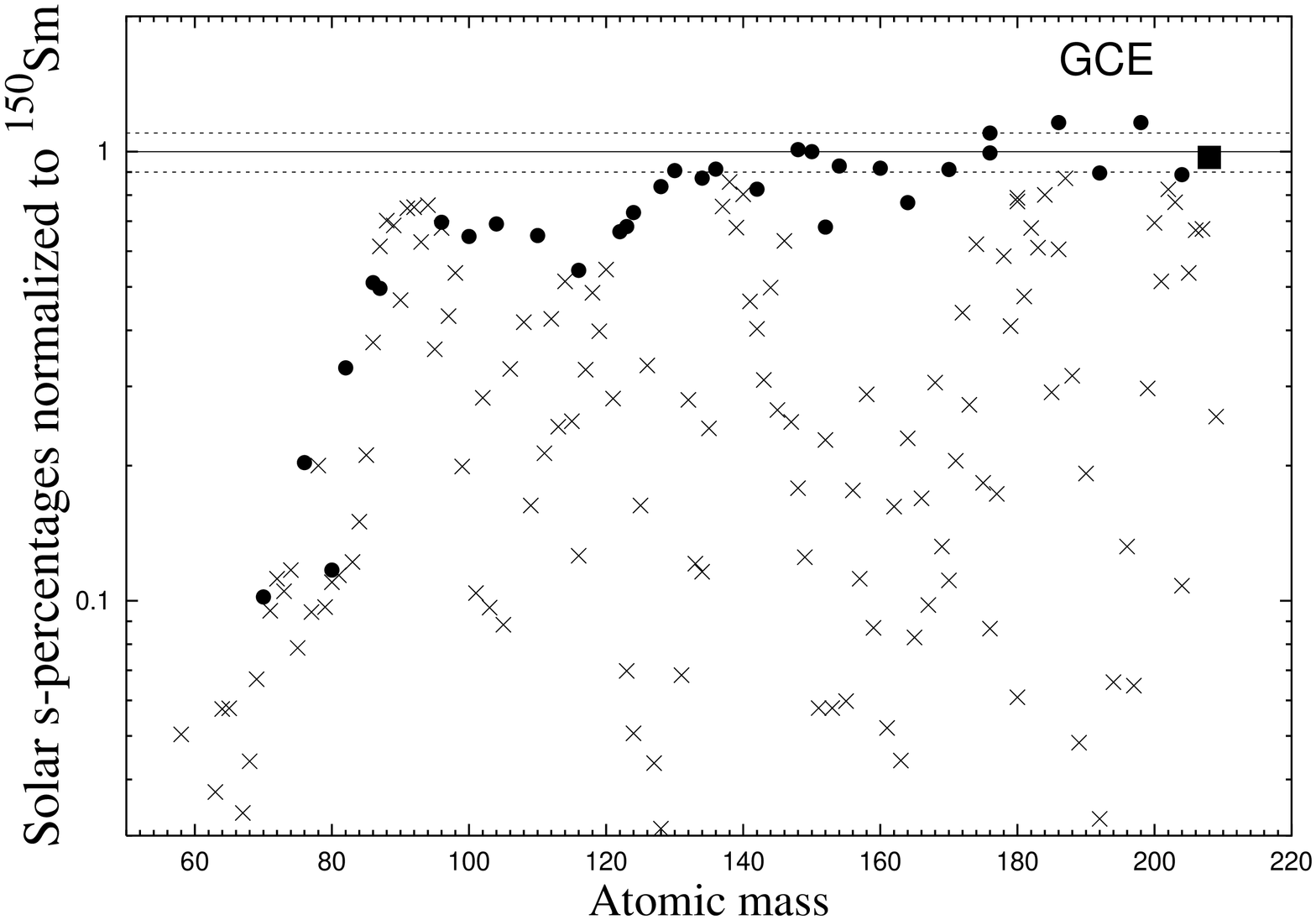}
\caption{Top panel: Solar $s$ abundances normalized to 
$^{150}$Sm versus atomic mass for the solar main component as 
in \citet{AKW99b}, but updated by \citet{BGS10}. 
Bottom panel: The predicted $s$-process distribution at the epoch 
of the Solar System formation obtained with a Galactic Chemical 
Evolution code \citep{TGA04} and AGB $s$-process yields at 
various metallicities. Note that $^{208}$Pb is boosted to about 
95\% of its solar abundance by the contribution from low metallicity 
AGB stars. Note also the 20-30\% deficit in the GCE distribution 
below magic neutron number N~=~82. These are the two major 
changes compared to the \citet{AKW99b} stellar~model with 
the ST pocket and [Fe/H]~=~$-0.3$, which are AGB models  
reproducing the main $s$-process component but only about
34\% of solar $^{208}$Pb.}
\label{fig:15}
\end{figure} 

It was shown by \citet{GAB98} and \citet{AKW99b} that the 
solar main component can be reproduced by assuming a standard $^{13}$C 
pocket, a metallicity of [Fe/H] = $-$0.3, and by averaging between 
stellar models of $M$ = 1.5 and 3 $M_\odot$. The distribution obtained 
with this prescription and updated nuclear input by \citet{BGS10} 
is plotted in the top panel of Fig. \ref{fig:15} normalized to 
the $s$-only isotope $^{150}$Sm. The set of $s$-only isotopes 
indicated by full circles is well reproduced. Open symbols have been 
used for $^{128}$Xe, $^{152}$Gd, and $^{164}$Er, which have a  
non-negligible $p$ contribution (10\% for Xe), for $^{176}$Lu, a long-lived 
isotope (3.8 $\times$ 10$^{10}$ yr), which decays into $^{176}$Hf, for 
$^{187}$Os, which is affected by the long-lived decay of $^{187}$Re (4.1 
$\times$ 10$^{10}$ yr), and for $^{180}$Ta, which receives also 
contributions from the $p$-process and from $\nu$-nucleus interactions in 
massive stars. The black full square corresponds to $^{208}$Pb, the 
only heavy isotope, which is clearly underproduced by the main $s$ component
\citep{ClR67}.

Employing a GCE code, in which the Galaxy is
subdivided into three zones (halo, thick, and thin disks), adopting the 
$s$-process yields from AGB stars of different mass and metallicity, 
and accounting for their respective lifetimes, \citet{TGA04} 
determined the temporal variation of the $s$-process abundances in 
the interstellar medium. Their results were recently updated by \citet{SGT09}.
The resulting $s$-process distribution at the epoch of the Solar System 
formation is plotted in the bottom panel of Fig. \ref{fig:15}. 
One finds that GCE calculation \citep{TGA04,SGT09} yield good agreement with the 
solar abundance values of the $s$-only isotopes between $^{134,136}$Ba 
and $^{204}$Pb. Moreover, also the solar abundance of $^{208}$Pb is well 
reproduced by the contributions from low metallicity AGB stars, thus
solving the long-standing problem of the origin of the strong $s$-component
in a natural way. Below magic neutron number N~=~82, however, there is a significant 
discrepancy between the abundance distribution obtained by the GCE 
approach and the Solar System values. It turned out that GCE models underproduce the $s$-process 
component of the solar-system abundances of Sr-Y-Zr by about 20-30\% \citep{TGA04}.
A similar deficit holds for the $s$-only isotopes from 
$^{96}$Mo up to $^{130}$Xe. This finding prompted 
\citet{TGA04} to postulate another source of neutron-capture 
nucleosynthesis in the Galaxy defined as Light Element Primary Process 
(LEPP). The LEPP process is different from the $s$ process in AGB stars 
and also different from the weak $s$-process component occurring in massive 
stars. \citet{TGA04} suggested that 8\% of solar Sr and 18\% of 
solar Y and Zr must come from the LEPP. There is general consensus for the need of an additional
LEPP source of yet unknown origin for the light isotopes, including Sr, Y, and Zr, 
an issue that is still highly debated \citep{FKP10,PGH10,QiW07}. 
\citet{MBC07} surveyed the possible ranges of parameters  (e.g. for the neutron density) 
that reproduce the abundance patterns of HD~122563, a very metal-poor star
representing the yields of the LEPP \citep{HAI06}.

Comparisons of the GCE models with observations of Galactic stars are 
discussed in Sec. \ref{sec4E}.

\section{Observational constraints \label{sec4}}

Models of nuclear reactions and enrichment of elements in the Universe
are examined by astronomical observations as well as by the analysis
of solar-system material. The mechanisms and astronomical sites of 
the $s$ process are constrained by the comparison of chemical abundance 
patterns with model predictions. Some key observational studies obtained 
until the middle of 1990s have been outlined in the review paper by
\textcite{WIP97}. In the following sections a summary of the
observational studies on the $s$ process obtained in the past 15 
years is presented.

Detailed chemical abundances have been determined for solar-system
material. As described before, the classical model of the $s$ process 
has been constructed by fitting the ${\sigma}N$ curve to those nuclei,
which are only produced by the $s$ process \citep{KBW89}. A more physical approach 
is to fit the abundances predicted by AGB models to the solar 
$s$-process component \citep{AKW99b}. 

The chemical abundances of the Solar System reflect the composition
of a particular site of the Galaxy at 4.6 billion years ago, to which
numerous nuclear processes in a variety of stars have contributed. 
Therefore, the abundance pattern of the $s$-process nuclei represents 
an average over the products from a variety of objects. In addition, 
information on individual nucleosynthesis events are provided by the 
rapid progress in the measurements on pre-solar grains, which provide 
accurate isotope ratios of important elements \citep{Nit09,ClN04,Zin98}. 

\subsection{Stellar abundances \label{sec4A}}

Useful information on chemical $s$-process yields by individual
objects is obtained by observations of stars and planetary nebulae. Stellar
abundances are derived from high resolution spectra of the
ultraviolet, optical and infrared ranges. In general, only elemental 
abundances can be determined in analyses of stellar spectral lines 
and the measurable elements are limited compared with the analyses of
solar-system material. However, observations for appropriately
selected targets (e.g. AGB stars) enable one to investigate directly
the products of individual processes, and to identify the respective
astrophysical sites. Moreover, measurements of isotope abundance 
ratios have been made for a few exceptional elements from detailed 
analyses of spectral line profiles (see Sec. \ref{sec4C}).

Measurable elements are dependent on objects. Stellar abundances are
usually derived from the absorption features formed at the surface of
stars in the so-called photosphere. In cool stars, neutral species 
of alkali metals (e.g. Na, K, Rb) and singly ionized species of 
alkaline earth metals (e.g. Mg, Ca) show strong resonance doublet 
lines. Singly ionized Sr and Ba atoms, which belong to elements of 
the $s$-process abundance peaks at neutron magic numbers 50 and 82,
exhibit strong absorption features in the optical range, which
makes it possible to determine their abundances even in very
metal-poor stars. Molecular absorption dominates in very cool
stars. In such cases, molecular bands, e.g for ZrO, can be used to
estimate the elemental and isotopic abundances of heavy
neutron-capture elements.

The abundances of noble gases (e.g. Ne, Ar) are very difficult to 
determine in cool stars as well as in the Sun because of the lack 
of useful spectral lines. Such elements are observable from emission 
lines in planetary nebulae, where these elements are ionized and 
excited by the ultraviolet photons from the hot central star. Hence, 
observations of planetary nebulae provide complementary information 
to the chemical yields obtained from stellar observations.

Considerable progress has been achieved in the past few decades in
stellar spectroscopy related to the $s$ process. They are promoted
by high resolution spectrographs mounted on large telescopes,
application of spectrum synthesis technique to high resolution spectra
of cool stars, and surveys of metal-poor stars that provide useful
samples of binary stars (see below).

\begin{figure}
\includegraphics[width=10.5cm]{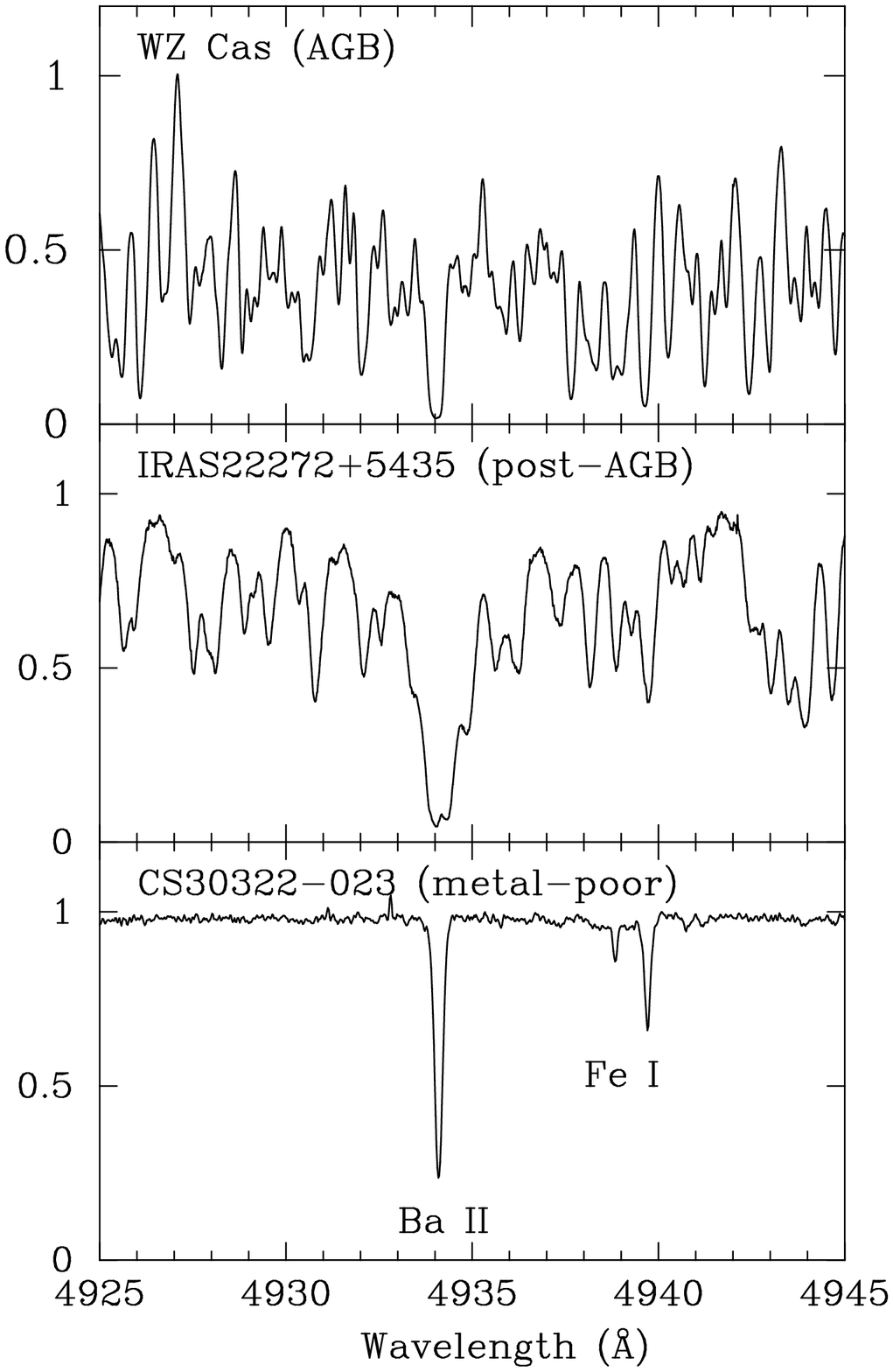}
  \caption{Comparison of spectra around the Ba {\small II} 4934~{\AA} line for
    AGB, post-AGB and CEMP stars. Spectral data were obtained with 
	the High Dispersion Spectrograph (HDS) of the Subaru Telescope with 
	a resolving power of 60,000 or higher. The spectrum
    of the metal-poor star is normalized to the continuum level
    (bottom). The atomic absorption lines are clearly seen in the
    spectrum and are useful for abundance analyses. In contrast,
    molecular absorption (mostly of C$_2$) is dominant in the AGB
    star (top), and the continuum level is very uncertain. Molecular
    absorption of the post AGB star is not as severe as in the AGB
    star, depending on stellar temperature.
\label{fig:17}}
\end{figure}

The main targets of $s$-process observations are (i) AGB stars, 
post-AGB stars and planetary nebulae evolving to white dwarfs, 
and (ii) binary companions accreting the products of primary
AGB stars. AGB stars are objects where the $s$ process is currently
taking place (or recently took place), giving direct 
information on the $s$-process nucleosynthesis in such stars. 
However, the surface of AGB stars is very cool and the spectra 
are crowded by numerous molecular lines (Fig. \ref{fig:17}). 
Accordingly, abundance analyses from such spectra are very 
difficult and the number of measurable elements is limited.

On the other hand, in binary systems with appropriate separation,
material from the surface of the primary star starts to accrete onto the
secondary companion when the primary evolves to an AGB star. Such mass
transfer across a binary system is expected by stellar winds from the
AGB primary or by forming a common envelope when the separation is 
sufficiently small. The secondary is usually an unevolved main-sequence 
star with a much longer timescale of stellar evolution compared to the 
AGB primary. After the primary evolves to a faint white dwarf, only the
companion, which preserves the material provided by the AGB primary, is
observable. In such cases, the $s$-process yields are easily measurable 
at the surface of the secondary, because the object is still relatively 
warm and molecular absorption is not severe (Fig. \ref{fig:17}).

\subsubsection{AGB, post-AGB, and planetary nebulae \label{sec4A1}}

As described in Sec. \ref{sec3} heavy $s$-process elements are enriched 
on the surface of AGB stars as the result of repeated TPs 
and TDU episodes. In such objects, the initial CNO abundances 
at the surface are significantly changed. In particular, carbon is enhanced 
and might become comparable or more abundant than oxygen. These stars are 
classified as spectroscopic types S or N, respectively. In massive AGB 
stars, which are affected by the hot bottom burning (HBB) process 
(\citet{Her05} and references therein), the surface composition changes 
toward the equilibrium values of the CNO cycle with a reduced C/O ratio. 
Some $s$-process elements might also be enriched in such stars.

The surface of AGB stars is cool ($\sim 3000$~K) and the optical and near 
infrared spectra are dominated by molecular absorption lines, e.g., from 
CO, TiO, and C$_{2}$, depending on chemical composition and temperature. 
While the abundances of C, N, O, and some other elements can be determined 
via molecular lines \cite[e.g. ][]{LGE86}, analyses of atomic spectra are 
very difficult. However, recent analyses based on the spectrum synthesis
technique provided important results on the surface abundances of heavy 
elements in cool AGB stars (see Sec. \ref{sec4D}). Molecular absorption features like ZrO are 
also useful to determine abundances of neutron-capture elements and their 
isotope ratios \cite{LSB95,Zoo85,PeB70}.

Excesses of heavy $s$-process elements are also found in post-AGB stars,
which are quickly losing their residual envelopes while evolving from AGB stars 
to planetary nebulae. The surface of such objects becomes warmer and 
molecular absorption becomes weaker as this transition proceeds, but 
the atmospheric structure is unstable and quite complicated as that of
supergiants, making spectral analyses difficult (Fig. \ref{fig:17}). 
The duration of this evolutionary stage is very short, indicating that 
one must observe apparently faint objects to increase the sample, even 
though the luminosity of such objects is high. Useful results have been
obtained by recent observations with large telescopes and detailed
analyses of high resolution spectra.

Planetary nebulae are formed by the material ejected from evolved low-
and intermediate-mass stars and ionized by the central star that is
evolving to a hot white dwarf. Hence, the spectra of nebulae provide direct
information on the yields that these stars contribute to the chemical 
enrichment of the Galaxy. The light elements (e.g., C, O) in planetary 
nebulae are accessible by X-ray observations \cite[e.g. ][]{MKM06}, 
while neutron-capture elements can be studied by optical spectroscopy
(Sec. \ref{sec4D}).

\subsubsection{Binary systems affected by mass transfer \label{sec4A2}}

Detailed information on the abundance patterns of heavy elements 
produced by AGB stars can be obtained by analyses of a binary
companion that is affected by mass accretion from the primary AGB 
star. In these cases, the target objects are main-sequence or red
giant stars, but they are distinguished from normal stars by excesses
of carbon and heavy $s$-process elements like Ba.

Such stars with high metallicity similar to the Sun are known as Ba 
stars \cite{BiK51}. Since the CNO abundances and, in particular, the oxygen
abundance in stars with solar metallicity are already high, mass
accretion from carbon-enriched AGB stars does not significantly change
the molecular features of their spectra. Instead, strong absorption
features of Ba can be a signature for the stellar classification. Periodic
variations of the radial velocity have been found for many Ba stars,
supporting the scenario that these stars experienced accretion
of $s$-process enhanced material from a primary AGB star in a
binary system \cite{Mcc84b, McW90}.

Mass accretion from AGB stars has stronger impacts on spectra of
metal-poor stars. Such objects are classified into CH stars
\cite{Kee42} or subgiant CH stars \cite{Bon74}. Excesses of heavy
elements have also been detected in such objects, and the binarity 
has been confirmed by radial velocity monitoring \cite{Mcc84b, McW90,PrS01}. 
Recent surveys of metal-poor stars based on the weakness of calcium 
absorption lines have detected a number of carbon-enhanced objects, 
e.g. the HK-survey \citep{BPS92,BSM07}, 
the Hamburg/ESO Survey \citep{Chr03}, the Sloan Digital Sky Survey 
(SDSS, \citealt{YAA00}) including the subprogram SEGUE  (Sloan Extension 
for Galactic Understanding and Exploration) and the SEGUE Stellar Parameter 
Pipeline (SSPP; \citealt{LBS08a,LBS08b}). Follow-up high-resolution 
spectroscopy has been made by the ESO Large Programme First Stars 
with VLT/UVES \cite[e.g. ][]{SBB06}, the Chemical Abundances of Stars 
in the Halo (CASH) Project \citep{RFS08}, and others 
\cite[e.g. ][]{ABC07}. Enhancements of heavy elements are found for most 
of these stars, indicating that their peculiar abundances originate from 
AGB nucleosynthesis and mass transfer in binary systems, as in the case of
CH stars. The sample of such objects contains stars with a variety
of metallicities, and enables one to investigate the metallicity
dependence of the $s$ process.

\subsection{Abundance patterns of heavy elements \label{sec4B}}

\subsubsection{Overall abundance patterns covering three magic neutron numbers \label{sec4B1}}

Measurements of abundance patterns of heavy elements in AGB
stars or objects affected by AGB nucleosynthesis and comparisons with model
predictions are important for understanding the $s$-process
mechanisms. The $s$-process abundance pattern exhibits three
peaks at $^{88}$Sr, $^{138}$Ba, and $^{208}$Pb, corresponding to the
magic neutron numbers 50, 82, and 126, respectively. The singly
ionized Sr and Ba have strong resonance lines in the optical range,
making such elements detectable even in spectra of very metal-poor 
stars. There are also many useful lines in the near UV and blue 
spectral range for the determination of e.g., Y, Zr, La, Nd, and Eu 
abundances. For this reason, measurements of light (ls, comprising Sr, 
Y, and Zr) and heavy (hs, with Ba, La, etc.) neutron-capture elements 
have been performed for many CH and Ba stars in order to estimate 
the neutron exposure from the ls/hs ratios.

Abundance studies of cool AGB stars are very difficult due to the 
dominant molecular absorption features. Important progress is, 
however, obtained by recent work. \textcite{ADG02} have determined 
chemical abundances of N type AGB stars. By means of the spectrum 
synthesis technique abundances of neutron-capture elements were determined 
for a large sample of AGB stars with excesses of such elements. The 
abundance ratios could be explained by $s$-process nucleosynthesis 
in AGB stars with $M\leq 3M_\odot$. However, the large scatter in 
the abundance ratios (e.g. [hs/ls]) suggests that the efficiency of 
the $s$ process is affected by yet uncertain model parameters
(Figs. \ref{fig:12} and \ref{fig:13}).

The $s$-process products of AGB stars are also recorded in Ba stars. 
\textcite{AlB06} studied neutron-capture elements for 26 Ba 
stars, separating the contribution of the main $s$ process by 
the progenitor (i.e. the former primary star in the binary system 
that provided the $s$-enhanced material) from other original 
components of the observed Ba stars and compared the observational 
results with the model predictions by \textcite{Mal87a} and 
\textcite{Mal87b} for single and exponential neutron-exposures of the
$s$ process. Also \textcite{SPS07} determined abundances of
neutron-capture elements in Ba stars. Excesses of heavy 
neutron-capture elements and the hs/ls ratios are discussed in 
their work, while no significant $s$ process effects were found 
for Cu, Mn, V and Sc. 

A recent important progress is that measurements have been extended to
Pb, the stable element at the third abundance peak of the $s$ process. 
The lack of strong spectral features in the optical range makes 
the abundance measurements of Pb from stellar spectra very difficult. 
The detection of Pb was reported for the post-AGB star FG Sge by 
\textcite{GLW98}. Measurements of Pb abundances have been made by 
\textcite{ANR00, ARN01} for CEMP stars, by \textcite{VGJ01} for CH 
stars, and by \textcite{AlB06} for Ba stars using the absorption line
of neutral Pb at 4058~{\AA}. These observations made it possible to
investigate the overall $s$-process abundance pattern more 
consistently. By now abundance studies for Pb have been made for 
more than 20 carbon-enhanced objects  \cite{ARN02, VGJ03,
LGC03, CCQ03, SBM04, SBB06, ABS08}. Moreover, \textcite{ISG05} 
reported the detection of Bi as well as Pb for the carbon-enhanced 
metal-poor star CS 29497-030. Fig. \ref{fig:14} shows the abundance 
pattern of this and another well studied carbon-enhanced star
(see below).

The Pb abundances determined for carbon-enhanced metal-poor stars
revealed that the abundance ratios of elements between the second and
third abundance peaks (e.g. [Pb/hs]) show large star-to-star scatter
(Fig. \ref{fig:12}) as discussed in Sec. \ref{sec3}. A similar 
scatter is also found in the abundance ratios of 
elements between the first and second abundance peaks. This scatter 
can be interpreted as due to a dispersion of the hydrogen mixing 
into the $^{13}$C pocket that determines the efficiency of the $s$ 
process in AGB stars \cite[e.g. ][]{BGW99}. The efficiency is 
expected to depend on the metallicity, which determines the ratio 
of neutrons to seed nuclei at the $s$-process site. However, Fig. 
\ref{fig:12} shows no clear correlation between the metallicity 
or iron abundance and the [Pb/hs] or [hs/ls] abundance ratios. Thus, 
the reason for the scatter of the abundance ratios is still unclear.

Using the technique described in Sec. \ref{sec3D} and AGB stellar
models of disk metallicity, \citet{HGB09} were able to fit the Ba
stars in a range of $^{13}$C-pocket efficiencies ST$\times$2 down 
to ST/3 and initial AGB masses between 1.5 and 3 $M_\odot$. The 
results are plotted in Fig.~\ref{fig:12} for [hs/ls] and for the 
somewhat more uncertain [Pb/hs]. 

\subsubsection{Detailed abundance patterns: CEMP-$s$ and -$s/r$ stars \label{sec4B2}}

In addition to the overall abundance patterns, which are represented 
by the abundance ratios between the three $s$-process peaks, detailed 
abundance measurements for CEMP stars with [C/Fe] $> 1$ provide useful 
constraints on the origin of the heavy elements and the role of the 
$s$ process.  

An illustrative case are the high Eu abundances,
relative to Ba and other $s$-process elements, in some CEMP stars, which
can not be explained by standard $s$-process models \cite{HBS00,
ARN02, CCQ03}. Although the $r$ process is the dominant source of Eu
in solar-system material (Eu is sometimes called an $r$-process
element), this element is also produced by the $s$ process. Therefore,
large enhancements of heavy elements in carbon-enhanced objects can
also provide some excess Eu. However, the measured abundance ratios of Eu
with respect to Ba, La and other elements around Eu are significantly
higher than in model predictions as well as in the $s$-process
component of solar-system material. For this reason, the Eu excess 
suggests a large $r$-process contribution in addition to the $s$-process
component, and such stars are sometimes called CEMP-$s/r$ stars \citep{BeC05}. 
As discussed in Sec. \ref{sec3D}, a preliminary comparison 
of CEMP-$s$ with the theoretical models  
by \citet{BGS06} has provided reasonable interpretations for all 
CEMP-$s$ and CEMP-$s/r$ stars published so far as illustrated for the $s$-process 
indexes [hs/ls] and [Pb/hs] in Fig.~\ref{fig:12}. A full analysis with 
updated AGB models is underway \citep{BGS10}. 

Examples of the detailed elemental abundance patterns of two CEMP-$s/r$ stars 
have already been presented in Fig.~\ref{fig:14}, where the top 
panel shows the situation for the blue metal-poor (BMP) star 
CS~29497--030 with [Fe/H]~=~$-$2.57 \citep{ISG05}. A large 
number of elements have been detected in that star besides the usual 
ls and hs elements and Pb, including Nb, Bi, some of the 
heaviest rare-earth elements, and a few useful upper limits. The 
abundance ratios with respect to Fe are normalized to the solar values, 
and compared to an AGB model of the same metallicity, for an inferred initial mass of 
1.3~$M_\odot$, and a rather small $^{13}$C pocket (case ST/9). Note 
that no dilution is required to fit this CEMP-$s$ star,
suggesting that the material from the AGB companion dominates at the 
surface of this object.
The very low ratio [hs/Eu]~$\sim$~0 indicates that the original 
composition of the binary system had a huge $r$-process enhancement 
of [Eu/Fe]~$\sim$2~dex, even higher than the known $r$-II 
stars\footnote{Such stars are labeled $r$-II by \citet{CBB04}, 
defined as those very metal-poor ([Fe/H] $<$ $-$2.5) stars with 
[Eu/Fe] $>$ 1.0 and [Eu/Ba] $<$ 0.0 (e.g., CS~31082-001 by \citet{HCB02}, 
CS~22183-031 by \citet{HAK04}, HE 1523-0901 by \citet{FCN07}, 
CS~29497-004 by \citet{CBB04}). A few other stars, such as 
BD+17 3248 \citep{CSB02} and CS~30306-132 \citep{HAK04}, would 
formally be classified in this system as $r$-II stars (+0.3 $\leq$ 
[Eu/Fe] $\leq$ +1.0 and [Eu/Ba] $<$ 0.0).}. 

CS~31062–-050 is a subgiant CH star with a large enhancement 
in $s$- and $r$-process elements. Lines of interesting elements 
such as Os and Ir \citep{ABG06} and Na \citep{ABC07} have been 
detected due to its relatively cool temperature and low gravity 
($T_{\rm eff}$ = 5600 $\pm$ 150 K and log$_{10}$ $g$ = 3.0 $\pm$ 0.3).
This star has probably undergone the first dredge-up episode, 
where the convective envelope extends over about 80\% of the 
stellar mass. This implies that the $s$-process rich material 
accreted from the AGB star by stellar winds needs to be diluted 
by a large extent with the original material of the observed star. 
In the bottom panel of Fig.~\ref{fig:14} the observed abundances
are compared with AGB yields from models of 1.5 $M_\odot$ (case 
ST/2.7) and $dil$ = 1.0 dex, which corresponds to ten parts of 
original material mixed with one part of accreted material from the 
AGB star. This star appears to be particularly highly enhanced 
in both $s$- and $r$-process elements. [Eu/Fe] observed is about 
1 dex higher than expected for a pure $s$-process AGB 
prediction (blue dashed line), indicating an important initial 
$r$-process contribution corresponding to [r/Fe]$^{ini}$ = 1.6.
Ba is possibly overestimated 
with respect to the other heavy-$s$ elements \citep{ABG06}.
 
\begin{figure}
  \centering
\includegraphics[angle=-90,width=13cm]{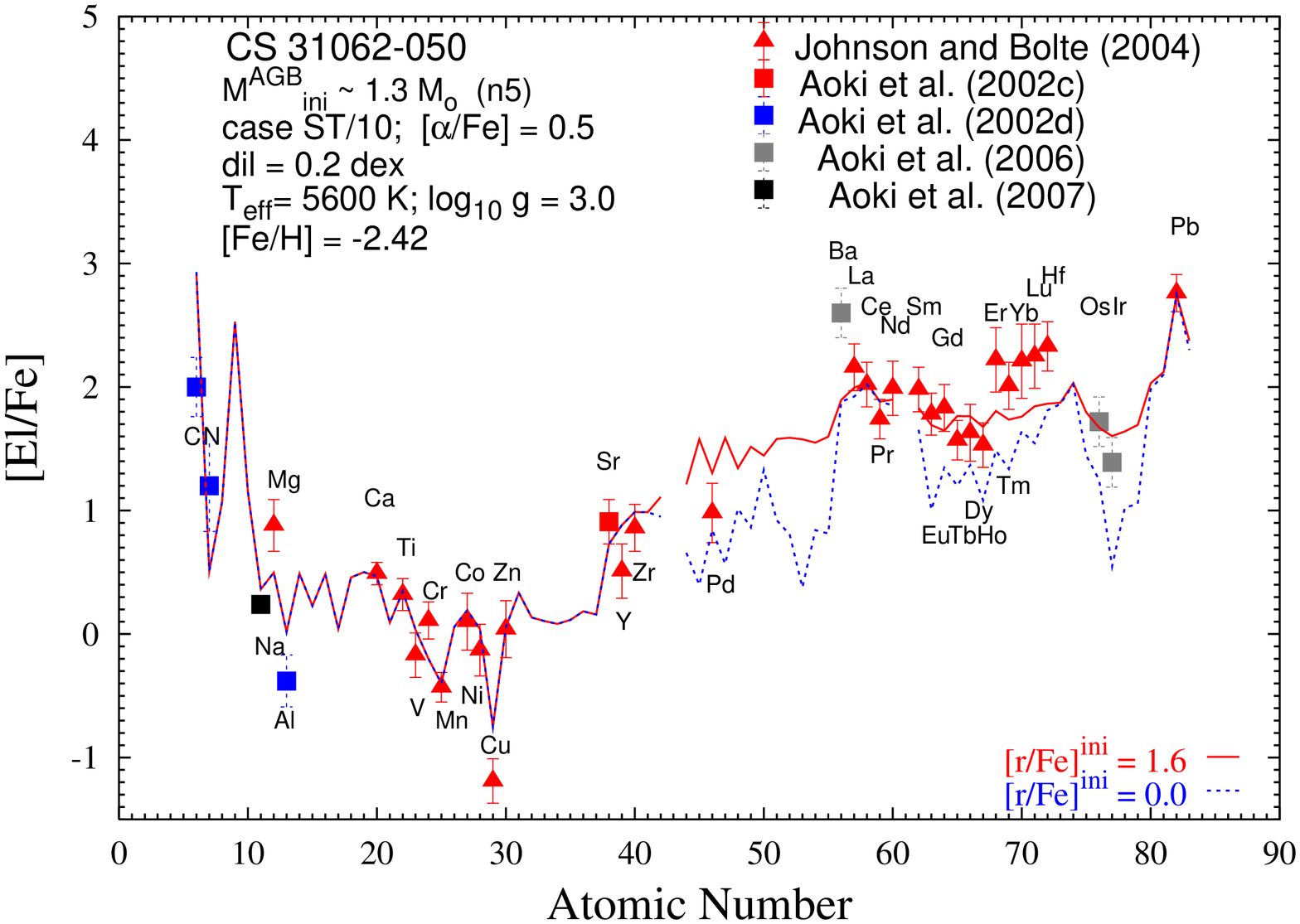}
\caption{(Color online)
The same as in the bottom panel of Fig. \ref{fig:14}, but for AGB models 
with initial mass M = 1.3 M$_{\odot}$, case ST/10, and $dil$ = 0.2 dex. This 
solution is obtained under the hypothesis of a turn-off star before the first 
dredge-up.}
\label{fig:18}
\end{figure}

Detailed abundance ratios provide other constraints on AGB models.
For instance, the observation of Zr and
Nb permits the immediate confirmation of the extrinsic AGB nature 
of CS 29497-030. In AGB stars Nb is produced by the radiogenic decay of 
the long-lived $^{93}$Zr ($\tau_{\rm 1/2}~=~1.5 \times 10^6$~yr), which 
will mostly decay later in the interstellar medium to $^{93}$Nb. In the envelope of an 
intrinsic AGB one would expect [Zr/Nb] $=$ 1, but in an extrinsic (mass 
receiving) star like CS~29497--030 all $^{93}$Zr should have already 
decayed to $^{93}$Nb, and thus one expects [Zr/Nb]~=~0. This ratio 
were potentially a powerful tool for determining the state of $s$-process
enhanced stars, but, unfortunately, Nb has only a single easily observable 
transition. The Nb II 3215.6~\AA\ line falls in the crowded near-UV 
spectral region, and no systematic survey of Nb abundances in CEMP-$s$ 
stars has been undertaken so far.

Na in CS 31062-050, for which a single line has been detected \citep{ABC07}, 
is lower by approximately 1 dex than the model predictions for the
$M = 1.5 M_\odot$ case (Fig. \ref{fig:14}). A lower Na yield is obtained with 
models of lower mass AGB stars and would agree with an AGB model of $M$ = 1.3 $M_\odot$ (case ST/10 and 
[r/Fe]$^{ini}$ = 1.6, and a negligible dilution of 0.2 dex, Fig. \ref{fig:18}). 
This solution would only be compatible with a subgiant star before 
the first dredge-up. Note that the exact evaluation of the effective 
temperature at which the first dredge-up occurs may be affected by 
the treatment of gravitational settling in binary systems with mass 
transfer as it is the case in CEMP stars. Another possibility is to consider 
low mass AGB models at {\it very} low metallicities. As discussed before, low 
metallicity AGBs with initial mass $M < 1.5 M_\odot$ may undergo a 
deep first TDU episode (Sec. \ref{sec3C}).

\subsection{Branchings \label{sec4C}}

Detailed abundance measurements, in particular determinations of
isotope ratios, are useful for the analysis of the branching points 
of the $s$ process to derive observational constraints on the
temperature and neutron density of the $s$-process sites as 
illustrated at the following examples.

\subsubsection{$^{85}$Kr branch \label{sec4C1}}

The ground state of the unstable isotope $^{85}$Kr has a half-life of
10.7 years.  At low neutron density, e.g. $n_n<10^{7}$ cm$^{-3}$,
the $\beta$-decay of $^{85}$Kr dominates over the neutron capture
rate and the $s$-process path runs from $^{84}$Kr to $^{85}$Rb to 
$^{86}$Sr. At high neutron density the neutron capture chain runs 
from $^{84}$Kr to $^{86}$Kr and on to $^{87}$Rb (see \textcite{KBW89}). These main paths 
are somewhat affected by the minor branching at $^{86}$Rb as well. 
As a result, a high Rb abundance ratio with respect to Sr, Y, or Zr 
signifies an $s$ process at high neutron density ($>10^{8}$~cm$^{-3}$), 
because of the small neutron capture cross section of the neutron magic 
isotope $^{87}$Rb. In thermally pulsing AGB stars low neutron densities 
are expected during the inter-pulse phase, where the neutron
source is the $^{13}$C($\alpha, n$)$^{16}$O reaction, while 
high neutron densities are obtained during the He shell flashes, when the
$^{22}$Ne($\alpha, n$)$^{25}$Mg reaction is activated at the higher 
temperatures at the bottom of the reaction zone.

In very cool stars Rb abundances are measurable via resonance lines of 
neutral Rb at 7800~{\AA} and 7947~{\AA}. This is not the case for warmer 
stars because the ionization potential of this element is very low, 
and ionized species have no measurable lines in the optical and 
the near infrared ranges.

Rb abundance ratios were obtained for AGB stars of spectral type
M, MS and S by \textcite{LSB95}, who derived values of Rb/Sr $\sim $
0.05 for $s$-processed material from stellar surface abundances and
taking the dilution by envelope material into consideration. The low
Rb/Sr ratio is consistent with the predicted low neutron density 
during the inter-pulse phase in low-mass AGB stars. These authors
also measured the Zr isotope ratios from ZrO molecular bands,
and investigated the branching at $^{95}$Zr, an isotope with a
$\beta$-decay half life of 65 days. No evidence of existence of $^{96}$Zr 
was found, supporting the low neutron density estimated from the 
Rb abundances.  A similar conclusion was obtained for carbon stars 
by \textcite{ABG01}, who measured the abundances of Rb and other 
elements for 21 N-type carbon stars and found that the [Rb/Sr, Y, Zr] 
abundances are better explained by AGB models for low-mass ($M\lesssim 3M_{\odot}$) 
than for intermediate-mass stars, although the determination of 
Rb is uncertain for carbon stars and some calibration of abundance
ratios to a non $s$-process-enhanced AGB star (the J-type carbon star WZ
Cas) was required.

On the other hand, \textcite{GGP07} investigated OH/IR
stars, oxygen-rich AGB stars that are believed to have high masses
(4--8~$M_{\odot}$). They found large enhancements of Rb in these
objects over a wide range ($-1.0<$[Rb/Fe]$<2.6$). Given the only 
mild excesses of $s$ material ([Zr/Fe]$<0.5$), the Rb/Zr
ratios are significantly higher than in low-mass AGB stars (S-type
and carbon stars) investigated in the above studies. Accordingly, the
high Rb/Zr ratios were interpreted as evidence for the
$^{22}$Ne($\alpha, n$)$^{25}$Mg reaction during thermal pulses 
in massive AGB stars.  
 
Measurements of Rb abundances were extended to metal-deficient
($-2.0<$[Fe/H]$<0.0$) stars in the Galactic disk and halo by
\textcite{ToL99}. Excluding CH stars, the Y, Zr, and Ba in their
sample with [Fe/H]$<-0.5$ are underabundant, while Rb is
overabundant ([Rb/Fe]$=0.21$) on average. The high Rb abundance ratio
with respect to Y and Zr are at least partially attributable to the
larger component of the $r$ process in metal-deficient stars than in
solar-system material. However, an alternative possibility
is that the $s$-process neutron density is higher in
metal-deficient AGB stars than that in stars with solar-like
composition. 

Rb abundances in globular cluster stars with $-1.7<$[Fe/H]$<-1.2$ 
have been studied by \textcite{YAL06, YLP08}. While Rb, as well
as other $s$-process elements, is overabundant in M4 (see Sec.
\ref{sec4E}), the [Rb/Fe] ratios in NGC~6752 and M13 are similar
to the solar value. The [Rb/Y] and [Rb/Zr] ratios are constant within the
uncertainty of the measurement, suggesting that the nature of the
$s$ process that has contributed to these clusters is similar (although
the $s$ contribution is significantly larger in M4 than in the other two
cases). The discussion might be, however, more complicated if the
contribution of the $r$ process is fully taken into consideration. Such
measurements are also important for the understanding of the abundance
variation of light elements (e.g., O, Na, Mg) found in some globular clusters
\cite{YAL06}.

\subsubsection{$^{151}$Sm branch \label{sec4C2}}

The branching at $^{151}$Sm is of particular interest because the 93 yr half-life  
of $^{151}$Sm decreases by about a factor of 30 at $s$-process 
temperatures \citep{TaY87}. This 
branch has been studied using the solar-system abundance ratios of Gd
isotopes (e.g. \textcite{WVK95b}). However, the physical conditions of
the $s$ process in an individual site (object) are not obtained from the
analysis of solar-system abundances. Moreover, the Gd isotopes are
also affected by contamination of the $p$ process.

\begin{figure}
\includegraphics[width=9.5cm, angle=0]{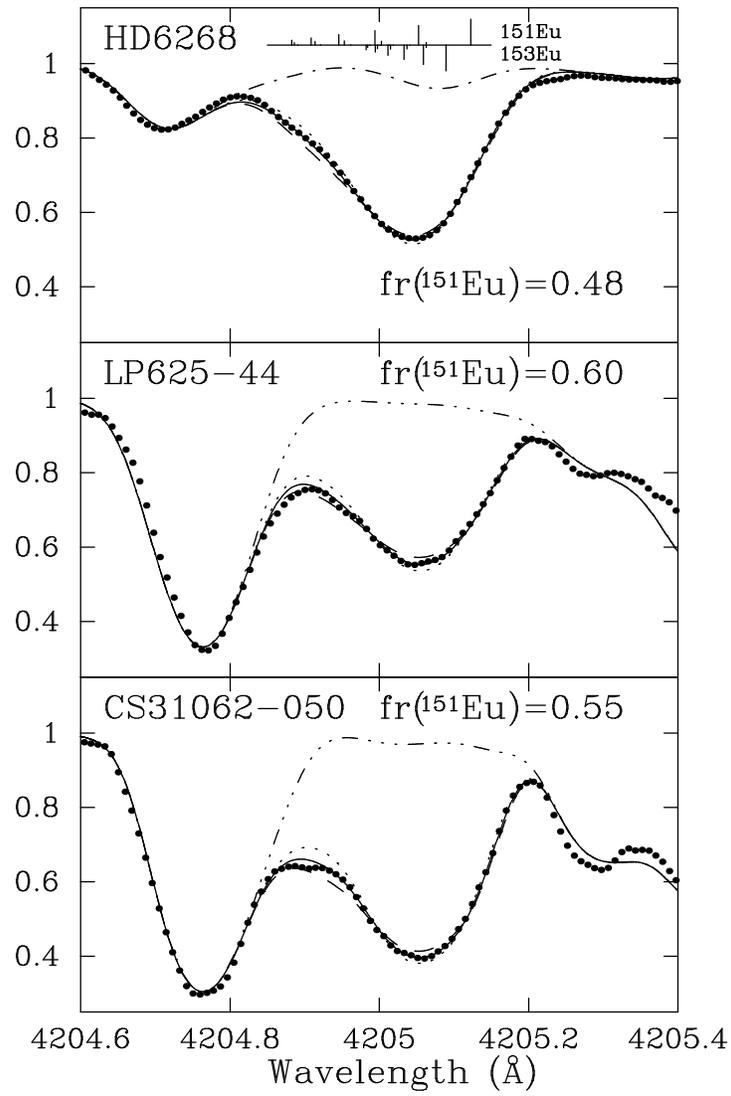}
\caption{Measurements of Eu isotope ratios from the Eu II absorption
line profile for an $r$-process-enhanced star (top) and CEMP-$s$
and CEMP-$s/r$ stars. Adapted from Aoki {\it et al.} (2003). 
\label{fig:16}}
\end{figure}

The effect of the $^{151}$Sm branching also appears in the abundance
ratio of the two stable europium isotopes $^{151}$Eu and $^{153}$Eu. 
Although measurements of isotope ratios from stellar spectra are 
difficult in general, Eu lines show relatively large hyperfine 
splitting, which is characteristic of the two isotopes (Fig. \ref{fig:16}). Thus, detailed 
analyses of Eu absorption line profiles enable one to estimate the 
Eu isotope ratios \cite{LWD01}. \textcite{SCL02} and \textcite{AHB03} 
determined the Eu isotope ratios for $r$-process-enhanced, metal-poor
stars. The results agree well with the Eu isotope ratio of the
$r$-process component of solar-system material (with a 48\% fraction of
$^{151}$Eu), confirming the usefulness of the line profile analysis.  

\textcite{ARI03} applied such an analysis to two CEMP
stars that show large excesses of $s$-process elements. The
fraction of $^{151}$Eu derived are 55 and 60\%, slightly
higher than was found for $r$-process-enhanced stars. These results are
consistent with the predictions of $s$-process models for wide ranges of
neutron density and temperature, given the uncertainties of the
measurements (3\%). However, the high neutron density case might be
preferable for explaining the results if recent measurements of
$^{151}$Sm neutron capture cross section are adopted
\cite{MAA06,WVK06c}. Further detailed analyses for a larger sample will give
stronger constraints on neutron density and temperature at $s$-process
sites of low metallicity.

\subsection{Stellar evolution \label{sec4D}}

Observations of characteristic $s$-process elements are important 
for probing the evolution of AGB stars (Sec. \ref{sec3C}). 

Tc has no stable isotope, and the first isotope was artificially
produced in 1937. While $^{98}$Tc has the longest half-life ($t_{1/2} = 4 \times
10^{6}$ yr), $^{99}$Tc ($t_{1/2} = 2 \times 10^5$ yr) is expected to 
be the most abundant isotope produced by the $s$-process in AGB stars. 
The discovery of Tc in the spectrum of an AGB star of spectral type 
S by \textcite{Mer52b} provided firm evidence for the synthesis of 
heavy elements in such evolved stars, and for stellar nucleosynthesis 
in general.

Abundance studies for Tc in AGB stars have been made by \textcite{LLB87},
\textcite{SmL88} for example. Recently, \textcite{VSL07} compiled the 
observational results for S stars and investigated the correlation of 
the spectral features between Tc and Li, because a strong Li line 
indicates the contribution of HBB in massive AGB stars. Tc line 
absorption is detected in 28 stars of their sample, and nine of 
these stars show also a strong Li line. Tc is expected to be observed 
in low-mass AGB stars, which are actively producing the $s$ abundances, 
the simultaneous detection of Tc and Li suggests the existence of a 
concomitant production mechanism for Li, although HBB is not expected
in AGB stars.


\textcite{GGP07} studied Zr as well as Li in Galactic OH/IR stars, 
which are oxygen-rich AGB stars with relatively high masses (3-4 $M_\odot$). 
These stars are expected to show HBB, the CNO cycle at the bottom of
hydrogen-rich envelope, that makes the C/O ratio lower than unity and
produces high Li abundances.  Although the effect of HBB is confirmed 
by the Li over-abundances in a half of their sample, no excess Zr was 
found in these cases. This led to the conclusion that the high
mass AGB stars in our Galaxy do not show any significant $s$-process
enrichments, in contrast to the results derived for AGB stars
in the Magellanic clouds, which are lower in metallicity by a factor of
two or three than Galactic objects. \textcite{PSL93} and \textcite{SPL95} 
found effects of HBB for luminous (massive) AGB stars in the Magellanic 
clouds that show $s$-process enhancement as well. These observations
confirm that the $s$ process in massive AGB stars depends strongly on 
metallicity.  

The heavy neutron-capture elements in planetary nebulae have been
recently studied by optical and near-infrared spectroscopy. 
\textcite{SZW07} measured weak emission lines in the optical range 
for five planetary nebulae. These observations included  
neutron-capture elements, for which the required atomic data 
for a reliable abundance determination became available recently. 
The discovered excesses of Kr and Xe in three objects can be
assigned to the first and second $s$-process abundance peaks, 
although a large correction for the $r$-process contribution is 
required for Xe. These two elements are enhanced by a similar factor 
in three planetary nebulae ([Xe/Kr]$\sim 0$), suggesting the 
effect of an $s$ process with a significant neutron-exposure.

\textcite{StD08} measured abundances of the light neutron-capture
elements Kr and Se for a large number of planetary nebulae (81
objects for Kr and 120 for Se). The abundances were determined from
emission features in near infrared (2.2~$\mu$m) and showed
that 44\% of the sample, which corresponds to 20\% of all
planetary nebulae in the Galaxy, are $s$-process-enriched 
([Kr, Se/Ar] $>0.3$). 

\subsection{Contribution to the Galactic chemical evolution} \label{sec4E}

The stars in the substructures of the Milky-Way, the thin and thick
disks, the bulge, and the halo, differ with respect to metallicity 
and kinematical properties, i.e. their orbital motion around
the Galactic center. The formation timescale of the Galactic
structure is usually estimated by the abundance ratios between 
the $\alpha$ elements, e.g. Mg, and iron \citep{Mcw97}.

The thin disk is the relatively new component of the Galaxy, and the
Sun is involved in this structure. The metallicity is similar to the
solar one, while the ages of stars range between zero to ten billion
years. The thick disk consists of old stars of lower
metallicity. The formation of this component is still debated,
a possible scenario being the burst-like star formation when a small
galaxy merged with the Milky Way at some early epoch of the Galactic 
history \cite[e.g. ][]{FrB02}.

The bulge consists mostly of old stars with a broad metallicity
distribution, including stars of solar metallicity. It has been 
suggested that the bulge was formed at very early times of the 
Universe and its evolution during Galactic history has been
discussed. So far, abundance measurements of neutron-capture
elements for the bulge are still quite limited because of its long
distance and severe interstellar extinction. Therefore, future studies 
with larger telescopes are required.

The halo structure consists of old, metal-deficient stars and
extends to distances of 100 kparsec around the disks. The formation timescale
is estimated to be one or at most a few billion years after the Big
Bang. About 150 globular clusters have been found in the Galaxy, which 
also belong to the old population of the Galaxy and are believed to be
related to the formation of the halo and bulge structures.

There are tens of satellite dwarf galaxies around the Milky Way, which
probably formed and evolved by interactions with
the Galaxy. Spectroscopy for individual stars in such dwarf galaxies
has revealed significant $s$-process contributions in some
cases, providing observational constraints on the $s$ process at
different metallicities and their roles in chemical enrichment
\citep{THT09}.

\subsubsection{$s$-Process contributions to Galactic field stars \label{sec4E1}}

Because the lifetimes of low- to intermediate-mass stars, the progenitors of
AGB stars that are responsible for the main component of the
$s$ process, are longer than for massive stars ($M\gtrsim 8-10 M_{\odot}$),
it is not expected that the main $s$ process deriving from AGB stars 
of all masses and metallicities contributed to the chemical evolution of 
the early Galaxy for  [Fe/H] $< -1.5$ \citep{TGA04}. 

As for  the 
weak $s$ process, which is to be ascribed to the presupernova evolution of 
massive stars, we point out that, besides the present MACS uncertainties 
in the region below $A = 90$ (see Sec. \ref{sec3B}), neutron capture in 
massive stars is driven by the $^{22}$Ne($\alpha, n$)$^{25}$Mg reaction, 
where $^{22}$Ne acts as a secondary-like source. In fact $^{22}$Ne 
comes from the original CNO abundances, which are transmuted into $^{14}$N 
during H burning and then converted to $^{22}$Ne by the chain
$^{14}$N($\alpha, \gamma$)$^{18}$F($\beta^+\nu$)$^{18}$O($\alpha, \gamma$)$^{22}$Ne 
in the early phase of core He burning (\citet{RGB93} and references therein). 
Consequently, also the  weak $s$ process is not expected to play any role in 
Galactic Halo stars. The weak $s$ process is also believed to occur in massive
stars (Sec. \ref{sec3C}). However, the process requires high metallicities and 
is, therefore, effective only in young, metal-rich stars. 
Although the abundance patterns produced by the weak $s$ 
process are not clearly found in stellar atmospheres, some 
constraints have been obtained from Cu, Zn, Ga, and Ge abundances 
in Galactic stars \citep{PGH10}. 

With respect to Zn, the most abundant isotope $^{64}$Zn (48.6\% of 
solar Zn) derives from SNe of Type II in the $\alpha$-rich freezout 
of neutrino winds \citep{WoH92} or in hypernovae \citep{UmN02} while 
the other 50\% of solar Zn are almost fully ascribable to the weak
$s$ process (Bisterzo et al. 2004). Assuming that SNII produce about 1/3
of solar Fe, this means that the ratio [Zn/Fe] in the Halo should be
a bit positive, about 0.2 dex on average, consistent with spectroscopic
observations \cite[e.g. ][]{CDS04}.

Concerning Cu, the weak $s$ process accounts for 90\% and the main 
$s$ component for 5\% of the solar abundance, whereas SN Ia are not 
predicted to contribute any Cu \citep{TNY86}. This implies that [Cu/Fe] 
in the halo should be constant and strongly negative, around $-$0.8 dex. 
The origin of this small primary Cu contribution may be ascribable to 
the explosive nucleosynthesis in massive stars. These expectations are 
confirmed by theoretical expectations of \citet{WoW95} for a range of massive 
stars with metallicities from 0 to solar. A large number of
spectroscopic observations of [Cu/Fe] versus [Fe/H] exist in the
literature, again confirming the above expectations 
\cite[e.g. ][]{BGP04,RoM07,PGH10}.

The nucleosynthetic origin of primary Zn and Cu is still a subject of
debate. This may imply different processes, e.g. the so-called $\nu p$
process \citep{FML06}. Other light elements beyond the Fe
group, like Ga and Ge, should behave like Cu, that is with a major weak
and secondary-like $s$-process contribution \citep{PGH10}.
For the few spectroscopic observations available in Halo stars, we
refer to \citet{CSB05}.

Although the astrophysical sites of the
$r$ process are not well identified, massive stars that terminate their
lives by core-collapse supernovae would be promising candidates for the 
progenitor. Due to the short lifetimes of massive stars it appears 
plausible that the $r$ process contributed significantly  
to the enrichment of the early Galaxy.
Apart from the $r$ process and weak $s$ process,
recent studies on light neutron-capture elements suggest a LEPP 
as an additional source of these elements in the very early 
Galaxy \cite{TCP02, TGA04, AHB05} as discussed in Sec. \ref{sec3}.

The abundances of Ba (or La) and Eu are used as indicators of the $s$-
and $r$-process contributions to the origin of the heavy elements. 
Analyses of the abundances in the Solar System showed that about 80\% 
of Ba \citep{TGG99,SGT09} and 70\% of La \citep{WDH06} originate from 
the $s$ process, while about 95\% of Eu comes from the $r$ process 
\cite{KBW89, AKW99b}.

Heavy neutron-capture elements of metal-poor stars in the halo
have been recently studied with high resolution spectroscopy
\cite{MPS95, RNB96, BPA00, SSC04, HAK04, AHB05, FDH07, SCG08}.  On the other hand, 
the chemical composition of a large sample of disk stars has been 
studied in the past two decades \cite[e.g. ][]{EAG93}. Recently, 
\citet{RTL03} and \citet{RLA06} determined abundances of many 
elements including Y, Ba and Eu for 181 thin disk stars and 95 
thick disk stars, respectively, based on high quality spectra. 

\begin{figure}
  \centering
\includegraphics[angle=0,width=13cm]{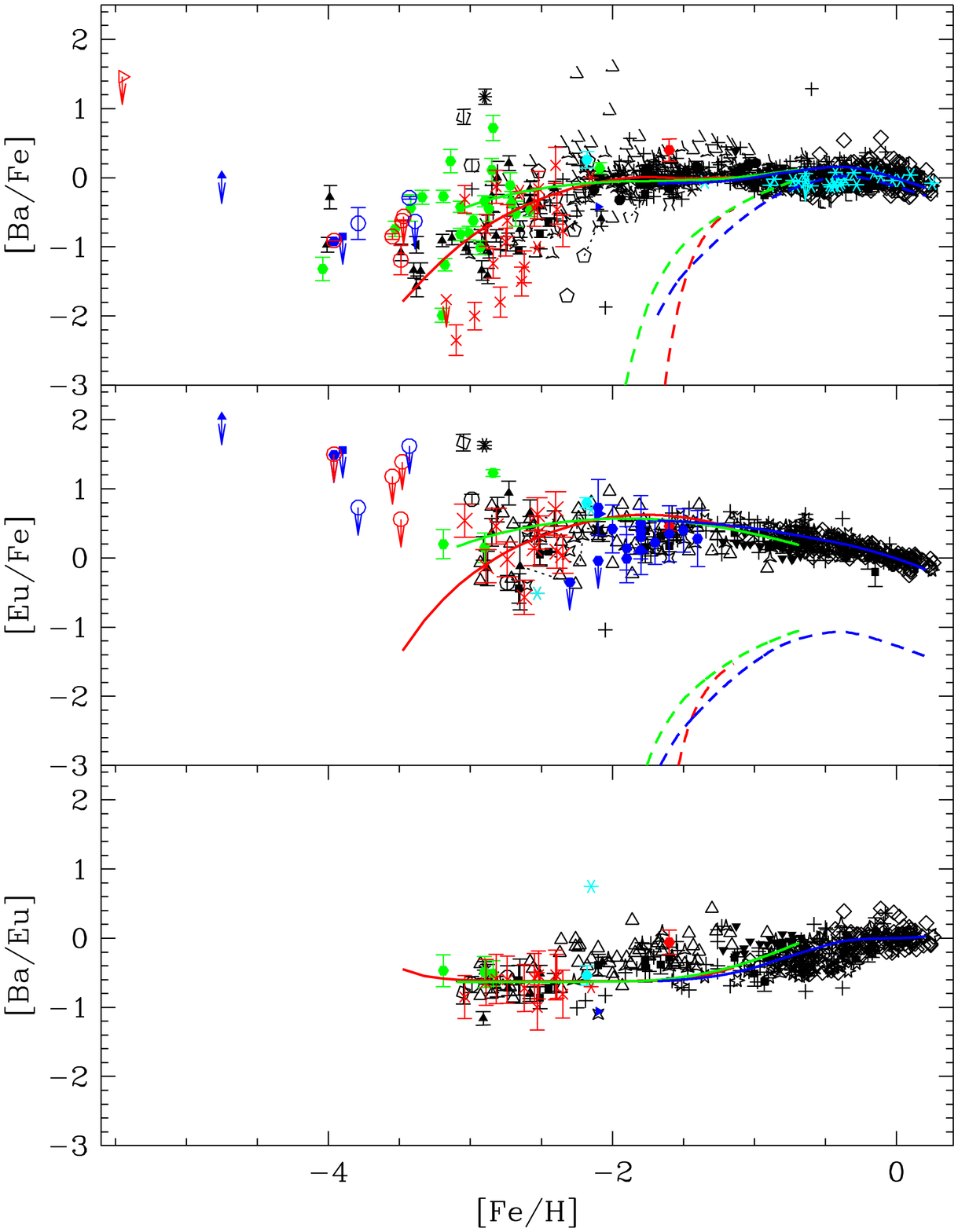}
\caption{(Color online) Top panel:  The evolution of the $s$-process 
fractions of [Ba/Fe] versus [Fe/H] in the Galactic halo as well as 
in the thick and thin disk (dashed lines) and theoretical predictions 
of the total $s/r$ abundances (solid lines) from \citet{TGA04,SGT09}. The spectroscopic 
data from observations of Galactic disk and halo stars are collected 
from the literature \citep{EAG93,GrS94,MPS95,Mcw98,JMN99,ToL99,BPA00,Ful00,
NRB01,MaG01,MiK01,MaZ06,ISS06,ABS08,AoH08,LBJ08,CMC07,NCK07,FCN07,MZG08,
RLS08,AHB05,FDH07,CCM08,AFC06,YGG05,VGJ03,CSB02}. Error 
bars are plotted only when reported for individual objects by the authors. 
The dotted line connects a star observed by different authors.
Analogous plots are shown for [Eu/Fe] (middle panel) and [Ba/Eu] 
(bottom panel).\label{fig:19}}
\end{figure}

The elemental abundance ratios for Ba and Eu, as a function of 
metallicity, are shown in Fig. \ref{fig:19}, which includes also 
examples for theoretical GCE expectations. The distributions were 
obtained by considering the $s$ and $r$ components separately. 
The Galactic abundances of these elements were computed from 
the sum of both processes by comparing model results \citep{SGT09,TGA04}
with the available spectroscopic observations of field stars at 
different metallicities. For the $s$ component of each isotope 
at the epoch of the formation of the Solar System only the 
contributions of AGB stars are considered. Subsequently, the 
$r$-process residual method ($r = 1- s$) was used to determine 
the respective solar system $r$-process fractions, assuming 
that the production of $r$ nuclei is a primary process occurring 
in Type II supernovae, independent of metallicity. Given the 
present theoretical uncertainties of the $r$-process modeling, the 
$r$-process residual method is to be considered as an approximation 
of the fractional $r$ abundances in Solar System material. It does 
not exclude, however, that some spread in the $r$-process ratios, 
e.g. [Ba/La]$_r$, [Ba/Eu]$_r$, [Eu/Pb]$_r$, [Eu/Th]$_r$, may exist 
in low metallicity stars, indicating a multiplicity of $r$-process 
components. Significant examples concerning [Eu/Th]$_r$ are 
already reported from spectroscopic observations \citep{PHC04,RKF09}.
Concerning the ratio [Ba/Eu]$_r$, a certain spread is apparent in 
Fig. \ref{fig:19}, but a more detailed analysis of the
spectroscopic data and the related uncertainties reported by different 
authors might be be necessary. From the theoretical point of view, a 
range of $r$-process predictions has been advanced in recent works 
by \citet{KFP07a,FKP10}.

The top and middle panels refer to the typical $s$- and $r$-process 
elements [Ba/Fe] and [Eu/Fe], whereas the bottom panel shows their 
ratio [Ba/Eu]. Theoretical GCE expectations 
using only the $s$-process products from AGB stars in the Galactic 
halo as well as in the thick and thin disk are separately indicated 
by dashed lines. Theoretical predictions of the total ($s + r$) yields 
are shown as solid lines. We recall that the elemental composition 
of the $r$ process has been obtained via the $r$-residual method 
described before. Below [Fe/H] $< -1.5$ the $r$ process dominates 
the theoretical expectations. According to our prescriptions the $r$ 
process is to be considered of primary origin, i.e. to originate 
from reactions starting from H and He. However, as discussed in 
\citet{TGA04}, calculations of the GCE trend versus metallicity have 
been made assuming that only a small range of massive stars, with initial 
masses of $8-10$ $M_\odot$ are involved in the $r$-process production. 
This implies that [Ba/Fe] as well [Eu/Fe] decrease below [Fe/H] $< -2.3$, 
but also other choices may be invoked to explain the general decrease of 
spectroscopic observations. As to the large observed spread of [Ba,Eu/Fe],
an easy explanation could be that at those metallicities the interstellar 
medium in the Halo was not fully homogenized (Travaglio, Galli and Burkert 2001, Raiteri et 
al. 1999,  Ishimaru and Wanajo 1999, Ishimaru et al. 2004). Notice that 
the observed ratio [Ba/Eu] stays almost flat (bottom panel of
Fig. \ref{fig:19}). In conclusion, the [Eu/Fe] shown in Fig. \ref{fig:19} is 
basically explained by the $r$ process. In the metal-rich range, a 
decreasing trend of [Eu/Fe] with increasing metallicity is found as 
in the case of [$\alpha$/Fe] \cite[e.g. ][]{Mcw97}. Actually, the 
decreasing trend of [Eu/Fe] in the disk versus higher metallicities 
is not due to a decreasing efficiency of the $r$ process, but to an 
increasing apport of Fe in the interstellar medium from the long-lived 
SNe of type Ia. By contrast, the model curves for [Ba/Fe]  
(top panel of Fig. \ref{fig:19}) indicate that the main $s$ process 
is the dominant contributor to Ba in the metal-rich range ([Fe/H]$>-1$). 
Indeed, there is no decreasing trend of [Ba/Fe] with increasing 
metallicity. The models also indicate that the Ba in metal-poor stars 
is provided by the $r$ process as in the case of Eu. 

[La/Eu] is another indicator for the relative contributions of the two 
processes. \citet{SSC04} suggested a correlation 
with kinematical properties, because it was found that stars with high 
[La/Eu] show low velocities with respect to the Galactic plane. 
Therefore, these stars possibly indicate the existence of a substructure in the 
halo to which the $s$ process has contributed. In view of the long 
time scale of the $s$-process effects, such a substructure has an 
impact on the formation history of the halo.

On average, stars in the thick disk have lower metallicity than thin 
disk stars. Low metallicity thick disk stars have 0.3-0.5 dex higher
[Eu/Fe] than the solar value, while [Ba/Fe] is slightly lower. The 
overabundance of Eu implies that the formation time scale of the thick 
disk is shorter than that of type Ia supernovae (approximately a few 
billion years), which is also supported by the abundance ratios 
between $\alpha$ elements and iron. The [Ba/Eu] ratios of $-$0.5 to $-$0.7 
agree with the $r$-process composition, indicating that the dominant 
source of heavy neutron-capture elements in thick disk stars is indeed 
the $r$ process. The absence or minor contribution of the $s$ process 
supports the rapid formation of the thick disk. However, there are 
several thick disk stars with higher metallicity than thin disk stars,
which have similar [Ba/Fe] to solar. Their existence indicates that star formation 
continued longer in the thick disk, although the fraction of such 
stars is quite small.

\begin{figure}
  \centering
\includegraphics[width=11cm]{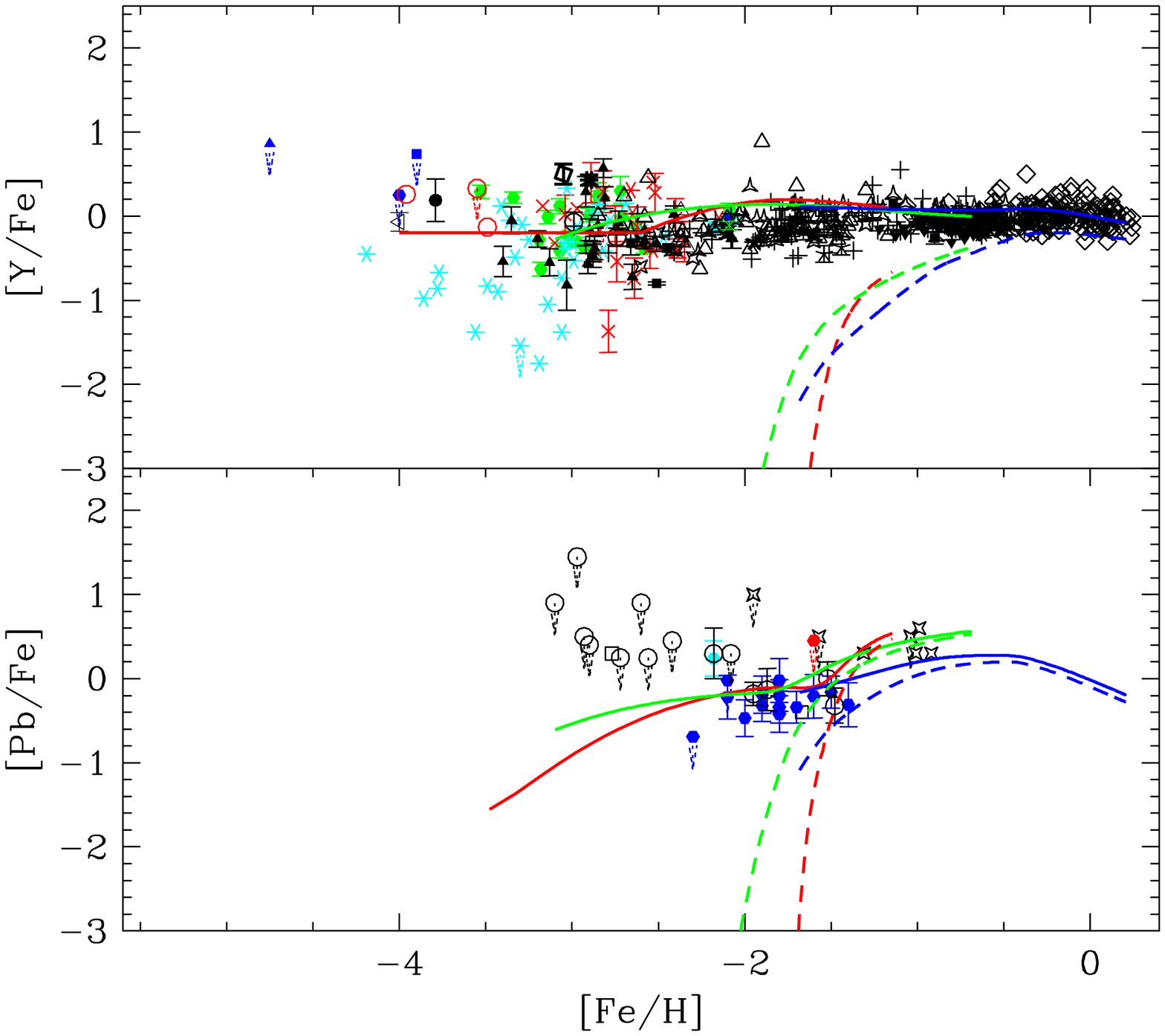}
\caption{(Color online) Same as Fig. \ref{fig:19}, but for [Y/Fe] (top)
and [Pb/Fe] (bottom). The observational data of Y are from
the references given in Fig. \ref{fig:19}. The Pb data are from 
\citet{SCB98,TGB01b,RLS08,RKF09,ISS06,AoH08}.}
\label{fig:20}
\end{figure} 

Fig. \ref{fig:20} shows the light and heavy $s$-process elements
[Y/Fe] and [Pb/Fe] as a function 
of metallicity. Pb, which is formally produced by the $s$ process in 
low metallicity AGB stars (Sec. \ref{sec3E}), represents a good $s$-process indicator. 
\citet{TGB01b} provided model predictions of Pb enrichments in the 
Galaxy compared to observations in several metal-poor stars, but the 
observational constraint was rather weak due to the small sample 
size. Recently, Pb abundances of 12 giants in the halo were determined 
by \citet{ABS08}. The [Pb/Fe] and [Pb/Eu] ratios of these stars are 
constant within the observational uncertainties, and the [Pb/Eu] 
values agree with the inferred $r$-process component in the Solar 
System, clearly indicating that there is no significant $s$-process 
contribution in these stars (see also \citet{RKF09} for the latest 
compilation). However, the sample for [Fe/H]$>-1.5$, where the 
$s$ process is expected to become significant, is still scarce.
Further measurements of Pb abundances for halo stars, particularly 
in those stars with high [La/Eu], will be useful for understanding 
of the formation time scale of the halo.

The situation is more complicated in the enrichment of the light
neutron-capture elements. Fig. \ref{fig:20} (top panel) shows [Y/Fe] as 
an example. As discussed in Sec. \ref{sec3E}, the Galactic chemical 
enrichment of Y by the apport of all previous generations of AGB stars 
at the epoch of the solar system formation explains only 67\% of the 
solar Y abundance \citep{TGA04, SGT09}. According to that discussion, 
the solid line in the plot includes contributions of 67\%, $5-10$\%,
about 8\%, and $15-20$ \% from the main and weak $s$ process, from 
the $r$ process, and from the LEPP, respectively. The flat theoretical 
GCE prediction of [Y/Fe] = $-$0.2 dex in the Halo is based upon the 
assumption that the primary LEPP component is obtained in all massive 
stars, which are exploding as SNII. We recall instead that the primary
$r$ process was assumed to derive from a small range of massive stars,
between 8 and 10 $M_\odot$ (see discussion for the [Eu/Fe] trend and
Fig. \ref{fig:19}. The [Y/Fe] in extremely metal-poor stars ([Fe/H]$<-3$) 
shows very large scatter, indicating a diversity of the relative 
contributions by the LEPP and the $r$ process, as well as incomplete 
mixing in the gas cloud from which these stars have formed.

The Y abundances of thick disk stars are similar to the stars in the
thin disk. However, the origin of Y would be significantly different, 
given the different [Y/Eu] ratios of the two structures. The dominant
source of this element in thin disk stars, as in the Solar System, is
the (main) $s$ process \cite{TGA04}. On the other hand, the $r$ process 
and the LEPP should be significant sources of Y in thick disk stars. 
The thick disk is currently assumed to have suffered a rapid evolution
and a high star formation rate, such that contributions from SNIa or 
from the main $s$ process by AGB stars should be absent. Due to the
high star formation rate, however, higher [Fe/H] values are observed 
compared to the Halo. Indeed, [Eu/Fe] is almost flat with an average 
of 0.5 dex, the same as observed in the Halo. A similar behavior is 
found for the so-called $\alpha$-enhancement. Therefore, we may
expect [Y/Fe] = $-$0.2 dex, as discussed before in the analysis of 
Fig. \ref{fig:20} (top panel). Consequently, the small variation 
of [Y/Fe] in thick disk stars with respect to the thin disk is not 
surprising.

\subsubsection{Globular clusters and galaxies in the local group \label{sec4E2}}

The Ba and Eu abundances measured in globular clusters have been 
summarized by \citet{GSC04}. Fig. 6 of their paper indicates
that there are no significant variations in the abundances of Ba and 
Eu compared to the observational uncertainties. One remarkable 
exception is the metal-poor cluster M15, which shows significant 
star-to-star scatter of Ba and Eu abundances \citep{SCK97}, presumably as the 
result of "local" massive star nucleosynthesis. In general, however,
globular clusters are modestly $r$-process-rich compared with 
Solar System material.  

There are several exceptional clusters that exhibit significant
$s$-process contributions. A few clusters show variations in metallicity 
and abundance ratios. Among those, $\omega$ Cen represents a remarkable
case. This cluster has a metallicity distribution between [Fe/H]$=-1.8$
and $-0.8$, implying chemical evolution inside the cluster. Studies of
neutron-capture elements for a significant sample of cluster
stars by \textcite{NoD95} and \textcite{SSC00} have shown that the 
[Ba/Fe] and [La/Fe] ratios increase with increasing metallicity. 
This behavior can be explained by the contributions of the $s$ process 
in 1--3~M$_{\odot}$ AGB stars. Accordingly, star formation in this 
cluster continues over a period, which is longer than the lifetime 
of these stars, i.e. longer than one billion years. 

Other exceptions are clusters without significant variation of chemical
abundances but with an enhancement in $s$-process elements. A well known
example is M4 ([Fe/H]$=-1.2$), where [Ba/Fe] is 0.6~dex higher than
in other clusters with similar metallicity \cite{ISK99}. The fact
that no variations of metallicity and chemical abundance ratios are observed
suggests that the origin of these heavy elements is primordial. In 
other words, the cluster forming cloud was already polluted by 
the ejecta of previous generations of stars. This is a strong constraint 
for the formation timescale of the cluster. 

A few other clusters are also suggested to have excesses of $s$-process
elements, though they are not as clear as in M4. A good probe of the
$s$-process contribution is Pb. Abundances of this element have been
measured in four clusters including M4 by \textcite{YAL06} and
\textcite{YLP08}. In M4, Pb was found to be enhanced as expected
from other $s$-process elements, but no such excess was found for 
three other clusters, which showed Pb abundances compatible with that in
halo stars \cite{ABS08}.
 
Heavy element abundances have been measured for bright stars
(supergiants and stars at the red giant branch tip) in the Magellanic
clouds, the irregular satellite galaxies of the Milky Way. Recently,
\textcite{PHS08} reported chemical abundances of a large sample of
disk stars in the Large Magellanic cloud. Y, Zr, Ba, and La abundances
were determined for 30--50 red giant stars covering a metallicity 
range of $-1.3<$[Fe/H]$<-0.3$. The heavy neutron-capture
elements Ba and La turned out to be overabundant, while the ls elements
Y and Zr are underabundant. Although the measurement of Eu was not 
available for the sample, Ba and La are likely of $s$-process origin. 
The high abundance ratios between the hs and ls elements are compatible 
with the model prediction for the $s$ process in metal-deficient AGB stars.

Measurements of heavy elements in stars of dwarf spheroidal galaxies
around the Milky Way have been made in the past decade 
\cite[e.g. ][]{SCS01,SVT03,SAI04}. Although the sample size is still 
small, in particular for the measurement of Eu abundances, the 
neutron-capture elements in metal-deficient ([Fe/H]$<-1.0$) stars 
in dwarf galaxies can be explained by the $r$ process, while 
$s$-process contributions are seen in metal-rich stars of some 
galaxies, e.g. in Fornax and Carina \citep{SVT03,THT09}. The $s$-process 
contributions from low- and intermediate-mass AGB stars are an 
important probe for the star formation history of galaxies. The 
$\alpha$/Fe abundance ratios of stars in dwarf galaxies are lower 
than those of halo stars in general, indicating the possible role 
of type Ia supernovae. This is unlikely the case in galaxies 
without significant $s$-process contribution, because the time scales 
for type Ia supernovae are longer than the evolution of  
intermediate-mass stars. 

\section{Summary}
 
The $s$-process, which is ascribed to low mass stars during the TP-AGB 
phase (main and the strong components), and to massive stars (the weak  
component), is discussed with respect to the underlying nuclear physics,
current stellar models, and the rapidly growing observational evidence.

From the nuclear physics side, there is an increasingly complete set of 
neutron capture measurements that provide the necessary Maxwellian averaged 
cross sections for detailed network calculation of the $s$-abundance patterns 
of the various scenarios. Over the last two decades considerable experimental 
progress was achieved on the basis of new and improved neutron facilities
and detector developments. Although the productive facilities at Oak Ridge 
and Karlsruhe have been closed recently, replacement became available through
intense pulsed neutron sources using spallation reactions (n\_TOF at 
CERN, J-PARC in Japan, and LANSCE at Los Alamos) or the $^{7}$Li($p,
n$) reaction (FRANZ at Frankfurt/Germany and SARAF at the Weizmann Institute 
in Jerusalem, both under construction). Apart from the high fluxes,
which these facilities have in common, they exhibit widely complementary 
characteristics, thus providing promising solutions for a variety of improved 
TOF measurements. Such measurements benefit also from developments in detector 
technology, aiming at higher efficiency (total absorption calorimeters) or 
minimized neutron sensitivity. Combined with new data acquisition systems 
and rapidly growing computing power, a new generation of experiments has
already provided a number of very accurate cross sections at astrophysically
relevant energies. In parallel, the activation method proved to play
an important role because of the superior sensitivity, which enabled
first measurements on unstable branch-point nuclei along the $s$-process 
path.

Future efforts in $s$-process experiments would be most useful in the following 
areas: Improvements in the accuracy of ($n, \gamma$) cross sections are needed 
in mass regions where present uncertainties are still exceeding the $3-5$\% level,
i.e. around magic neutron numbers, in the Fe-Sr region, and for the lighter 
elements. The persisting problem of the cross sections for the neutron source 
reactions $^{13}$C($\alpha, n$)$^{16}$O and $^{22}$Ne($\alpha, n$)$^{25}$Mg in 
the respective Gamow windows is still unsolved. Together with the abundant 
light elements, which act as neutron poisons, the source reactions 
determine the $s$-process neutron balance and represent, therefore, important 
constraints for stellar models, i.e. for the role of the $^{13}$C pocket in 
thermally pulsing low-mass AGB stars. The scarce information for ($n, \gamma$)
cross sections of unstable isotopes, which are crucial for the analysis 
of $s$-process branchings, need to be completed. The feasibility of such 
measurements will benefit from progress in neutron facilities and advanced
experimental techniques. Ultimately, they will also be needed to treat the 
extended reaction paths into the neutron rich region, which follow from 
the high neutron densities during C shell burning in massive stars and 
during the first, strong pulse in low metallicity AGB stars.

Last but not least, theory remains indispensable for complementing the
experimental ($n, \gamma$) information, either by closing gaps in the 
data, where measured cross sections are not (yet) available or by 
providing stellar enhancement factors to correct the laboratory results
for the effect of thermally populated excited nuclear states in the hot
stellar plasma. Correspondingly, the even more pronounced enhancement of
the weak interaction rates as a function of neutron and electron density
at the $s$-process sites remains an important domain of theoretical
studies, especially because experimental work in this field had been long 
neglected. 

The weak $s$ process, which produces a large fraction of the $s$ isotopes  
between Fe and Sr during convective core He burning and subsequent 
convective shell C burning, is of secondary nature. The neutron source 
is driven by ($\alpha, n$) reactions on $^{22}$Ne deriving 
from the conversion of initial CNO nuclei to $^{14}$N during core H 
burning via the sequence $^{14}$N($\alpha, \gamma$)$^{18}$F($\beta^+ \nu$)$^{18}$O, 
and subsequently by $^{18}$O($\alpha, \gamma$)$^{22}$Ne reactions at 
the beginning of convective core He burning, when the central temperature 
raises above $2.5 \times 10^8$ K. The weak $s$ 
contribution to the solar abundances is not easy to estimate in a 
quantitative way due to the present uncertainties of the stellar cross
sections in the range from Fe to Se and to the physical uncertainties 
in the treatment of the pre-explosive and explosive nucleosynthesis in 
supernovae. However,  half of solar Zn and about $70-80$\% of solar Cu, 
Ga, Ge, and As are to be ascribed to the weak $s$ process. The interplay 
between theory and spectroscopic observations is briefly discussed.  
 
All $s$ isotopes beyond $A = 90$ and of about half of solar Pb are 
contributed by the main $s$ process. The second half of solar Pb is 
produced by low mass AGB stars at low metallicities (strong component).
Below $A=90$ the contribution of the main component to the $s$-process 
abundances decreases rapidly. The main $s$ process is not a unique 
process, but depends on the initial mass, metallicity, the strength 
of the $^{13}$C-pocket, the efficiency of the TDU, and the choice of 
the mass loss rate. Stellar models could 
be verified by comparison with a large body of data obtained from analyses 
of presolar material in form of circumstellar dust grains and by the 
conspicuous number of observations of MS, S, C(N) and Ba stars of the 
Galactic disk as well as of CH stars in the halo. The $s$-process
contribution to the cosmic abundances in the interstellar medium in the 
mass range $A > 90$ is the result of all previous generations of AGB 
stars that polluted the interstellar medium before the formation of the 
Solar System. 

The impact of the chemical evolution of the Galaxy is analyzed for the 
representative elements Y, Ba, and Pb corresponding to the $s$-abundance 
peaks at magic neutron numbers 50, 82, and 126. The origin of the heavy 
neutron-capture elements is partly due to the main $s$ process and partly 
to the primary $r$ process. Usually the $r$-process contribution to each 
isotope in the solar system is estimated using the so-called  $r$-residual 
method by subtracting the well defined $s$-process components from the
respective solar abundances. This approach provides a fair basis for 
comparison with the still uncertain predictions from current
$r$-process models.
 
Progress in stellar modeling depends on a continuous interplay between 
theory and observation. This is particularly true for recent observations 
of the rare class of carbon-enhanced metal-poor stars with $s$-process 
enhancements, the CEMP-$s$ stars, which are main sequence, turnoff, or 
giant stars of low mass ($M \sim 0.8 M_\odot$) in close binary systems. 
The primary more massive companion (now a white dwarf) evolved along the 
AGB and polluted the envelope of the observed star with C and
$s$-process elements when it lost its entire envelope at the end of 
the AGB phase.  
 
A strongly debated issue is the subclass of CEMP-$s/r$ stars, showing  
$s$- and $r$-process contributions at the same time, although both 
processes are of completely different astrophysical origin. Some of them 
are highly enhanced in Ba, Ce, and La, which belong to the second 
$s$-process peak at $N=82$, as well as in Eu, which is a typical 
$r$-process element. In fact, the $s$ and $r$ elements beyond Ba are 
enhanced at the same level in these very metal-poor stars. A plausible 
scenario considers the formation of binary systems in giant clouds that 
were locally polluted by the ejecta of type II supernovae. 
 
Apart from the contributions from the weak and main $s$ process and from
the $r$ process, the light $s$-process elements Sr, Y, and Zr in the solar 
system contain an additional component contributed by a primary source of 
still unknown origin, the Light Element Primary Process (LEPP). The different 
proposed hypotheses, all related to the most advanced phases 
of pre-explosive and explosive nucleosynthesis in massive stars, represent
a most relevant issue of present nucleosynthesis research. 

These intriguing problems have been recognized by the rapid increase of 
observational data in the past decade, using high-resolution and high 
signal-to-noise spectra. Further chemical abundance studies, in particular 
isotope abundance measurements for key elements, will provide useful hints 
and constraints for understanding the physical processes behind these 
unsolved problems.

Further improvement of the models related to the weak, main, and strong 
$s$-process components coming from massive and intermediate-to-low mass stars, 
will have a strong impact on studies of the Milky Way and surrounding 
smaller galaxies. Great efforts have been made to understand the chemical 
evolution and formation history of these galaxies, which are  traced by 
chemical abundance ratios as well as kinematic properties of individual 
stars. Abundance ratios of $s$-process elements provide useful 
constraints for the chemical evolution models, i.e. on the time scale of the 
star formation history and on the initial mass function.

\section{Acknowledgements}

We are deeply indebted to O. Straniero and S. Cristallo for 
continuous clarifying discussions concerning the modeling of 
AGB stars. We would like to thank also the referees for their
suggestions, which led to a substantial improvement of the 
manuscript.

\newcommand{\noopsort}[1]{} \newcommand{\printfirst}[2]{#1}
  \newcommand{\singleletter}[1]{#1} \newcommand{\swithchargs}[2]{#2#1}

\begin{table}[htb]
\caption{MACS results at $kT$=30 keV for elements with a pair of 
$s$-only isotopes. (Data are from KADoNiS \citep{DHK05}, 
see Sec. \ref{sec2D}).
\label{tab1}}
\begin{ruledtabular}
\begin{tabular}{c c c c c } 
$s$-only isotopes & \multicolumn{2}{c}{Maxwellian averaged cross section (b)} 
		 		  & Abundance ratio \cite{RoT98}    & Branching ratio \\
\hline
$^{80,82}$Kr       & 267$\pm$14    &  90$\pm$6     & 2.28/11.58  & 0.61$\pm$0.05  \\  
$^{122,124}$Te     & 295$\pm$3     & 155$\pm$2     & 2.55/4.74   & 1.06$\pm$0.02  \\  
$^{128,130}$Xe     & 262.5$\pm$3.7 & 132.0$\pm$2.1 & 1.92/4.08   & 0.96$\pm$0.02  \\  
$^{134,136}$Ba     & 176.0$\pm$5.6 & 61.2$\pm$2.0  & 2.417/7.854 & 0.94$\pm$0.04  \\  
$^{148,150}$Sm     & 241$\pm$2     & 422$\pm$4     & 11.24/7.38  & 0.88$\pm$0.01  \\  
\end{tabular}
\end{ruledtabular}
\end{table}

\newpage
\begin{table}[htb]
\caption{Recent MACS results at $kT$=30 keV obtained via TOF techniques 
and with the activation method. (Data are from KADoNiS \citep{DHK05}, 
see Sec. \ref{sec2D}).
\label{tab2}}
\begin{ruledtabular}
\begin{tabular}{c c l c l} 
Target isotope & \multicolumn{4}{c}{Maxwellian averaged cross section (b)}      \\
               & \multicolumn{2}{c}{TOF technique}  & \multicolumn{2}{c}{Activation method}\\
\hline
$^{58}$Fe      & 12.1$\pm$1.1  & \citep{KWH83}       & 13.5$\pm$0.7  & \citep{HKU08a}      \\  
$^{59}$Co      & 38$\pm$3  	   & \citep{SpM76} 		 & 39.6$\pm$2.7  & \citep{HKU08a}  	   \\  
$^{62}$Ni      & 25.8$\pm$3.7  & \citep{ABE08} 		 & 20.2$\pm$2.1  & \citep{NPA05,Dil09a}\\  
      		   & 37.0$\pm$3.2  & \citep{TTS05} 		 & 23.4$\pm$4.6  & \citep{DFK10}  	   \\  
$^{87}$Rb      & 15.5$\pm$1.5  & \citep{JaK96b}		 & 15.8$\pm$0.9  & \citep{HKU08b} 	   \\  
$^{88}$Sr      & 6.01$\pm$0.17 & \citep{KWG00,KWG01a} & 6.13$\pm$0.18 & \citep{KZB90} 	   \\  
$^{89}$Y       & 21$\pm$3      & \citep{MAM78c}      & 19.0$\pm$0.6  & \citep{KZB90} 	   \\  
$^{139}$La     & 32.4$\pm$3.1  & \citep{TAA07} 		 & 31.6$\pm$0.8  & \citep{BDH03}  	   \\  
$^{146}$Nd     & 91.2$\pm$1.0  & \citep{WVK98a}		 & 87.1$\pm$4.0	 & \citep{TDK95}  	   \\  
$^{148}$Nd     & 146.6$\pm$1.9 & \citep{WVK98a}		 & 152$\pm$9  	 & \citep{TDK95}  	   \\  
$^{176}$Lu     & 1639$\pm$14   & \citep{WVK06a} 	 & 1599$\pm$85$^*$& \citep{BeK80}  	   \\  
$^{180}$Hf     & 156.5$\pm$1.9 & \citep{WVK06b} 	 & 168$\pm$9     & \citep{BKW82}  	   \\  
$^{197}$Au     & 588$\pm$20    & \citep{MHW75,Mac82a}& 592$\pm$9     & \citep{RaK88}  	   \\  
\end{tabular}
\end{ruledtabular}
$^*$Original value renormalized by \citet{BBK00}.
\end{table}

\begin{table}[htb]
\caption{Feasibility of future TOF measurements on unstable branch 
point isotopes at the FRANZ facility.
\label{tab3}}
\begin{ruledtabular}
\begin{tabular}{c cc l} 
Sample       & Half life      & $Q$ value        & Comment      \\
             & (yr)           & (MeV)            & 				\\
\hline
$^{63}$Ni      & 100.1        & $\beta^-$, 0.066 & TOF work in progress \cite{Cou09}, sample with low enrichment              		   \\  
$^{79}$Se      & 2.95 $10^5$  & $\beta^-$, 0.159 & important branching, constrains $s$-process temperature in massive stars   		   \\  
$^{81}$Kr      & 2.29 $10^5$  & EC, 0.322 	     & part of $^{79}$Se branching 	   			   			   	  		  		  		   \\  
$^{85}$Kr      & 10.73 	 	  & $\beta^-$, 0.687 & important branching, constrains neutron density in massive stars   		  		   \\  
$^{95}$Zr      & 64.02 d 	  & $\beta^-$, 1.125 & not feasible in near future, but important for neutron density low mass AGB stars   \\  
$^{134}$Cs     & 2.0652  	  & $\beta^-$, 2.059 & important branching at $A=134,135$, sensitive to $s$-process temperature   		   \\
               &              &                  & in low-mass AGB stars, measurement not feasible in near future   		  		   \\  
$^{135}$Cs     & 2.3 $10^6$   & $\beta^-$, 0.269 & so far only activation measurement at $kT=25$ keV by \citet{PAK04}  	   	  		   \\  
$^{147}$Nd     & 10.981 d 	  & $\beta^-$, 0.896 & important branching at $A=147/148$, constrains neutron density in low-mass 		   \\
               &              &                  & AGB stars 		   	  			   			  		  		  	 		  		   \\  
$^{147}$Pm     & 2.6234  	  & $\beta^-$, 0.225 & part of branching at $A=147/148$											  		   \\
$^{148}$Pm     & 5.368 d 	  & $\beta^-$, 2.464 & not feasible in near future  	   										  		   \\  
$^{151}$Sm     & 90    		  & $\beta^-$, 0.076 & existing TOF measurements, full set of MACS data available				  		   \\ 
			   &	 		  &  			   	 & \cite{WVK06c,AAA04c} 	   	   	   	  	   								  		   \\  
$^{154}$Eu     & 8.593  	  & $\beta^-$, 1.978 & complex branching at $A=154,155$, sensitive to temperature and neutron density 	   \\  
$^{155}$Eu     & 4.753  	  & $\beta^-$, 0.246 & so far only activation measurement at $kT=25$ keV by \citet{JaK95b}  	   		   \\  
$^{153}$Gd     & 0.658 		  & EC, 0.244 	     & part of branching at $A=154,155$  	   		 	 								   \\  
$^{160}$Tb     & 0.198 		  & $\beta^-$, 1.833 & weak temperature-sensitive branching, very challenging experiment 				   \\  
$^{163}$Ho     & 4570  		  & EC, 0.0026 	     & branching at $A=163$ sensitive to mass density during $s$ process,				   \\ 
               &              &                  & so far only activation measurement at $kT=25$ keV by \citet{JaK96a}  	   		   \\  
$^{170}$Tm     & 0.352 		  & $\beta^-$, 0.968 & important branching, constrains neutron density in low-mass AGB stars    	   	   \\  
$^{171}$Tm     & 1.921  	  & $\beta^-$, 0.098 & part of branching at $A=170,171$  	   		   	  		   	   					   \\  
$^{179}$Ta     & 1.82  		  & EC, 0.115        & crucial for $s$-process contribution to $^{180}$Ta, nature's rarest stable isotope  \\  
$^{185}$W      & 0.206 		  & $\beta^-$, 0.432 & important branching, sensitive to neutron density and $s$-process temperature 	   \\
               &              &                  & in low-mass AGB stars  	   	  	 		 		 	 			 				   \\  
$^{204}$Tl     & 3.78  		  & $\beta^-$, 0.763 & determines $^{205}$Pb/$^{205}$Tl clock for dating of early solar system    	   	   \\  
\end{tabular}
\end{ruledtabular}
\end{table}

\end{document}